\documentclass[10pt,brazil,english,american,aps,prb,floatfix,superscriptaddress,twocolumn,showpacs,amsmath,amssymb,reprint]{revtex4-1}
\usepackage[T1]{fontenc}
\usepackage[latin9]{inputenc}
\usepackage{color}
\usepackage{bm}
\usepackage{amsmath}
\usepackage{amssymb}
\usepackage{graphicx}
\usepackage{setspace}
\usepackage{esint}

\makeatletter

\providecommand{\tabularnewline}{\\}

\PassOptionsToPackage{caption=false}{subfig} 
\usepackage{hyperref}
\hypersetup{
breaklinks=true,
colorlinks=true,
citecolor=blue,
linkcolor=blue,
filecolor=blue,
urlcolor=blue
}
\IfFileExists{lmodern.sty}{\usepackage{lmodern}}{}

\makeatother

\usepackage{babel}
\begin{document}

\title{Random SU(2)-symmetric spin-$S$ chains}

\author{V. L. Quito}

\affiliation{Instituto de Física Gleb Wataghin, Unicamp, Rua Sérgio Buarque de
Holanda, 777, CEP 13083-859 Campinas, SP, Brazil}

\author{José A. Hoyos}

\affiliation{Instituto de Física de São Carlos, Universidade de São Paulo, C.P.
369, São Carlos, SP 13560-970, Brazil}

\author{E. Miranda}

\affiliation{Instituto de Física Gleb Wataghin, Unicamp, Rua Sérgio Buarque de
Holanda, 777, CEP 13083-859 Campinas, SP, Brazil}

\date{\today}
\begin{abstract}
We study the low-energy physics of a broad class of time-reversal
invariant and SU(2)-symmetric one-dimensional spin-$S$ systems in
the presence of quenched disorder via a strong-disorder renormalization-group
technique. We show that, in general, there is an antiferromagnetic
phase with an emergent SU($2S+1$) symmetry. The ground state of this
phase is a random singlet state in which the singlets are formed by
pairs of spins. For integer spins, there is an additional antiferromagnetic
phase which does not exhibit any emergent symmetry (except for $S=1$).
The corresponding ground state is a random singlet one but the singlets
are formed mostly by trios of spins. In each case the corresponding
low-energy dynamics is activated, i.e., with a formally infinite dynamical
exponent, and related to distinct infinite-randomness fixed points.
The phase diagram has two other phases with ferromagnetic tendencies:
a disordered ferromagnetic phase and a large spin phase in which the
effective disorder is asymptotically finite. In the latter case, the
dynamical scaling is governed by a conventional power law with a finite
dynamical exponent.
\end{abstract}


\maketitle

\section{Introduction}

Systems with random interactions comprise an important research field
in condensed matter physics. At the same time that some disorder is
unavoidable in experiments, its presence can completely change the
long wavelength behavior of the system, driving it through quantum
phase transitions (for a review see, e. g., references\ \onlinecite{LeeRMP1985,igloi-review,vojta-review06}).
Remarkably, the novel low-energy behavior of matter that appears in
the strong-disorder limit is typically very distinct from that of
the clean systems. A striking example is the low-energy behavior of
one-dimensional random spin chains. For sufficiently strong randomness,
the ground state of the antiferromagnetic (AF) Heisenberg spin-$S$
chain becomes a collection of nearly independent singlets of spin
pairs {[}see Fig.\ \foreignlanguage{english}{\hyperref[fig:random-singlets]{\ref{fig:random-singlets}(a)}}{]}:
the so-called random singlet state\textit{.\ \cite{fisher94-xxz}}
The energy spectrum associated with these singlets is extremely broad
and is responsible for singular response functions. The magnetic susceptibility
$\chi$, for instance, diverges as $\sim(T\ln^{1/\psi}T)^{-1}$, with
a universal (disorder-independent) tunneling exponent $\psi=1/2$.
In addition, the typical and average spin-spin correlation functions
behave quite differently. While the former one decays as a stretched
exponential $\sim e^{-r^{\psi}}$, where $r$ is the distance between
the spin in convenient units, the latter decays much more slowly,
as a power law $\sim r^{-4\psi}$. The fact that typical and average
values differ so much is the hallmark of phases governed by infinite-randomness
fixed points, a concept that could only be grasped after the development
of a strong-disorder renormalization-group (SDRG) method.\cite{madasguptahu,madasgupta,bhatt-lee}
In this method, one keeps track of the entire effective distribution
of energy and length scales under the renormalization group (RG) coarse
graining. In the vicinity of an infinite-randomness fixed point these
distributions tend to become infinitely broad along the RG flow. 

Later on, it was realized that very similar behavior would also appear
in other random spin chains, namely, at multicritical points of dimerized
spin-$S$ chains,\ \cite{PhysRevLett.89.277203} in AF SU($N$) spin
chains,\ \cite{HoyosMiranda} and in non-abelian anyonic SU(2)$_{k}$
spin chains.\ \cite{fidkowski-etal-prb09} The difference is that
$\psi$ now depends on other model details. In the first model, the
tunneling exponent is $\psi=1/N_{1}$, where $N_{1}$ is the number
of dimerized phases meeting at the multicritical point (the maximum
being $2S+1$). In the second model, $\psi=1/N_{2}$ where $N_{2}$
is the number of different spin representations describing the effective
spin degrees of freedom (the maximum being $N$). In the third model,
$\psi=1/N_{3}$ where $N_{3}=k$. In addition, the ground state of
these models is distinct from the usual pairwise random singlet state
because the singlets are now formed by a larger number of spins, such
as spin trios, quartets and so on {[}see Fig.\ \foreignlanguage{english}{\hyperref[fig:random-singlets]{\ref{fig:random-singlets}(b)}}{]}.

Recently,\ \cite{QuitoHoyosMiranda} we have shown that random spin-1
chains with bilinear and biquadratic SU(2)-symmetric interactions
harbor two types of random singlet phases: one in which the spin singlets
are formed by spin pairs and characteristic tunneling exponent $\psi=1/2$,
and another in which the spins are made in the great majority of spin
trios. In addition, $\psi=1/3$. More interestingly, these phases
exhibit emergent SU(3) symmetry.

Despite all these developments, we still do not have a simple criterion
to decide whether a given random spin chain model is in a certain
random singlet phase, or whether the random singlet state has an emergent
symmetry. In this paper, we investigate the most general SU(2)-symmetric
random spin-$S$ model with nearest-neighbor time-reversal-symmetric
interactions only. We show that, in the strong disorder limit, these
random singlet states are realized in this model. In general there
is a random singlet phase where the singlets are formed by spin pairs
only and the corresponding universal exponent $\psi=\frac{1}{2}$.
Strikingly, this pairwise random singlet exhibits an emergent SU($2S+1$)
symmetry. We show that this phase is characterized not by one, but
rather by $2S$ stable fixed points associated with the same pairwise
random singlet state. The difference between them stems only from
the structure of the low-energy excitations, which are all deformations
of the spectrum with exact SU($2S+1$) symmetry. Furthermore, the
presence of additional SU(2)-invariant couplings (with increasing
powers of the scalar products of spin operators) helps stabilize these
phases, providing them with a large basin of attraction.

In addition to this pairwise random singlet phase, we show that integer
spin-$S$ chains possess also another random singlet phase in their
phase diagram. This state is formed by spin trios and other multiples
of three, although in much less quantity, see Fig.\ \foreignlanguage{english}{\hyperref[fig:random-singlets]{\ref{fig:random-singlets}(b)}}.
Besides, it exhibits a universal exponent $\psi=\frac{1}{3}$. For
$S=1$, this state exhibits emergent SU(3) symmetry. For higher integer
spins, there is no such symmetry enhancement.

The experimental relevance of systems with a large number of internal
degrees of freedom stems from the possibility of their realization
in condensed matter systems or, more promisingly, in cold-atom systems
loaded in optical lattices. Candidates include alkali bosons, such
as $^{23}$Na and $^{87}$Rb, as well as alkali ($^{132}$Cs) or alkaline-earth
fermions ($^{9}$Be, $^{135}$Ba, $^{137}$Ba, $^{87}$Sr or $^{173}$Yb).
In the Mott insulating limit, the importance of spin-spin interactions
beyond the Heisenberg term has been discussed for these systems.\cite{PhysRevA.68.063602,wuetal2003,PhysRevLett.93.250405,Wumodphys06}
Furthermore, in some cases the systems possess a symmetry larger than
the usual SU(2). For example, in the case of spin-$\frac{3}{2}$ fermions,
an exact SO(5) symmetry has been discovered which requires no fine-tuning.\cite{wuetal2003,Wumodphys06}
In the case of alkaline-earth atoms, the total spin is of purely nuclear
origin and is decoupled from the remaining dynamics. It thus provides
the internal degrees of freedom that realize an SU($N$) symmetry,\cite{cazalillaetal2009,gorschkovetal2010}
as in the cases of $^{87}$Sr ($N=10$)\cite{zhangetal2014} or $^{173}$Yb
($N=6$).\cite{scazzaetal2014} An SU(6) Mott insulating state of
$^{173}$Yb atoms has been achieved,\cite{taieetal2012} although
lowering the temperature below the spin-exchange scale remains a challenge.
Two of us have already analyzed disordered SU($N)$-symmetric chains.\cite{HoyosMiranda}
The phases we will discuss include these enhanced symmetry points,
as well as others, as we will discuss.

The present manuscript is structured as follows. First, in Section~\ref{sec:Model},
we present the model and introduce the irreducible spherical tensor
notation in which our RG treatment is more natural. We also make a
connection with the projector notation which is often used by the
cold-atom community. The decimation steps of the SDRG procedure are
derived in Section~\ref{sec:Method}. This generalizes the methods
of Ref.\ \onlinecite{westerbergetal} from the generic random spin-$S$
Heisenberg chain to the generic random spin-$S$ time-reversal and
SU(2)-symmetric chain. We then provide a summary of the results of
the RG flow in Sec.\ \ref{sec:Technical-summary}. In Section~\ref{sec:Fixed-Points},
we discuss the RG fixed point structure. We determine that an phase
exists which has a random singlet ground state with emergent SU($2S+1$)
symmetry. We also find the other fixed point which happens only for
integer spin-$S$ chains. Finally, we briefly mention other fixed
points with FM tendencies. In Secs.\ \ref{sec:Spin-3/2-chain} and
\ref{sec:Spin-2-Chain}, we apply in detail our framework to the particular
cases of spin-$\frac{3}{2}$ and spin-$2$ chains, respectively. A
less technical summary of our main results is given in Sec\ \ref{sec:summary}.
In Sec.\ \ref{sec:Conclusions} we give a final summary and discuss
future directions that remain to be explored. Much of our more technical
developments are described in several Appendices.

\section{Model \label{sec:Model}}

A broad class of disordered time-reversal and SU(2) symmetric spin-$S$
Hamiltonians of a linear chain of $N_{{\rm sites}}$ sites (with periodic
boundary conditions) can be written as

\begin{equation}
H=\sum_{i=1}^{N_{{\rm sites}}}\sum_{J=0}^{J_{{\rm max}}}\alpha_{i}^{\left(J\right)}\left(\mathbf{S}_{i}\cdot\mathbf{S}_{i+1}\right)^{J},\label{eq:Hamiltonian_SU2}
\end{equation}
where $\mathbf{S}_{i}=\left(S_{i}^{x},S_{i}^{y},S_{i}^{z}\right)$
are the usual spin-$S$ operators at site $i$, $\alpha_{i}^{\left(J\right)}$
are exchange couplings that are taken to be independent random variables
distributed with the probability distributions $\overline{P}_{J}\left(\alpha\right)$.
The maximum power $J_{{\rm max}}$ for a spin-$S$ system is $2S$,
since larger powers of the spin operators can be written as linear
combinations of smaller ones.

For our renormalization-group transformations, it will be convenient
to rewrite the Hamiltonian in Eq.\ \eqref{eq:Hamiltonian_SU2} in
terms of irreducible spherical tensors (ISTs) $Y_{J,M}\left(\mathbf{S}_{j}\right)$
instead of powers of spin operators $S^{k}$.\cite{YangBhatt} The
ISTs can be defined via their commutation relations with the spin
operators (page 71 of Ref.~\onlinecite{Edmondsbook}) 

\begin{eqnarray}
\left[S_{i}^{\pm},Y_{J,M}\left(\mathbf{S}_{j}\right)\right] & = & \sqrt{J\left(J+1\right)\mp M\left(1\pm M\right)}\nonumber \\
 &  & \times Y_{J,M\pm1}\left(\mathbf{S}_{i}\right)\delta_{i,j},\label{eq:tensordef1}\\
\left[S_{i}^{z},Y_{J,M}\left(\mathbf{S}_{j}\right)\right] & = & M\,Y_{J,M}\left(\mathbf{S}_{i}\right)\delta_{i,j}.\label{eq:tensordef2}
\end{eqnarray}
Here, $Y_{J,M}(\mathbf{S}_{i})$ is an IST of rank $J$ with $2J+1$
components ($M=-J,-J+1,\ldots,J-1,J$) and they are functions of the
spin operators $\mathbf{S}_{i}$. From these commutation relations,
a recipe to construct them immediately follows. The idea is to start
with the operator of highest $M$,

\begin{equation}
Y_{J,M=J}\left(\mathbf{S}_{i}\right)\propto\left(S_{i}^{+}\right)^{J},\label{eq:highestIST}
\end{equation}
and make use of the commutation relations~\eqref{eq:tensordef1}
to lower the component index

\begin{equation}
Y_{J,M-1}\left(\mathbf{S}_{i}\right)=\frac{\left[S_{i}^{-},Y_{J,M}\left(\mathbf{S}_{i}\right)\right]}{\sqrt{J\left(J+1\right)-M\left(M-1\right)}}.\label{eq:irred_spherical_harm}
\end{equation}
Note that these ISTs resemble the spherical harmonics ${\cal Y}_{J,M}\left(\mathbf{r}\right)$.
There is, however, an important difference: if one wishes to ``promote''
spherical harmonics to ISTs, symmetrization is often required. For
instance, ${\cal Y}_{1,0}\propto\frac{z}{r}$ is promoted to $Y_{1,0}\propto S^{z}$,
but the $\frac{xz}{r^{2}}$ term of the spherical harmonic ${\cal Y}_{2,1}$
is promoted to $\frac{1}{2}\left(S^{z}S^{x}+S^{x}S^{z}\right)$. Since
the operator in Eq.~\eqref{eq:highestIST} is already symmetrized,
the recursive application of Eq.~\eqref{eq:irred_spherical_harm}
automatically leads to symmetrized IST operators. We will also use
the same IST normalization of Ref.~\onlinecite{Edmondsbook} (page
23).

There is another interesting aspect we wish to point out. The spin
operators themselves are obviously rewritten as first-rank ISTs. For
example, $S^{z}\propto Y_{1,0}$ and $S^{x}\propto\left(Y_{1,-1}-Y_{1,1}\right)$.
The dyadic term $S^{x}S^{x}$, however, is not, as it mixes ISTs of
different ranks. It has one component $\propto\left[Y_{1,-1},Y_{1,1}\right]\propto Y_{1,0}$
and another component $\propto Y_{1,1}^{2}\propto Y_{2,2}$. Evidently,
the former component transforms as a vector while the latter one transforms
as a second-rank tensor. Therefore, each power $(\mathbf{S}_{i}\cdot\mathbf{S}_{i+1})^{J}$
appearing in the Hamiltonian \eqref{eq:Hamiltonian_SU2} transforms
as linear combinations of tensors of ranks $J^{\prime}\leq J$. Rewriting
\eqref{eq:Hamiltonian_SU2} in terms of ISTs allows us to untangle
the different ranks. As we shall see, this is of fundamental importance
for the analysis of the SDRG flow.

We now come to the point of rewriting the Hamiltonian with this new
set of operators. The important step is to build rotation-invariant
two-site terms. We define the operator $\hat{O}_{J}$ as the scalar
product of IST operators of the same rank (page 72 of Ref.~\onlinecite{Edmondsbook})

\begin{equation}
\hat{O}_{J}\left(\mathbf{S}_{i},\mathbf{S}_{i+1}\right)\equiv\sum_{M=-J}^{J}\left(-1\right)^{M}Y_{J,M}\left(\mathbf{S}_{i}\right)Y_{J,-M}\left(\mathbf{S}_{i+1}\right).\label{eq:irred_tensor}
\end{equation}
One can explicitly check using Eqs.~\eqref{eq:tensordef1} and \eqref{eq:tensordef2}
that 

\begin{equation}
\left[S_{i}^{\pm}+S_{i+1}^{\pm},\hat{O}_{J}\right]=\left[S_{i}^{z}+S_{i+1}^{z},\hat{O}_{J}\right]=0,
\end{equation}
which shows that $\hat{O}_{J}$ is indeed rotation-invariant. For
example, the operator $\hat{O}_{2}$ is\cite{YangBhatt}

\begin{align}
\hat{O}_{2}\left(\mathbf{S}_{i},\mathbf{S}_{i+1}\right)= & \frac{15}{16\pi}\left(\mathbf{S}_{i}\cdot\mathbf{S}_{i+1}\right)+\frac{15}{8\pi}\left(\mathbf{S}_{i}\cdot\mathbf{S}_{i+1}\right)^{2}\nonumber \\
 & -\frac{5}{8\pi}\mathbf{S}_{i}^{2}\mathbf{S}_{i+1}^{2}.
\end{align}
In Appendix~\ref{sec:Appendix:notation}, we list all the $\hat{O}_{J}$'s
needed in this paper. The Hamiltonian in Eq.~\eqref{eq:Hamiltonian_SU2}
can thus be rewritten as

\begin{align}
H & =\sum_{i}H_{i,i+1}=\sum_{i=1}^{N_{{\rm sites}}}\sum_{J=0}^{J_{{\rm max}}}K_{i}^{\left(J\right)}\hat{O}_{J}\left(\mathbf{S}_{i},\mathbf{S}_{i+1}\right).\label{eq:Hamilt_tensors}
\end{align}
Obviously, the new coupling constants $K_{i}^{\left(J\right)}$ are
linear combinations of the original $\alpha_{i}^{\left(J\right)}$
and vice-versa (see Appendix~\ref{sec:Appendix:notation}). Additionally,
the distributions $\overline{P}_{J}\left(\alpha\right)$ determine
the distributions of $K_{i}^{\left(J\right)}$, ${\cal P}_{J}\left(K\right)$.
In this paper, we always work directly with ${\cal P}_{J}\left(K\right)$.

A third important form of the Hamiltonian involves the use of projection
operators onto states of well-defined total angular momentum $\tilde{S}$
of each pair of sites. The latter can be written for a pair of spins
as\ (page 38 of Ref.~\onlinecite{Edmondsbook})

\begin{align}
P_{\tilde{S}}(\mathbf{S}_{i}, & \mathbf{S}_{i+1})=\nonumber \\
 & \prod_{\sigma\ne\tilde{S}}\frac{2\mathbf{S}_{i}\cdot\mathbf{S}_{i+1}+2S\left(S+1\right)-\sigma\left(\sigma+1\right)}{\tilde{S}\left(\tilde{S}+1\right)-\sigma\left(\sigma+1\right)}\label{eq:projector_multiplets}\\
 & \phantom{\mathbf{S}_{i+1})}=\sum_{M=-\tilde{S}}^{\tilde{S}}\left|\tilde{S}M\right\rangle \left\langle \tilde{S}M\right|,
\end{align}
where $\sigma=\left|S_{i}-S_{i+1}\right|,\ldots,S_{i}+S_{i+1}$. It
is clear that $P_{\tilde{S}}\left(\mathbf{S}_{i},\mathbf{S}_{i+1}\right)$
selects, from all possible states of total angular momentum of the
pair, only the one equal to $\tilde{S}$. The generic SU(2)-symmetric
term for a pair of spins can thus be written as a linear combination
of these projectors and

\begin{equation}
H=\sum_{i}\sum_{J=0}^{J_{{\rm max}}}\epsilon_{i}^{\left(J\right)}P_{J}\left(\mathbf{S}_{i},\mathbf{S}_{i+1}\right).\label{eq:Hamilt_projectors}
\end{equation}
This third form is common when the spin Hamiltonian describes the
low-energy sector of cold-atom systems in optical lattices at commensurable
fillings. In such cases, if only $s$-wave scattering is retained,
then $\epsilon_{i}^{\left(J\right)}=0$ for odd $J$, as required
by the (anti-)symmetry of the wave function of a pair of (fermionic)
bosonic atoms \cite{CiobanuPRA2000,Wumodphys06}. Linear transformations
between $\epsilon_{i}^{\left(J\right)}$, $K_{i}^{\left(J\right)}$
and $\alpha_{i}^{\left(J\right)}$ are given in Appendix~\ref{sec:Appendix:notation}. 

For convenience, in Table~\ref{tab:Notation}, we summarize the notation
we use throughout the paper.

\begin{center}
\begin{table*}[t]
\begin{centering}
\begin{tabular}{|c|c|}
\hline 
Description & Notation\tabularnewline
\hline 
\hline 
Coupling constant of the term $\left(\mathbf{S}_{i}\cdot\mathbf{S}_{i+1}\right)^{n}$
{[}see Eq.~\eqref{eq:Hamiltonian_SU2}{]} & $\alpha_{i}^{\left(n\right)}$\tabularnewline
\hline 
Irreducible spherical tensor of rank $J$, component $M$, for a spin
$\mathbf{S}$ & $Y_{J,M}\left(\mathbf{S}\right)$\tabularnewline
\hline 
Coupling constant of the term $\hat{O}_{J}\left(\mathbf{S}_{i},\mathbf{S}_{i+1}\right)$
{[}see Eq.~\eqref{eq:Hamilt_tensors}{]} & $K_{i}^{\left(J\right)}$\tabularnewline
\hline 
Projector onto a multiplet of total angular momentum $\tilde{S}$
of the pair of spins $\mathbf{S}_{i},\mathbf{S}_{i+1}$ {[}see Eq.~\eqref{eq:projector_multiplets}{]} & $P_{\tilde{S}}\left(\mathbf{S}_{i},\mathbf{S}_{i+1}\right)$\tabularnewline
\hline 
Coupling constant of the term $P_{J}\left(\mathbf{S}_{i},\mathbf{S}_{i+1}\right)$
{[}see Eq.~\eqref{eq:Hamilt_projectors}{]} & $\epsilon_{i}^{\left(J\right)}$\tabularnewline
\hline 
\end{tabular}
\par\end{centering}

\caption{A summary of the notation used in this paper. \label{tab:Notation}}
\end{table*}

\par\end{center}

\section{Method: Strong Disorder Renormalization Group \label{sec:Method}}

In this section, we derive the RG decimation procedure for the Hamiltonian
in Eq.~\eqref{eq:Hamilt_tensors}. The basic idea of the strong-disorder
RG is to progressively eliminate \emph{local} high-energy degrees
of freedom while at the same time keeping the low-energy physics unchanged\ \cite{madasgupta,madasguptahu,bhatt-lee}.
The local energy scales $\Delta_{i}$ are the local gaps of the Hamiltonian
$H_{i,i+1}$ describing the coupling between spins $S_{i}$ and $S_{i+1}$.
We first find the largest gap $\Omega=\max\left\{ \Delta_{i}\right\} $,
which sets the RG scale. Let us say that $\Omega=\Delta_{2}$. We
keep the ground multiplet of spins $S_{2}$ and $S_{3}$ and remove
the higher-energy ones. Focusing on the four-spin Hamiltonian 

\begin{equation}
H_{4sites}=H_{1,2}+H_{2,3}+H_{3,4},\label{eq:H4sites}
\end{equation}
we then treat $H_{1,2}+H_{3,4}$ as a perturbation to $H_{2,3}$.
The procedure is iterated until we reach the low energy scale of interest.
Essentially two cases must be distinguished, according to whether
the eliminated multiplet is degenerate or not. Next, we outline these
two possible RG decimation steps. Some details of the derivation are
relegated to Appendix~\ref{sec:Appendix:RGstep}.

\subsubsection{First-Order Perturbation Theory \label{sub:First-Order-Perturbation}}

In the case when the lowest energy multiplet of the two-site problem
is not a singlet, it is generally sufficient to treat the the effect
of $H_{1,2}+H_{3,4}$ via first-order perturbation theory. If the
ground multiplet of spins $S_{2}$ and $S_{3}$ has total angular
momentum $\tilde{S}$ we can replace $S_{2}$ and $S_{3}$ by a new
effective spin $\tilde{S}$. The renormalized couplings $\tilde{K}_{1,3}^{\left(J\right)}$
between $\tilde{S}$ and $S_{1,4}$ are then obtained by projecting
$H_{1,2}+H_{3,4}$ onto this degenerate ground state. It is important
to note that the projected Hamiltonian has the same functional form
as the unperturbed one. See Fig.\ \eqref{fig:decimation} (``1st
order'' case) for a graphical representation of the procedure in this
case.

Let us show in more detail how $\tilde{K}_{1}^{\left(J\right)}$ and
$\tilde{K}_{3}^{\left(J\right)}$ can be found. Projecting $H_{1,2}$
onto the multiplet $\tilde{S}$ (similarly for $H_{3,4}$), we obtain

\begin{eqnarray}
\tilde{H}_{1,2} & = & P_{\tilde{S}}H_{1,2}P_{\tilde{S}}\nonumber \\
 & = & \sum_{J=0}^{J_{{\rm max}}}K_{1}^{\left(J\right)}\sum_{M=-J}^{J}\left(-1\right)^{M}\\
 &  & \qquad\times Y_{JM}\left(\mathbf{S}_{1}\right)P_{\tilde{S}}Y_{J-M}\left(\mathbf{S}_{2}\right)P_{\tilde{S}}\label{eq:first_order_projection}\\
 & = & \sum_{J=0}^{J_{{\rm max}}}\tilde{K}_{1}^{\left(J\right)}\sum_{M=-J}^{J}\left(-1\right)^{M}Y_{JM}\left(\mathbf{S}_{1}\right)Y_{J-M}\left(\tilde{\mathbf{S}}\right)\label{eq:first-order-projectionB}\\
 & = & \sum_{J=0}^{J_{{\rm max}}}\tilde{K}_{1}^{\left(J\right)}\hat{O}_{J}\left(\mathbf{S}_{1},\tilde{\mathbf{S}}\right),
\end{eqnarray}
where $P_{\tilde{S}}$ is the projector in Eq.\ \eqref{eq:projector_multiplets}.
The step from \eqref{eq:first_order_projection} to \eqref{eq:first-order-projectionB}
involves the application of the Wigner-Eckart theorem: $P_{\tilde{S}}Y_{J-M}\left(\mathbf{S}_{2}\right)P_{\tilde{S}}=f^{(J)}\left(S_{2},S_{3},\tilde{S}\right)Y_{J-M}\left(\tilde{\mathbf{S}}\right)$,
where the constant $f^{(J)}$ does not depend on $M$. The last feature
preservers the SU(2) symmetry. Therefore, the neighboring couplings
$K_{1}^{\left(J\right)}$ and $K_{3}^{\left(J\right)}$ are renormalized
to 

\begin{align}
\tilde{K}_{1}^{\left(J\right)} & =f^{\left(J\right)}\left(S_{2},S_{3},\tilde{S}\right)K_{1}^{\left(J\right)},\nonumber \\
\tilde{K}_{3}^{\left(J\right)} & =f^{\left(J\right)}\left(S_{3},S_{2},\tilde{S}\right)K_{3}^{\left(J\right)},\label{eq:first-order-results}
\end{align}
while the set of couplings $\left\{ K_{2}^{\left(J\right)}\right\} $
is removed. This is a generalization to higher-order ISTs of the RG
step derived by Westerberg \textit{et al.} in Ref.\ \onlinecite{westerbergetal}.
In\textbf{ }Appendix~\eqref{sec:Appendix:RGstep}, we obtain closed-form
expressions for the functions $f^{(J)}$ in terms of Wigner's \emph{6-j}
symbols.

In the lowest rank case $J=1$, i. e., for the usual Heisenberg Hamiltonian,
the effective couplings are always nonzero (except in the obvious
non-degenerate case $\tilde{S}=0$). Interestingly, this is not always
true in the higher-rank cases. In fact, there are two cases where
the above effective couplings can vanish:

\emph{Case (a): If the IST rank $J$ is larger than $2\tilde{S}$.}
In this case, the projection is zero simply because one cannot construct
a large rank-$J$ IST out of small angular momentum operators. A general
derivation can be found in Appendix~\ref{sec:Appendix:RGstep}. For
example, when $S_{2}=S_{3}=3/2$ (thus, $J_{{\rm max}}=3$) and the
ground multiplet has $\tilde{S}=1$, the IST $Y_{J=3,M}(\tilde{\mathbf{S}})$
vanishes identically. We will come back to this point in Sections\ \ref{sec:Spin-3/2-chain}
and \ref{sec:Spin-2-Chain} when we study the spin-$\frac{3}{2}$
and spin-2 chains;

\emph{Case (b): If the function $f^{\left(J\right)}$ vanishes for
some specific combinations of $S_{2}$, $S_{3}$ and $\tilde{S}$
not predicted by case (a).} For example, when $S_{2}=S_{3}=\frac{3}{2}$
and $J=\tilde{S}=2$, $f^{\left(J\right)}\left(S_{2},S_{3},\tilde{S}\right)=0$.
Note that $J<2\tilde{S}$, so case (a) does not apply. We can explicitly
show that this case never happens for ISTs with $J=1$. It is a feature
that only happens when higher rank ISTs are included. Since there
is no general rule to predict when it happens, its consequences have
to be analyzed on a case-by-case basis.

Evidently, these two cases configure a failure of the usual RG decimation
procedure since the renormalized constants vanish and the chain becomes
disconnected from that spin pair and is effectively broken up. The
remedy is to include corrections in higher orders of perturbation
theory. This would introduce new types of terms in the effective Hamiltonian
(such as 3-spin couplings), which makes the problem much harder to
treat. In this paper, we do not implement this remedy in full generality,
although we discuss some of its features in Appendix~\ref{sec:appendix-zeros}.
Nonetheless, as we discuss in Section~\ref{sub:otheraxes}, the effects
of such peculiar decimations in the RG flow are not important in the
great majority of flows. In general, first-rank couplings $K^{(1)}$
are present and the chains never become disconnected.

Finally, we emphasize that we assume that two or more multiplets of
well-defined angular momentum are not degenerate in the ground multiplet.
If this were not the case, the projection from Eq.~\eqref{eq:first_order_projection}
to Eq.~\eqref{eq:first-order-projectionB} would have to incorporate
the projection onto the additional multiplets. In general, this procedure
would change the functional form of the projected Hamiltonian and
a more elaborate RG procedure would be needed. Nevertheless, these
accidental degeneracies appear only by fine tuning of the coupling
constants $K_{2}^{(J)}$ and can be safely ignored. Such high-symmetry
cases are unstable in the sense that the RG flow is always away from
them. We will come back to this point later when we analyze a few
cases of physical importance.

\subsubsection{Second-Order Perturbation Theory for singlets}

\begin{figure}
\begin{centering}
\includegraphics[clip,width=0.8\columnwidth]{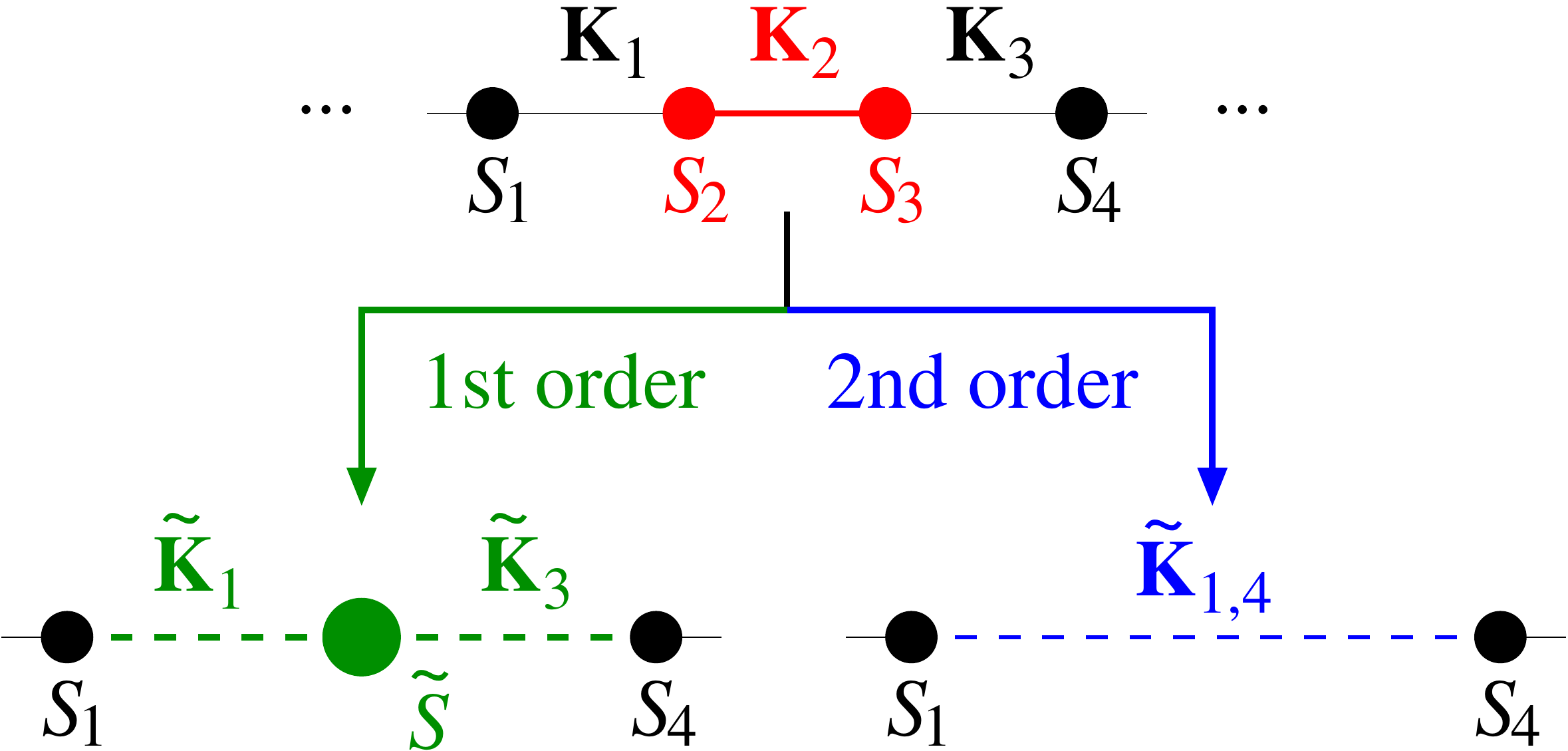}
\par\end{centering}

\caption{(Color online) Schematic decimation procedure. The decimated spins
$S_{2}$ and $S_{3}$ are either replaced by an effective spin $\tilde{S}$
(``1st order'' case) or removed from the system (``2nd order'' case),
depending on whether the local ground state of $H_{2,3}$ has a degeneracy
of $2\tilde{S}+1$ or is a non-degenerate singlet, respectively. The
set of renormalized couplings $\tilde{\mathbf{K}}=\left(K^{(1)},\dots,K^{(J_{{\rm max}})}\right)$
are given by Eqs.\ \eqref{eq:first-order-results} and \eqref{eq:second-order-effec-coupl},
respectively. \label{fig:decimation}}
\end{figure}

When the two-spin problem has a singlet ground state, the first-order
perturbation theory term vanishes and a second-order perturbation
analysis is necessary. In this case, the spin pair $S_{2}$ and $S_{3}$
is frozen into a singlet and decouples from the chain, while the neighboring
spins $S_{1}$ and $S_{4}$ become connected (due to virtual excitations
of the singlet) via new effective couplings not present in the initial
Hamiltonian. This is depicted in Fig.\ \eqref{fig:decimation} as
the ``2nd order'' case.

Let us define $S\equiv S_{2}=S_{3}$, since a necessary condition
for singlet formation is that the spins of the two sites are equal.
The second-order perturbation-theory renormalization of the Hamiltonian
is given by

\begin{eqnarray}
\tilde{H}_{1,4} & = & P_{0}\left(H_{1,2}+H_{3,4}\right)P_{\bar{0}}\frac{1}{E_{0}-H_{2,3}}P_{\bar{0}}\left(H_{1,2}+H_{3,4}\right)P_{0},\label{eq:second-order-correc-1}
\end{eqnarray}
where $E_{0}$ is the energy of the singlet, $P_{0}$ is the projector
onto the singlet state, and $P_{\bar{0}}=1-P_{0}$ is the projector
onto all other multiplets $J'=1,\ldots,2S$. Neglecting unimportant
constant terms, we get

\begin{align}
\tilde{H}_{1,4} & =P_{0}H_{1,2}P_{\bar{0}}\frac{1}{E_{0}-H_{2,3}}P_{\bar{0}}H_{3,4}P_{0}+\mbox{H.c.}\\
 & =2P_{0}H_{1,2}P_{\bar{0}}\frac{1}{E_{0}-H_{2,3}}P_{\bar{0}}H_{3,4}P_{0}.
\end{align}
Writing $H_{1,2}$ and $H_{3,4}$ explicitly, we are left with the
challenge of computing terms like $P_{0}Y_{J-M}\left(\mathbf{S}\right)P_{\bar{0}}$.
Let us denote by $J'\ne0$ an arbitrary total angular momentum present
in the projection operator $P_{\bar{0}}$. The Wigner-Eckart theorem\ (page
74 of Ref.~\onlinecite{Edmondsbook}) ensures the only value of $J'$
yielding a non-zero matrix element is $J'=J$, that is, $\left\langle 00\left|Y_{J-M}\left(\mathbf{S}\right)\right|J'M'\right\rangle \propto\delta_{J,J'}\delta_{M,M'}$.
Therefore, only the total angular momentum equal to the IST rank gives
a non-vanishing contribution to the perturbation theory. By defining
$\tilde{J}_{{\rm max}}=\min\left(J_{{\rm max}},2S\right)$, and $\Delta E\left(0,J\right)<0$
to be the energy difference between the ground state and the excited
state of total angular momentum $J$, we get

\begin{align}
\tilde{H}_{1,4}= & 2\sum_{J=1}^{\tilde{J}_{{\rm max}}}K_{1}^{\left(J\right)}K_{3}^{\left(J\right)}\times\nonumber \\
 & \sum_{M=-J}^{J}\frac{\left(-1\right)^{M}g\left(S,J\right)}{\Delta E\left(0,J\right)}Y_{JM}\left(\mathbf{S}_{1}\right)Y_{J-M}\left(\mathbf{S}_{4}\right),\label{eq:second_order_PT}
\end{align}
where

\begin{align}
\left(-1\right)^{M} & g\left(J,S\right)=\nonumber \\
 & \left\langle 00\left|Y_{J-M}\left(\mathbf{S}\right)\right|JM\right\rangle \left\langle JM\left|Y_{JM}\left(\mathbf{S}\right)\right|00\right\rangle .\label{eq:def-g}
\end{align}
Note that neither the energy denominator $\Delta E\left(0,J\right)$
nor the function $g\left(J,S\right)$ depends on $M$, since the Hamiltonian
is SU(2) symmetric. A closed-form expression for the function $g\left(S,J\right)$
can be found in Appendix~\ref{sec:Appendix:RGstep}, where the values
needed in this paper are also listed. In conclusion, the effective
Hamiltonian has the form

\begin{align}
\tilde{H}_{1,4} & =\sum_{J=1}^{\tilde{J}_{{\rm max}}}\tilde{K}_{14}^{\left(J\right)}\hat{O}_{J}\left(\mathbf{S}_{1},\mathbf{S}_{4}\right),\label{eq:second-order-eq}
\end{align}
where the renormalized couplings are

\begin{equation}
\tilde{K}_{1,4}^{\left(J\right)}=2\frac{g\left(J,S_{2}\right)}{\Delta E\left(0,J\right)}K_{1}^{\left(J\right)}K_{3}^{\left(J\right)}.\label{eq:second-order-effec-coupl}
\end{equation}

Eqs.\ \eqref{eq:first-order-results} and \eqref{eq:second-order-effec-coupl}
highlight the most important feature of the decimation procedure:
under the RG flow, \emph{the renormalized couplings $\tilde{K}_{i,j}^{(J)}$
depend only on coupling constants of the same rank $J$}, which is
a direct consequence of the SU(2) symmetry. This is why working with
the ISTs is a natural choice for these systems. This will have profound
consequences for the identification of the stable fixed points, as
will become clear later.

In summary, we have determined the decimation procedure for our generic
SU(2)-symmetric quantum spin-$S$ chain (see schematics in Fig.\ \eqref{fig:decimation}),
which generalizes the one obtained by Westerberg \emph{et. al}.\ \cite{westerbergetal}
devised to describe generic spin-$S$ Heisenberg chains (i.e., with
$J_{{\rm max}}=1$). We search for the strongest coupled spin pair
in the chain (which defines our RG cutoff energy scale $\Omega$)
and decimate it. If the local ground state is a singlet, the spin
pair is removed and the renormalized coupling constants between the
neighbor spins are given by Eq.\ \eqref{eq:second-order-effec-coupl}.
Otherwise, the spin pair is replaced by an effective spin $\tilde{S}$
given by the total angular momentum of the ground-state multiplet.
Moreover, this new effective spin degree of freedom interacts with
the nearest-neighbor spins via the renormalized couplings given by
Eq.\ \eqref{eq:first-order-results}. Upon decimation, the coupling
constants and the spins change. For a complete description of the
RG flow, one then needs to keep track of the joint distribution of
coupling constants and spin sizes at the cutoff energy scale $\Omega$:
${\cal Q}\left(\tilde{\mathbf{K}},\tilde{S};\Omega\right)$, where
$\tilde{\mathbf{K}}=\left(K^{\left(1\right)},\dots,K^{\left(J_{{\rm max}}\right)}\right)$.
The distribution of a particular variable can be obtained by integrating
out the other ones. For instance, the distribution of the $I$-th
coupling is ${\cal P}_{I}\left(K;\Omega\right)=\sum_{S}\int\prod_{J\neq I}{\rm d}K^{(J)}{\cal Q}\left(\tilde{\mathbf{K}},\tilde{S};\Omega\right)$.
In what follows, we analyze the fixed points of our SDRG flow.

\section{Technical summary of the RG flow and the corresponding zero-temperature
phases}

\label{sec:Technical-summary}

Given the prescriptions of the strong-disorder renormalization-group
method derived in Sec.\ \ref{sec:Method}, we are now set to apply
the SDRG decimation procedure to our system Hamiltonian \eqref{eq:Hamilt_tensors}
and analyze the general features of the RG flow. A complete characterization
involves (i) finding all the fixed points, (ii) classifying their
stability, and (iii) determining their basins of attraction in the
case of totally attractive fixed points. The fixed points are characterized
by the joint distribution of spin sizes and the $J_{{\rm max}}$ coupling
constants ${\cal Q}^{*}\left(\mathbf{K},S\right)$ (we denote a fixed-point
distribution by the superscript $^{*}$). Accomplishing these three
tasks allows us to determine the phase diagram of the system and the
low-energy physical behavior. As one can guess, this is not an easy
task and simplifications are needed. Below, we give a summary of the
structure of the RG flow and the simplifications that can be made
after one knows the fixed point distributions. We will focus on the
fully stable fixed points, which determine the stable phases of the
system.

One way of thinking about the problem is the following. Consider the
set of vectors $\left\{ \mathbf{\bm{\omega}}_{i}\right\} $ where
$\mathbf{\bm{\omega}}_{i}=\left(S_{i},\mathbf{K}_{i},S_{i+1}\right)$
and $\mathbf{K}_{i}=\left(K_{i}^{\left(1\right)},...,K_{i}^{\left(J_{{\rm max}}\right)}\right)$
is the vector of coupling constants. The set $\left\{ \mathbf{\bm{\omega}}_{i}\right\} $
defines the Hamiltonian \eqref{eq:Hamilt_tensors}. In the RG framework,
it defines an initial condition of the RG flow. In the $\mathbb{R}^{J_{{\rm max}}+2}$
space, this initial condition is just a set of vectors sharing the
same origin. Under the RG flow, these vectors change their lengths
and directions until they converge to a fixed-point distribution,
which can be viewed as another set of vectors $\left\{ \mathbf{\bm{\omega}}_{i}^{\ast}\right\} $.
We have carried out this detailed analysis and found the possible
fixed-point distributions of this system. As will be shown latter,
the fully stable fixed point distributions can be classified in two
major groups: one that has essentially AF correlations (but without
AF long range order) and another characterized by strong FM tendencies.
We now discuss their generic features.

The group of AF stable fixed points is characterized by a single spin
size, ${\cal Q}^{*}\left(\mathbf{K}_{i},S_{i}\right)={\cal R}^{*}\left(\mathbf{K}_{i}\right)\delta_{S,S_{i}}$.
Furthermore, the distribution of coupling constants is such that its
support lies strictly along a single coordinate axis $I$ in the $J_{\mathrm{max}}$-dimensional
space of vectors $\mathbf{K}_{i}$, $K_{i}^{\left(J\right)}=K_{i}^{\left(I\right)}\delta_{J,I}$.
In other words, 
\begin{eqnarray}
{\cal R}^{*}\left(\mathbf{K}_{i}\right) & = & \delta\left[K_{i}^{\left(1\right)}\right]\cdots\delta\left[K_{i}^{\left(I-1\right)}\right]{\cal P}^{*}\left[K_{i}^{\left(I\right)}\right]\nonumber \\
 &  & \times\delta\left[K_{i}^{\left(I+1\right)}\right]\cdots\delta\left[K_{i}^{\left(J_{\mathrm{max}}\right)}\right].
\end{eqnarray}
There are two possibilities for the distribution function ${\cal P}^{*}\left(x\right)$.
In one case, $K_{i}^{\left(I\right)}$ is strictly positive and 
\begin{equation}
{\cal P}^{*}\left(x\right)=\theta\left(x\right)\frac{\psi^{-1}-1}{\Omega\ln\Omega}\left(\frac{\Omega}{x}\right)^{1-\frac{\psi^{-1}-1}{\ln\Omega}},\label{eq:P(x)-IRFP}
\end{equation}
with $\psi=1/2$ being a universal tunneling exponent. The distribution
in Eq.~\eqref{eq:P(x)-IRFP} represents an infinite-randomness fixed
point, since its relative width (the ratio of its standard deviation
to its average value) grows without bounds as $\Omega\to0$. This
form is familiar from the well-studied case of disordered AF Heisenberg
chains\cite{madasguptahu,madasgupta,fisher94-xxz}. The relation between
energy $\Omega$ and length $L$ scales is activated, i.e., $\ln\Omega\sim-L^{\psi}$.
This has implications to many low-energy thermodynamic observables
such magnetic susceptibility $\chi\sim T^{-1}\left|\ln T\right|^{-1/\psi}$
and specific heat $C\sim\left|\ln T\right|^{-1/\psi}$. Moreover,
they are associated with an emergent SU($2S+1$) symmetry. Note that
the form of the distribution in this case varies as the cutoff $\Omega$
is reduced. It can be regarded as describing a \emph{fixed} point,
however, if one rescales the variables appropriately by the cutoff:\cite{fisher94-xxz}
if $\zeta=\ln\left(\Omega/x\right)$ and $\Gamma=\ln\left(\Omega_{0}/\Omega\right)$
(where $\Omega_{0}$ is the initial value of the cutoff), then the
$\zeta$ distribution has the form $\tilde{{\cal P}}^{\ast}\left(\zeta\right)=q^{\ast}\left(\zeta/\Gamma\right)/\Gamma$,
where $q^{\ast}\left(x\right)=e^{-x}$ is indeed fixed. Since the
fixed point is uniquely specified by a semi-axis direction in $\mathbf{K}_{i}$
space, it can represented by a point on the surface of the unit $d$-dimensional
hypersphere, where $d=J_{\mathrm{max}}-1$.

In the case of the other AF fixed point, the coupling constants along
the axis direction are either positive or negative \emph{with equal
probability}. This can only happen for chains with integer $S>1/2$.
The fixed-point distribution is still given by Eq.~\eqref{eq:P(x)-IRFP}
(with $x\to\left|x\right|$) but with the important difference in
the tunneling exponent: $\psi=1/3$. The other physical properties
have the same form as above but with $\psi=1/3$. The value $I=J_{S}\neq1$
depends on the spin size $S$. For the case of spin $S=1$, $J_{S}=2$
and this fixed-point is associated with an emergent SU(3) symmetry\cite{QuitoHoyosMiranda}.
For other spin sizes, the symmetry is only the bare SU(2) symmetry.
Strictly speaking, this AF fixed point cannot be represented as a
single point on the surface of the unit $d$-dimensional hypersphere,
since the coupling constants can have either sign.

Finally, there are two stable fixed points with strong FM instabilities.
The first one is the usual Heisenberg FM fixed point for which $K_{i}^{(1)}<0$
and $K_{i}^{(J>1)}=0$ for all sites $i$. This model has been studied
before\cite{alexander-bernasconi-jpc79,ziman-prl82,theodorou-jpc82,cieplak-ismail-jpc87,evangelou-katsanos-pla92}
and we will not consider it in this paper. The other fixed point is
characterized by \emph{finite} effective disorder and only Heisenberg
coupling constants with \emph{both} FM and AF signs are present, \textcolor{black}{namely},
$K_{i}^{(1)}\neq0$ (with both signs) and $K_{i}^{(J>1)}=0$ for all
sites $i$. This fixed point was thoroughly studied in Ref.\ \onlinecite{westerbergetal}
and is related to the so-called Large Spin phase. For weak disorder,
the system is governed by a universal finite-disorder fixed point.
For stronger disorder, the system flows to a line of finite-disorder
fixed points. In any case, the distributions of FM and AF couplings
are power laws ${\cal P}^{\ast}\left(x\right)\sim\left|x\right|^{z^{-1}-1}$,
with a non-universal exponent $z$ which depends on the disorder strength
but does not depend on the cutoff $\Omega$. The spin size distribution
is\textcolor{black}{{} expected to be a half normal w}hose width increases
as the energy scale $\Omega$ is lowered.

\section{Fixed Points and Their Stability}

\label{sec:Fixed-Points}

We now analyze in more detail the general features of the RG flow,
find all the AF fixed points and classify their stability. Our first
result is obtained straightforwardly. If initially $K_{i}^{(I)}\neq0$
and $K_{i}^{\left(J\ne I\right)}=0$ (with $1\leq I\leq J_{{\rm max}}=2S+1$),
then all the $K_{i}^{\left(J\ne I\right)}$ remain zero throughout
the entire RG flow since the couplings of ISTs of a given rank never
generate couplings of ISTs of other ranks {[}see Eqs.\ \eqref{eq:first-order-results}
and \eqref{eq:second-order-effec-coupl}{]}. Describing the stability
and the corresponding fixed-point distribution of spin sizes and coupling
constants is a task we accomplish in what follows.

\subsection{Pairwise random singlet states\label{sub:PairwiseRSP}}

\begin{figure}
\begin{centering}
\includegraphics[clip,width=1\columnwidth]{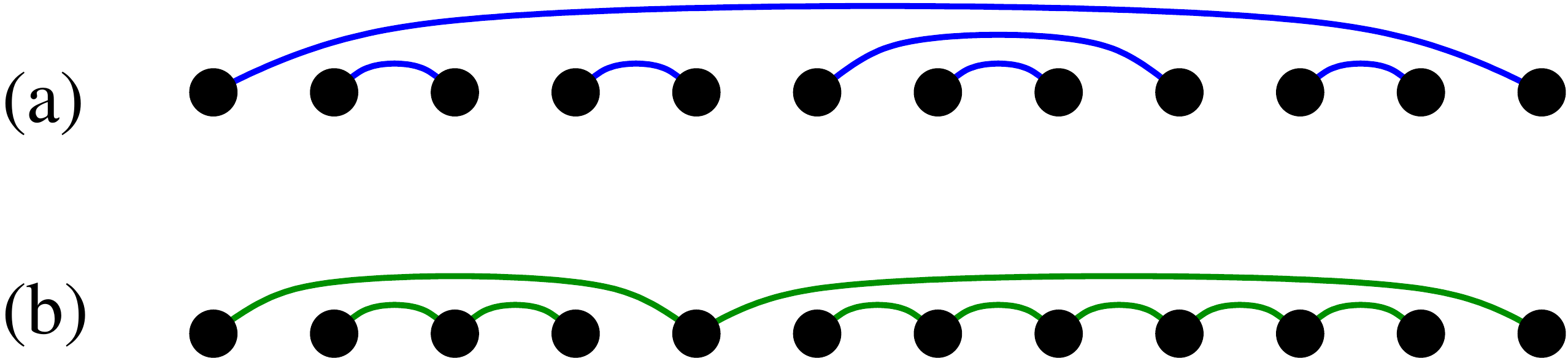}
\par\end{centering}

\caption{(Color online) A schematic depiction of two possible random singlet
states. (a) A pairwise random singlet state in which each spin forms
a singlet state with another spin indicated by the connecting line.
(b) A triplewise random singlet state in which spin trios (connected
by the lines) form a singlet state. Occasionally, a group of six (or
other multiple of 3) spins can also from a singlet state. Most of
the (pairs or trios of) singlets are formed by nearby spins, but there
are also singlets formed by spins that are arbitrarily far apart from
each other. \label{fig:random-singlets}}
\end{figure}

For simplicity, let us start our discussion focusing on the simplest
case: initially $K_{i}^{\left(I\right)}\ne0$, $K_{i}^{\left(J\ne I\right)}=0$,
and the ground state of the local Hamiltonian $H_{2,3}$ (see Fig.\ \ref{fig:decimation})
is always a singlet. If this is the case, every decimation step involves
second-order perturbation theory, as given by Eq.~\eqref{eq:second-order-effec-coupl}.
The RG flow is well understood and is just like the one of the random
AF spin-1/2 chain.\cite{fisher94-xxz} The spin size $S$ remains
fixed and the fixed-point coupling constant distribution ${\cal P}^{*}\left(K^{\left(I\right)}\right)$
is given by Eq.\ \eqref{eq:P(x)-IRFP} with $\psi=1/2$. This is
an infinite-randomness fixed point since the relative width of the
distribution increases without bonds at low energy scales, namely,
$\sigma_{K_{I}}/\left\langle K_{I}\right\rangle \rightarrow\infty$
as $\Omega\rightarrow0$. This provides \emph{a posteriori} justification
of the perturbative RG treatment yielding asymptotically exact results.
Its thermodynamics and the correlation functions can be computed straightforwardly.
For a review, see Ref.\ \onlinecite{igloi-review}. Moreover, the
corresponding ground state is a collection of nearly independent singlets
each of which is formed by only 2 spins {[}see Fig.\ \foreignlanguage{english}{\hyperref[fig:random-singlets]{\ref{fig:random-singlets}(a)}}{]},
hence the title of this subsection. We stress that, in principle,
it is possible to form singlet states with 2 or more spins as depicted
in Fig.\ \foreignlanguage{english}{\hyperref[fig:random-singlets]{\ref{fig:random-singlets}(b)}}.
These will be relevant later.

Naturally, we must inquire about the conditions under which such a
random singlet state is obtained, i.e., what are the values of spin
size $S$ and IST coupling rank $I$ which ensure that the local ground
state is always a 2-spin singlet. Although we could not obtain a rigorous
analytical proof, our extensive numerical verification (see Appendix\ \ref{sec:S=00003D0-condition})
indicates that such singlet state is the ground state of $H_{i,i+1}$
{[}see Eq.\ \eqref{eq:Hamilt_tensors}{]} whenever $K_{i}^{(J\neq I)}=0$
and $\left(-1\right)^{I}K_{i}^{(I)}<0$ for any spin value $S_{i}=S_{i+1}=S$.
Moreover, the sign of the renormalized $\tilde{K}_{i}^{(I)}$ is a
constant of the flow (see App.\ \ref{sec:Appendix:RGstep}), which
is a necessary condition to ensure that all decimations will be of
singlet-formation type (``second-order'' case in Fig.\ \ref{fig:decimation}).

Finally, we are now able to completely describe our first set of fixed
points. It has the following features: (i) all spins have the same
size $S_{i}=S$; (ii) all the coupling constants are zero except for
the ones corresponding to ISTs of rank $I$, and all of the latter
have the same sign, equal to $\left(-1\right)^{I+1}$; (iii) the fixed-point
distribution of non-zero couplings is ${\cal P}^{*}\left(x\right)$
as given by Eq.\ \eqref{eq:P(x)-IRFP} with $\psi=1/2$ and $0<x=\left(-1\right)^{I+1}K^{\left(I\right)}<\Omega$.
There is a caveat, though. The cutoff energy scale $\Omega$ is defined
as the maximum value of the local gaps. As the local gap is proportional
to $K_{i}^{(I)}$, the corresponding numerical pre-factor can be absorbed
in the definition of $x$.

Having found this set of $J_{{\rm max}}$ fixed points of infinite-randomness
type, the natural questions that arise are whether they are stable
or not and what is the size of their basins of attraction. We now
discuss their stability properties. The other question will be dealt
with in Secs.\ \ref{sec:Spin-3/2-chain} and \ref{sec:Spin-2-Chain}.

For each fixed-point of rank $I$, there are $J_{{\rm max}}-1$ independent
perpendicular directions. Let us call $\delta_{i}^{\left(J\right)}$
the relative deviation from the $I$-th axis in the $J$-direction
with $J\neq I$ at site $i$, namely, $\delta_{i}^{\left(J\right)}=\frac{K_{i}^{\left(J\right)}-K_{i}^{\left(J\right)*}}{K_{i}^{(I)}}$,
where $K_{i}^{\left(J\right)*}=0$ is the fixed-point value. To leading
order in $\delta^{(J)},$ the recursion relations \eqref{eq:second-order-effec-coupl}
become 

\begin{equation}
\tilde{\delta}_{1,4}^{\left(J\right)}\propto\delta_{1}^{\left(J\right)}\delta_{3}^{\left(J\right)},\label{eq:tilde-delta-stable}
\end{equation}
where the numerical pre-factor (whose magnitude is of order unity)
is irrelevant for our purposes. Following the steps of Ref.\ \onlinecite{fisher94-xxz},
it is easy to show that the mean value of $\ln\left|\tilde{\delta}_{i}^{(J)}\right|$
goes as $\sim-c\Gamma^{\phi}$, where $c$ is a non-universal constant
and $\phi=\frac{1+\sqrt{5}}{2}$ is the golden mean, which implies
a vanishing typical value $\tilde{\delta}_{typ}^{(J)}\sim\exp\left(-c\Gamma^{\phi}\right)$.
Thus, weak deviations in any of the perpendicular directions are strongly
irrelevant and all the $J_{{\rm max}}$ fixed points mentioned above
are stable.

We now search for other fixed points. Let us keep focusing on fixed
points for which the only possible decimations are of the 2-spin singlet-formation
variety. It is useful to rewrite the transformation rule \eqref{eq:second-order-effec-coupl}
in terms of ratios of coupling constants, ``angular variables'' in
the unit hypersphere in $\mathbf{K}$ space. For concreteness, let
us consider the case where $K_{i}^{\left(1\right)}$ is non-zero and
define $s_{i}^{\left(J\right)}=\frac{K_{i}^{\left(J\right)}}{K_{i}^{\left(1\right)}},~\left(2\leq J\leq J_{{\rm max}}\right)$.
The generalization to the other cases is straightforward. The recursion
relations~\eqref{eq:second-order-effec-coupl} can be rewritten as

\begin{eqnarray}
\tilde{K}_{1,4}^{\left(1\right)} & = & \Sigma^{(1)}\left(S,\mathbf{s}_{2}\right)\frac{K_{1}^{\left(1\right)}K_{3}^{\left(1\right)}}{K_{2}^{\left(1\right)}}.\label{eq:angular-K}\\
\tilde{s}_{1,4}^{\left(J\right)} & = & \Xi^{(J)}\left(S,\mathbf{s}_{2}\right)s_{1}^{\left(J\right)}s_{3}^{\left(J\right)},~\left(J=2,\ldots,J_{{\rm max}}\right)\label{eq:angular-renorm}
\end{eqnarray}
where $S=S_{2}=S_{3}$ is the spin size, $\Sigma^{(J)}\left(S,\mathbf{s}_{2}\right)=2\frac{g\left(J,S\right)}{\Delta E\left(0,J\right)}K_{2}^{\left(1\right)}$,
$\mathbf{s}_{i}=\left\{ s_{i}^{\left(2\right)},\dots,s_{i}^{\left(J_{{\rm max}}\right)}\right\} $
denotes the set of $J_{{\rm max}}-1$ angular variables, and $\Xi^{\left(J\right)}\left(S,\mathbf{s}_{2}\right)=\Sigma^{(J)}\left(S,\mathbf{s}_{2}\right)/\Sigma^{(1)}\left(S,\mathbf{s}_{2}\right)$.
Note that the local energy scale at site $2$ is essentially given
by $K_{2}^{\left(1\right)}$. Thus, the functions $\Sigma^{(1)}\left(S,\mathbf{s}_{2}\right)$
and $\Xi^{\left(J\right)}\left(S,\mathbf{s}_{2}\right)$ are just
geometric functions which are independent of $K_{2}^{\left(1\right)}$.
This separation between the energy variable $K^{\left(1\right)}$
and the set $\mathbf{s}_{2}$ is what allows us to find the conditions
under which $\tilde{\mathbf{s}}$ is kept constant under the RG flow.

At a FP, the set $\mathbf{s}_{i}$ becomes site-independent. Denoting
its fixed point value by $\mathbf{s}^{*}$, then 

\begin{equation}
s^{\left(J\right)*}=\Xi^{\left(J\right)}\left(S,\mathbf{s}^{*}\right)s^{\left(J\right)*}s^{\left(J\right)*}.\label{eq:aFPcondition}
\end{equation}
Thus, if $s^{(J)*}\neq0$, we must have $\Xi^{\left(J\right)}\left(S,\mathbf{s}^{*}\right)s^{\left(J\right)*}=1$.
Solving the $J_{{\rm max}}-1$ coupled Eqs.\ \eqref{eq:aFPcondition}
gives us all the FPs corresponding to the usual random 2-spin singlet
states. Since the geometric prefactor $\Xi^{\left(J\right)}\left(S,\mathbf{s}^{*}\right)$
depends nontrivially on $S$ and $\mathbf{s}^{*}$, we will have to
solve Eqs.\ \eqref{eq:aFPcondition} on a case by case basis. This
is done for $S=3/2$ in Section\ \ref{sec:Spin-3/2-chain} and $S=2$
in Appendix~\ref{sec:Appendix-RG-equations-spins}. Here, we will
assume that such fixed points are known and provide general stability
criteria for them.

Defining $\delta^{\left(J\right)}=s^{\left(J\right)}-s^{\left(J\right)*}$
and expanding Eq.~\eqref{eq:angular-renorm} up to quadratic order
in $\delta$, we obtain

\begin{eqnarray}
\tilde{\delta}_{1,4}^{\left(J\right)} & = & s^{\left(J\right)*}\Xi^{\left(J\right)*}\left(\delta_{1}^{\left(J\right)}+\delta_{3}^{\left(J\right)}\right)+s^{\left(J\right)*}\sum_{K=2}^{J_{{\rm max}}}\gamma^{\left(K\right)*}\delta_{2}^{\left(K\right)}\nonumber \\
 &  & +\Xi^{\left(J\right)*}\delta_{1}^{\left(J\right)}\delta_{3}^{\left(J\right)},\label{eq:tilde-delta14}
\end{eqnarray}
where $\Xi^{\left(J\right)*}=\Xi^{\left(J\right)}\left(S,\mathbf{s}^{*}\right)$
and $\gamma^{\left(K\right)*}=\left.\frac{\partial\Xi^{\left(J\right)}}{\partial s_{2}^{\left(K\right)}}\right|_{\mathbf{s}_{2}=\mathbf{s}^{*}}$.
Using $\Xi^{\left(J\right)*}s^{\left(J\right)*}=1$ for $s^{\left(J\right)*}\ne0$,
and keeping only the leading-order terms, we rewrite Eq.\ \eqref{eq:tilde-delta14}
as

\begin{equation}
\tilde{\delta}_{1,4}^{\left(J\right)}=\begin{cases}
\delta_{1}^{\left(J\right)}+\delta_{3}^{\left(J\right)}+s^{(J)*}\sum_{K=2}^{J_{{\rm max}}}\gamma^{\left(K\right)*}\delta_{2}^{\left(K\right)}, & \mbox{if }s^{(J)*}\neq0,\\
\Xi^{\left(J\right)}\delta_{1}^{\left(J\right)}\delta_{3}^{\left(J\right)}, & \mbox{otherwise}.
\end{cases}\label{eq:delta-renorm}
\end{equation}
Therefore, for $s^{\left(J\right)*}\neq0,$ the iterations of $\delta^{\left(J\right)}$
correspond to a random walk and this quantity grows without bounds.
More precisely, the typical value $\left|\tilde{\delta}_{\mathrm{typ}}^{\left(J\right)}\right|\sim\delta_{0}\Gamma^{\alpha_{{\rm asym}}}+\sigma_{0,\delta}\Gamma^{\alpha_{{\rm sym}}}$,
with $\alpha_{{\rm asym}}=\frac{1}{2}\left(1+\sqrt{5+4s^{\left(J\right)*}\gamma^{\left(J\right)*}}\right)$
and $\alpha_{{\rm sym}}=\frac{1}{4}\left(1+\sqrt{5+4\left(s^{\left(J\right)*}\gamma^{\left(J\right)*}\right)^{2}}\right)$,
where $\delta_{0}$ and $\sigma_{0,\delta}$ are the mean and the
width of the bare distribution of $\delta^{\left(J\right)}$.\ \cite{fisher94-xxz}
This means that perturbations in both the positive and the negative
$J$-th directions are relevant. On the other hand if $s^{\left(J\right)*}=0,$
then \eqref{eq:delta-renorm} becomes identical to \eqref{eq:tilde-delta-stable}
and the typical value of $\left|\delta^{\left(J\right)}\right|\sim\exp\left(-c\Gamma^{\phi}\right)$,
meaning the perturbations in both the negative and positive $J$-th
direction are irrelevant.

\begin{figure}
\begin{centering}
\includegraphics[clip,width=0.7\columnwidth]{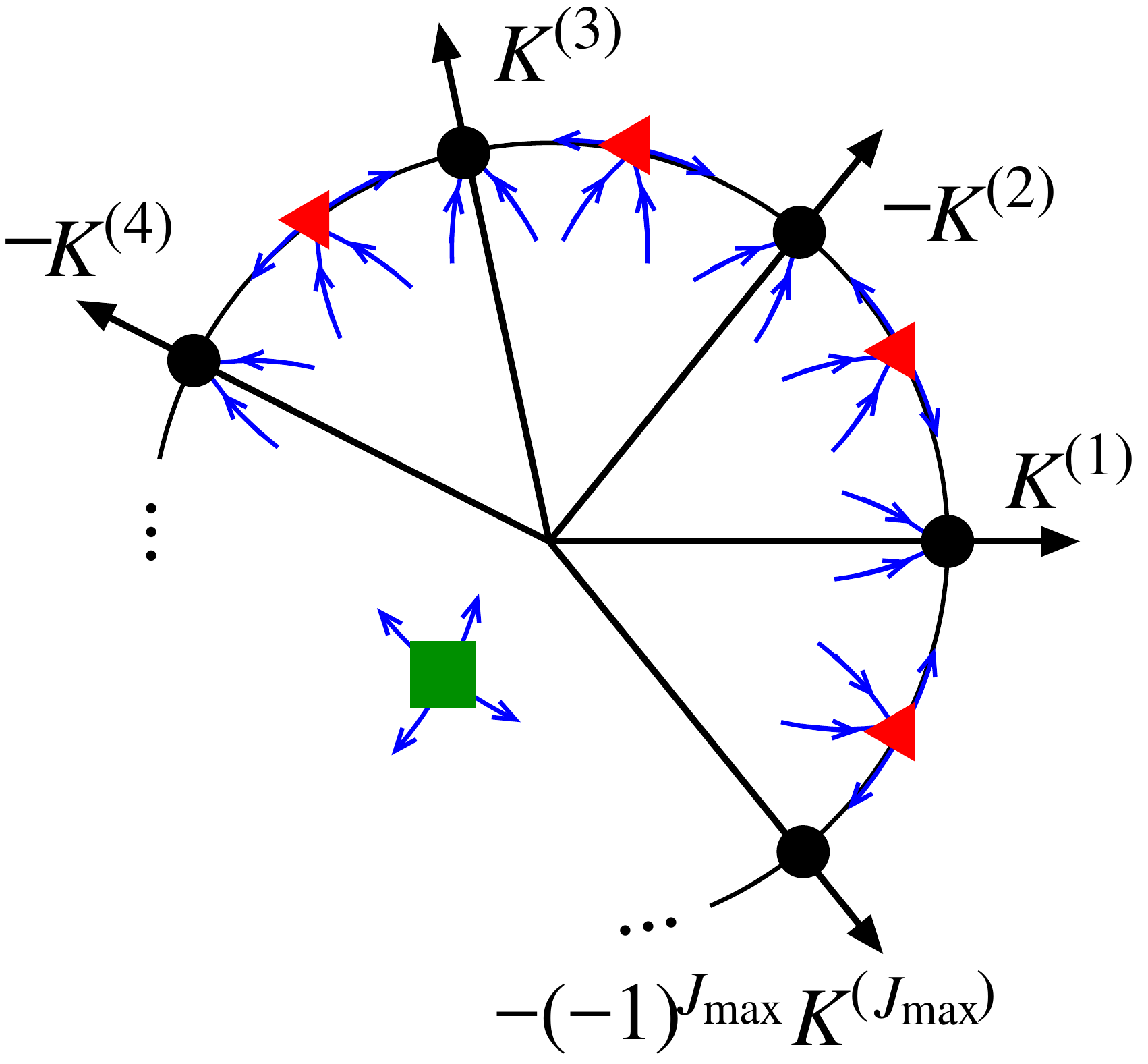}
\par\end{centering}

\caption{(Color online) Schematic drawing of the hyper-octant formed by all
the fully stable AF fixed points of pairwise singlet nature {[}black
circles on the $\left(-1\right)^{J+1}K^{(J)}$ semi-axes{]}. Semi-stable
AF fixed points exist on the hyperplanes formed by two (red triangles)
or more (not shown) coordinate axis directions. There is also a totally
unstable fixed point (green square) somewhere in the middle of the
hyper-octant. This totally unstable fixed point is exactly SU($2S+1$)
symmetric while the other ones have emergent SU($2S+1$) symmetry.\label{fig:Hyper-octant}}
\end{figure}

We are now able to state a clear criterion for the stability of the
2-spin-singlet fixed points reported here. Let such a fixed point
be located at $\mathbf{s}^{*}$, which defines a point on the surface
our unit hypersphere in $\mathbb{R}^{J_{{\rm max}}}$. Recall we are
assuming $K_{i}^{\left(1\right)}\neq0$ and thus $\mathbf{s}^{*}=0$
means the fixed point is on the first cartesian axis in $\mathbf{K}$
space. Then, the fixed point $\mathbf{s}^{*}$ is stable with respect
to any SU(2)-symmetric local perturbation in the $\pm J$-th directions
provided $s^{(J)*}=0$; otherwise, it is unstable in that direction.
This is depicted schematically in Fig.\ \ref{fig:Hyper-octant} where
we try to mimic a hyper-octant formed by the directions of the stable
AF fixed points reported here. The stable fixed points are drawn as
black circles. Fixed points which lie on coordinate hyper-planes (such
as the $K^{\left(1\right)}\times-K^{\left(2\right)}$ one) have both
stable (out of the hyper-plane) and unstable (in hyper-plane) directions.
These are drawn as red triangles. We will call them \emph{planar fixed
points}. Finally, a totally unstable fixed point (drawn as a green
square) lies somewhere inside the hyper-octant.

Our first result is already very significant: the large number of
completely stable AF fixed points along each coordinate axis in $\mathbf{K}$
space define a hyper-octant on the surface of the unit hypersphere
in $\mathbb{R}^{J_{{\rm max}}}$ in which only pairwise singlet-formation
decimations takes place (the 2nd-order route in Fig.\ \ref{fig:decimation}).
Therefore, all of these fixed-points (including the unstable ones
inside the hyper-octant), despite being different, are characterized
by a unique pairwise random singlet state {[}see Fig.\ \hyperref[fig:random-singlets]{\ref{fig:random-singlets}(a)}{]}.
As will be shown in detail for the cases of $S=3/2$ and $S=2$ in
Secs.\ \ref{sec:Spin-3/2-chain} and \ref{sec:Spin-2-Chain}, respectively,
they all exhibit an \emph{emergent} SU($N$) symmetry with $N=2S+1$,
as reported before for the particular case of spin-1 systems.\cite{QuitoHoyosMiranda}
Indeed \textcolor{black}{(see Appendix~\ref{sec:appendix-SU(N)-invariant-Hamil})},
the completely unstable fixed point in the middle of this hyper-octant
is \emph{exactly} SU($N$)-symmetric, i.e., at that particular point,
the SU(2) symmetric Hamiltonian \eqref{eq:Hamiltonian_SU2} can be
recast as a random AF Heisenberg SU($N$) chain where the ``spin''
operators at odd (even) sites are generators of the fundamental (anti-fundamental)
representation of the SU($N$) group.

Recently,\ \cite{QuitoHoyosMiranda} we have shown that SU(2)-symmetric
random spin-1 chains realize distinct random singlet phases with emergent
SU(3) symmetry, which we called ``mesonic'' and ``baryonic''. Our
first result shows that the pairwise random singlet state of the usual
spin-$S$ random Heisenberg chain displays emergent SU($2S+1$) symmetry.
This ground state is the same in all other fixed points inside the
AF hyper-octant of Fig.~\ref{fig:Hyper-octant}. Therefore, we can
conclude that the totally unstable fixed point in the middle of this
hyper-octant, which is exactly SU($2S+1$) symmetric, governs the
low-energy physics of entire hyper-octant. In this sense, the hyper-octant
is the region in parameter space where deformations of the SU($2S+1$)
symmetric FP does not destroy the SU($2S+1$) symmetry of the ground
state of the local Hamiltonian $H_{i,i+1}$. Furthermore, we can view
all the pairwise random singlet states as a generalization to higher
spins of the ``mesonic'' random singlet state found in the spin-1
case,\ \cite{QuitoHoyosMiranda} in which the singlets are formed
by a ``particle''---``anti-particle'' pair, corresponding to the fundamental
and the anti-fundamental representations of SU($2S+1$).

\subsection{Triplewise random singlet states}

\label{sub:TriplewiseRSP}

We know that $K_{i}^{(I)}\neq0$ and $K_{i}^{\left(J\ne I\right)}=0$
is a fixed point of the RG flow but, so far, we have only explored
the semi-axes with $\left(-1\right)^{I}K_{i}^{(I)}<0$ which define
the AF hyper-octant of Fig.~\ref{fig:Hyper-octant}. What are the
corresponding fixed points when we have the opposite signs $\left(-1\right)^{I}K_{i}^{(I)}>0$,
or when the signs are mixed? In general, since there will be no obvious
singlet formation, the spin sizes will tend to increase and the flow
becomes much more involved, a point we deal with later on. Here, however,
we focus on a peculiar fixed point in which the spin size does not
grow either, but instead remains fixed throughout the chain. How can
this be possible? It is possible if the following requirement is fulfilled:
whenever there is a decimation of ``1st-order'' type (see Fig.\ \ref{fig:decimation}),
then the new renormalized spin size $\tilde{S}$ must equal $S$.
Hence, under the two conditions that (i) $\tilde{S}=S$ and (ii) that
all the couplings are of the same rank, i.e., $K_{i}^{(I)}\neq0$
and $K_{i}^{\left(J\ne I\right)}=0$, a different AF fixed point is
realized. 

We now inquire whether there are rank values $I$ that satisfy the
first requirement that $\tilde{S}=S$. Evidently, $S$ is necessarily
integer, as it must appear as the sum of two equal spins $S$. We
have explicitly verified up to $S=9$ that there is always one and
only one rank value $I_{S}$ that fulfills this requirement, namely,
$I_{1}=2$ and $I_{S}=3$ for $S=2,\ldots,8$. For $S>9$, there is
no $I_{S}$ with $\tilde{S}=S$ as a ground state. Therefore, the
fixed point here reported only appears for integer spins $S\leq8$.

Let us now discuss the corresponding fixed-point distribution. Suppose
that only 1st-order decimations occur. This necessarily requires that
$\left(-1\right)^{I_{S}}K_{i}^{(I_{S})}>0$ for all $i$ along the
RG flow. For all of the rank values $I_{S}$ we have found, however,
the signs of the renormalized couplings $\tilde{K}_{i}^{(I_{S})}$
are reversed after ``1st-order''-type renormalizations. Therefore,
we conclude that this new fixed-point has coupling constants $K_{i}^{(I_{S})}$
with mixed signs. The fractions of positive and negative values can
be obtained straightforwardly. Since a 1st-order decimation always
reverses the sign of the renormalized coupling constant, it favors
equal fractions of positive and negative signs. As these fractions
are preserved by 2nd-order decimations {[}see Eq.\ \eqref{eq:second-order-effec-coupl}{]},
we then conclude that there is an equal fraction of 1st-order and
2nd-order decimation steps at this fixed point. As shown in Ref.\ \onlinecite{HoyosMiranda},
this leads to an infinite-randomness fixed-point distribution ${\cal P}^{*}(\left|K^{\left(I_{S}\right)}\right|)$
given by Eq.\ \eqref{eq:P(x)-IRFP} but with a different universal
tunneling exponent $\psi=\frac{1}{3}$. Furthermore, due to the 1st-order
decimation steps, the singlets are not formed by spin pairs, but rather
acquire a more complex structure which depends on the distribution
of couplings constant signs. If in the bare Hamiltonian all the couplings
are such that $\left(-1\right)^{I_{S}}K_{i}^{(I_{S})}>0$, then all
the singlets formed have a number of spins that is a multiple of three
as depicted in Fig.\ \hyperref[fig:random-singlets]{\ref{fig:random-singlets}(b)}.
If on the other hand the signs of the bare couplings are random, then
the singlets can be formed by any number of spins. In any case, the
probability of finding a singlet formed by $m$ spins decays $\sim m^{-x}$,
with $x\approx3.8$, as we verified numerically.

For the case of the spin-1 chain, it was shown that the corresponding
triplewise random singlet state possesses an emergent SU(3) symmetry.\ \cite{QuitoHoyosMiranda}
The difference with respect to the spin-1 pairwise random singlet
state, which also possesses SU(3) symmetry, is the representation
of the spin operators. While in the latter case the spin operators
on odd (even) sites are the generators of the fundamental (anti-fundamental)
representation of the SU(3) group, in the former case they are all
generators of the fundamental representation. In the general case
of spin-$S$ chains, however, this triplewise random singlet state
does not possess an obvious enlarged symmetry.

Finally, we report that this peculiar AF fixed point is also stable
against small SU(2) perturbations along the other transverse directions.
This can be shown by a similar analysis as was done for the previous
case of pairwise random singlet states. The analysis, however, is
more involved due to the presence of 1st-order decimation steps as
well. In addition, we have verified it numerically for the cases of
$S=1$ (see Ref.\ \onlinecite{QuitoHoyosMiranda}) and $S=2$ (see
Sec.\ \ref{sec:Spin-2-Chain}).

\subsection{Large Spin phase }

\label{sub:Large-Spin-Phase}

In Sec.\ \ref{sub:PairwiseRSP} we have considered fixed points in
which all the coupling constants have the same sign. In Sec.\ \ref{sub:TriplewiseRSP},
we have considered the case in which both signs are present in the
fixed-point Hamiltonian. For the latter case, however, a particular
rank value $I_{S}$ is required. In both cases, the spin size $S$
remained constant. We now consider another particular case, namely,
the case of the Heisenberg chain, i.e., $K_{i}^{(1)}\neq0$ and $K_{i}^{(I>1)}=0$.

The first case we discuss is the Heisenberg chain with both AF and
FM couplings. This is a very important special case because, as we
will see, it has a large basin of attraction. This fixed point was
thoroughly studied in Ref.\ \onlinecite{westerbergetal} and we now
summarize some of what is known. The fixed-point distribution of local
gaps $\Delta_{i}$ obeys the following scaling 
\begin{equation}
{\cal Q}_{1}^{*}\left(\Delta,S\right)=\frac{x{\cal Q}_{{\rm AF}}\left(\frac{\Delta}{\Omega},\frac{S}{\Omega^{\alpha}}\right)+\left(1-x\right){\cal Q}_{{\rm FM}}\left(\frac{\Delta}{\Omega},\frac{S}{\Omega^{\alpha}}\right)}{\Omega^{1+2\alpha}},\label{eq:P(x)-LSP}
\end{equation}
where $\alpha$ and $x$ are constants. The latter is the fraction
of AF couplings $K_{i}^{(1)}>0$. In addition, it was shown that ${\cal Q}_{1}^{*}$
is not of infinite-randomness type as in Eq.\ \ref{eq:P(x)-IRFP}
but rather of a finite-disorder variety. As a consequence, the relation
between energy and length scales is not activated but a more usual
power law $\Omega\sim L^{-z}$ where $z=-1/\left(2\alpha\right)>0$
is the critical dynamical exponent. The relation between $\alpha$
and $z$ comes from the fact that the average spin size grows as $\left\langle S\right\rangle \sim\sqrt{L}$,
which can be viewed as a consequence of the decimations leading to
a random walk in spin space. Because the effective spin increases
without bonds, the corresponding phase is called a Large Spin Phase.
Although the thermodynamics is relatively well understood (the magnetic
susceptibility is Curie like $\chi\sim T^{-1}$ and the specific heat
vanishes as $C\sim T^{1/z}\left|\ln T\right|$), the ground-state
spin-spin correlations are not so.\ \cite{hikihara}

The fact that $\alpha<0$ implies that the width of the spin size
distribution grows without bounds along the RG flow and, therefore,
the fraction of 2nd-order decimations vanishes, since it requires
an AF coupling shared by spins of the same size. Without the multiplicative
structure of the 2nd-order decimation (notice that the length scales
always renormalize additively), the scaling is no longer activated
and the effective disorder does not grow indefinitely. Hence, a finite-disorder
fixed point.

Finally, it was also found that there is a universal finite-disorder
fixed point (with $\alpha\approx-0.22$ and $x\approx0.63$) which
attracts all systems whose bare disorder is below a critical value.
Systems whose the bare disorder is greater than this critical value
are attracted by a line of finite-disorder fixed points where the
corresponding critical exponents are non-universal.

\subsection{Higher symmetry fixed points}

\label{sub:unknownI}

So far, we have described: (i) the fixed points of the AF hyper-octant
(see Sec.\ \ref{sub:PairwiseRSP}) which involve coupling constants
with uniform signs $\left(-1\right)^{J}K_{i}^{(J)}<0$; the stable
fixed points lie on the hyper-octant coordinate semi-axes while all
the remaining ones are unstable fixed points; (ii) the stable AF fixed
point in which the coupling constant sign is random and along the
axis with rank $I_{S}$ (see Sec.\ \ref{sub:TriplewiseRSP}); (iii)
the stable fixed point where the couplings belong to the first rank
$J=1$ and also have random signs, which was extensively studied before
\cite{westerbergetal} (see Sec.\ \ref{sub:Large-Spin-Phase}). We
now list some other fixed points that have higher symmetry than SU(2)
and discuss their implications on the RG flow.

It is always possible to fine-tune the IST couplings in order to realize
a higher SU($2S+1$) symmetry in the bare Hamiltonian \eqref{eq:Hamilt_tensors}.
In Appendix~\ref{sec:appendix-SU(N)-invariant-Hamil} we show how
to construct these Hamiltonians. In this case, we can rewrite the
SU(2)-symmetric Hamiltonian as a Heisenberg chain of SU($2S+1$) spins
which are nothing but irreducible representations of the SU($2S+1$)
group. One of these higher-symmetric fixed points is the unstable
one in the middle of the AF hyper-octant. Here, the higher-symmetric
spins on even (odd) sites are generators of the fundamental (anti-fundamental)
representation of the SU($2S+1$) group. As we have discussed in Sec.\ \ref{sub:PairwiseRSP},
our SDRG method is well suited for treating this case.

Another case is the one in which all the higher-symmetric spins are
generators of the fundamental irreducible representation of the SU($2S+1$)
group. Here, our SDRG method only works for the $S=1$ case.\ \cite{QuitoHoyosMiranda}
The reason is very simple. Consider for instance the $S=2$ case.
The Clebsch-Gordan series of the product of two fundamental irreducible
representations of the SU($N$) group, with $N=2S+1=5$, is $5\otimes5=10\oplus15$.
Applying our SDRG method in the AF case, we then have to keep the
$10$-fold degenerate manifold. How such a degenerate manifold can
be recast as an SU(2) spin of our Hamiltonian \eqref{eq:Hamilt_tensors}?
We need to match the local dimensions $2\tilde{S}+1=10$. But the
Clebsch-Gordan series of the product of two $S=2$ SU(2) spins is
$5\otimes5=1\oplus3\oplus5\oplus7\oplus9$. The only way to obtain
the 10-fold degenerate manifold of the SU(5) case is by fine-tuning
to degeneracy either the $\tilde{S}=0$ and $\tilde{S}=4$ multiplets
or the $\tilde{S}=1$ and $\tilde{S}=3$ multiplets. In either case,
we cannot use the SDRG idea of replacing both spins $S_{2}$ and $S_{3}$
by a single effective spin $\tilde{S}$ as depicted in Fig.\ \ref{fig:decimation}.
In this case, one needs to replace $S_{2}$ and $S_{3}$ by two other
spins. Furthermore, one will need to introduce new operators in order
to keep the structure of the low-energy spectrum. The SDRG method
then becomes considerably more involved and we will not deal with
these complications in the present study. Instead, we point out that
a generalization of the SDRG method to SU($N$) symmetry is capable
of handling this special case, as was done in Ref.\ \onlinecite{HoyosMiranda}. 

Are there other higher-symmetric fixed points? Certainly there are
as, for instance, exemplified by the FM counterpart of the two cases
mentioned above. But we will not concern ourselves with them because
of their instability against SU(2)-symmetric perturbations. Thus,
they have little consequence for the determination of the phases of
our model Hamiltonian \eqref{eq:Hamiltonian_SU2}. Naturally, they
may govern the low-energy physics at phase transitions but the study
of these is out of the scope of the present paper. For the cases of
AF SU($N$)-symmetric fixed points, nonetheless, the low-energy behavior
is known.\ \cite{HoyosMiranda}

\subsection{Unknown fixed points: breakdown of perturbation theory?}

\label{sub:unknownII}

As already pointed out, there are fixed points whenever the coupling
vectors $\mathbf{K}_{i}$ point in one of the rank directions. For
the rank direction $J=1$, our method recovers the one of Ref.\ \onlinecite{westerbergetal}
and we can completely characterize two possible fixed points: the
AF infinite-randomness fixed point (see Sec.\ \ref{sub:PairwiseRSP})
for $K_{i}^{(1)}>0$ and $S_{i}=S$ throughout the chain and the finite-disorder
fixed point (see Sec.\ \ref{sub:Large-Spin-Phase}) in which the
sign of $K_{i}^{(1)}$ as well as the spin sizes $S_{i}$ are random
variables. For the special rank direction $J=I_{S}$, our method can
also describe the corresponding fixed points as long as $S_{i}=S$
in the bare Hamiltonian (see Sec.\ \ref{sub:TriplewiseRSP}). Finally,
for the cases in which $S_{i}=S$ and $\left(-1\right)^{J}K_{i}^{(J)}<0$,
we also can describe the corresponding fixed point (see Sec.\ \ref{sub:PairwiseRSP}).
Are there other fixed points?

We have studied the RG flow in all cases via a numerical implementation
of the SDRG method (see Sec. \ref{sec:Method}). In order to do so,
we start with a chain of $\sim10^{6}$ spins with random $K_{i}^{\left(J\right)}$
couplings, such that the ratios $\frac{K_{i}^{\left(J\right)}}{K_{i}^{\left(1\right)}}$
are the same at all sites. In other words, our initial Hamiltonian
has uniform initial angles and only radial disorder in $\mathbf{K}$-space.
During the numerical flow, we follow the distributions of $K_{i}^{\left(J\right)}$
and spin sizes, which allows us to fully characterize the RG flow
numerically. We have found that there are a few cases in which the
flow is ``pathological'' because all the renormalized couplings between
sites vanish, as discussed in Sec.\ \ref{sub:First-Order-Perturbation}.
Note that, since the rank-1 renormalized coupling $\tilde{K}_{i}^{\left(1\right)}$
never vanishes, a necessary condition for this ``pathological'' flow
to occur is $\tilde{K}_{i}^{\left(1\right)}=0$ for all sites. It
turns out that when this latter condition is met, then one or both
of the routes for the appearance of vanishing coupling constants between
sites is unavoidably generated along the RG flow. Once such a ``special''
bond is decimated, the corresponding renormalization leads to a broken
chain. In those cases (see App.\ \ref{sec:appendix-zeros}), we have
performed degenerate second-order perturbation theory and new operators
are introduced. Nonetheless, we have found that upon further decimations,
other zeroes appear requiring a treatment that goes to higher orders
in perturbation theory. We have not pursued this further. Instead,
we believe that other zeroes will appear and this is an intrinsic
aspect of the problem. Within our theoretical framework, we are unable
to decide whether these zeroes are the manifestation of the true low-energy
physics of the problem or whether it is a simple artifact of the method,
indicating its breakdown and pointing to fundamentally new physics
in these cases. We then leave as an open question the true low-energy
behavior of these fixed points.

Nevertheless, in general, there will always be couplings of $J=1$
rank (the Heisenberg term) in the bare Hamiltonian. In this case,
we argue that the flow is in general towards the FM or AF fixed points
with only $J=1$ couplings. Indeed, in this case, only couplings of
rank $J=1$ will survive and the flow is naturally towards the Large-Spin-phase
fixed point. Furthermore, even if we were to carry out the perturbation
theory to higher orders, the corresponding renormalized couplings
would be typically much weaker than the $J=1$ renormalized couplings
(which are finite in first order of perturbation theory). Therefore,
these couplings will become sub-leading and the flow is towards the
Large-Spin-phase fixed point.

In Secs.\ \eqref{sec:Spin-3/2-chain} and \eqref{sec:Spin-2-Chain},
we explore in more detail the RG flow for the cases of the spin $S=3/2$
and $S=2$ chains.

\section{Spin-$\frac{3}{2}$ chain \label{sec:Spin-3/2-chain}}

In this section we apply the SDRG method derived in Sec.\ \ref{sec:Method}
to study the strong-disorder limit of the SU(2)-symmetric random spin-$\frac{3}{2}$
chain. For concreteness, we study the Hamiltonian \eqref{eq:Hamilt_tensors}
in which $S_{i}=\frac{3}{2}$ for all $i$ and the coupling vectors
\begin{eqnarray}
\mathbf{K}_{i} & = & \left(K_{i}^{(1)},K_{i}^{(2)},K_{i}^{(3)}\right)\nonumber \\
 & = & K_{i}\left(\sin\theta_{i}\cos\phi_{i},\sin\theta_{i}\sin\phi_{i},\cos\theta_{i}\right)
\end{eqnarray}
are distributed in the following way. The magnitude of the couplings
$K_{i}$ is distributed according to 
\begin{equation}
P_{0}(K)=\frac{1}{\Omega_{0}D}\left(\frac{\Omega_{0}}{K}\right)^{1-\frac{1}{D}},
\end{equation}
with $0<K_{i}<\Omega_{0}$. Here $\Omega_{0}$ is a microscopic high-energy
cutoff and $D$ parameterizes the bare disorder strength. First, we
consider the angles $\theta_{i}=\theta$ and $\phi_{i}=\phi$ to be
uniform throughout the chain. Note that in this case the unit hyper-sphere
is the 2-sphere in $\mathbb{R}^{3}$. We will thus determine the phase
diagram on the surface of the 2-sphere. At the end of this section,
we analyze the case of random initial angles.

\subsection{The antiferromagnetic octant}

\label{sub:afoctant}

Particularizing the results of Sec.\ \ref{sub:PairwiseRSP} to the
case $S_{i}=3/2$ we know that there must be $J_{{\rm max}}=2S=3$
stable AF fixed points along the three coordinate semi-axes defined
by $K_{i}^{\left(1\right)}>0$, $K_{i}^{\left(2\right)}<0$, and $K_{i}^{\left(3\right)}>0$.
Furthermore, these semi-axes define an octant on the surface of the
2-sphere where only 2nd-order decimations occur since the ground state
of the local Hamiltonian (\ref{eq:Hamilt_tensors}) is always a singlet
{[}see Fig.\ \ref{fig:decimation}{]}. In this case, the renormalized
coupling constants are given by Eqs.\ (\ref{eq:second-order-effec-coupl}),
which we rewrite as 
\begin{equation}
\tilde{K}^{(J)}=\frac{K_{1}^{(J)}K_{3}^{(J)}}{\mathbf{v}^{(J)}\cdot\mathbf{K}_{2}},\label{eq:2nd-S32}
\end{equation}
with the vectors $\mathbf{v}^{(1)}=\frac{1}{5}\left(2,-30,147\right)$,
$\mathbf{v}^{(2)}=-\frac{1}{10}\left(4,-40,49\right)$, and $\mathbf{v}^{(3)}=\frac{1}{35}\left(16,-40,126\right)$.
We explicitly verified these results by numerically implementing the
SDRG decimations.

\begin{figure}
\begin{centering}
\includegraphics[clip,width=0.9\columnwidth]{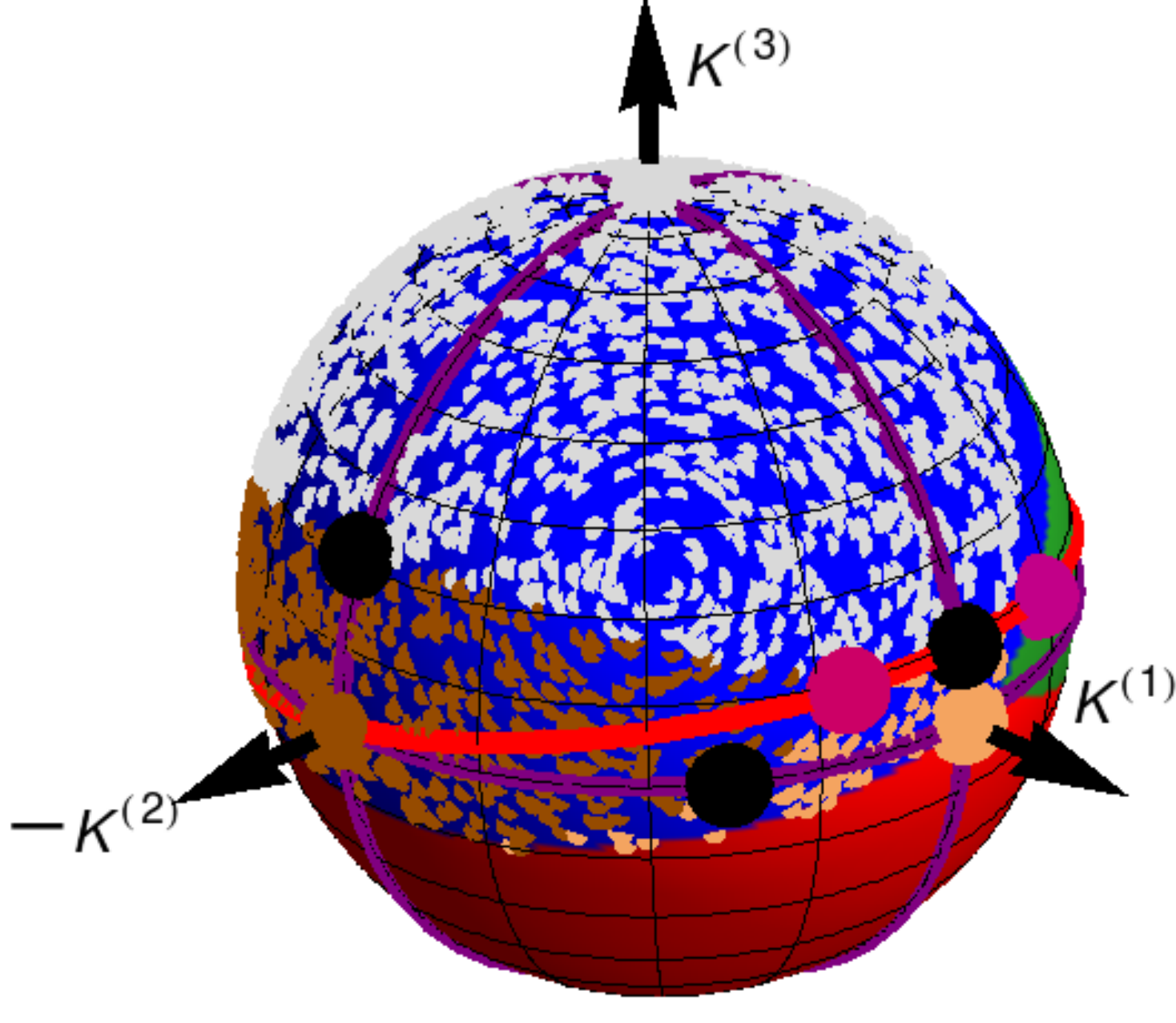}
\par\end{centering}

\caption{(Color online) Schematic phase diagram of the disordered spin-$\frac{3}{2}$
chain on and around the AF octant $K^{(1)}\times-K^{(2)}\times K^{(3)}$.
The thick dots on the surface of the sphere represent initial conditions
of the numerical RG flow. The dots are colored according to their
fate in the flow and, therefore, map out the distinct basins of attraction.
The beige dots flow to the $K^{\left(1\right)}>0$ (Heisenberg) fixed
point (beige circle), the brown ones to the $K^{\left(2\right)}<0$
fixed point (brown circle), while the white ones have only $K^{\left(3\right)}>0$
couplings at the fixed-point Hamiltonian (white circle on the North
pole). The pink fixed point in the middle of the octant is totally
unstable and has an exact SU(4) symmetry. The other semi-stable fixed
points are represented as black circles. The ground state of the \emph{local}
Hamiltonians {[}$H_{23}$ in Eq.~\eqref{eq:H4sites}{]} is indicated
by the background color, see Table \ref{tab:Color-scheme} for the
color scheme (the FM $\tilde{S}=3$ orange region is not visible from
this viewing angle). The red and green regions (not marked by any
dot for clarity) correspond to the LSP fixed-point. On the red line,
the system has an exact SO(5) symmetry.\label{fig:AFoctant} }
\end{figure}

In Fig.\ \ref{fig:AFoctant} we plot the resulting flow diagram on
and around the AF octant. We have chosen many initial conditions,
parameterized by the initial angles $\theta,\,\phi$ (represented
as thick dots) and disorder strength $D$. We confirm the flow to
be independent on $D$ (as long as $D$ is sufficiently large). For
small $D$, one has to be careful. A consistent approach would be
to treat the perturbations to the local Hamiltonian (\ref{eq:H4sites})
to higher orders of perturbation theory than the second.\cite{hoyos08}
This approach becomes much more involved and we do not do it the present
study. Nonetheless, it is worth mentioning that, even at the Heisenberg
point this issue is still controversial. In this case, the recursion
relation (\ref{eq:2nd-S32}) reduces to $\tilde{K}^{(1)}=\frac{5K_{1}^{(1)}K_{3}^{(1)}}{2K_{2}^{(1)}}$.
The numerical prefactor $\frac{5}{2}>1$, which means the renormalized
coupling can be bigger than the decimated ones if the bare disorder
is weak. Therefore, the decimation procedure is internally inconsistent.
As a result, it has been argued that the weak-disorder regime corresponds
to a random singlet state with spin-1/2 rather than 3/2 excitations\ \cite{RefaelFisher}.
Another view is that weak disorder is irrelevant\ \cite{Saguia2003,Carlon2004}.
In all of these studies, however, there is general agreement that
at strong disorder the ground state is a spin-3/2 random singlet state
as reported here.

The dots on the surface of the sphere are colored according to the
corresponding fixed point to which they flow (beige, brown and white
circles on the axes). As expected, all initial conditions inside the
AF octant flow to one of the stable fixed points on the axes of the
octant. Note that they attract even other initial conditions outside
(but close to) the AF octant. The corresponding pairwise RSP has,
therefore, a large basin of attraction. In addition, we display the
semi-stable planar fixed points (black circles) as well as the totally
unstable SU(4) symmetric fixed point (pink circle) inside the octant.
Their locations are determined analytically, as shown below, and agree
with our numerical data.

In order to find the location of the fixed points, we need to solve
Eq.\ (\ref{eq:aFPcondition}) for the $S=3/2$ case. Defining the
vector $\mathbf{s}_{i}=\left(s_{i}^{\left(2\right)},s_{i}^{\left(3\right)}\right)$
where $s_{i}^{\left(2,3\right)}=\frac{K_{i}^{\left(2,3\right)}}{K_{i}^{\left(1\right)}}$,
then from Eq.\ (\ref{eq:2nd-S32}) we have 
\begin{align}
\tilde{s}^{(2)} & =-2s_{1}^{\left(2\right)}s_{3}^{\left(2\right)}\left(\frac{2-30s_{2}^{(2)}+147s_{2}^{\left(3\right)}}{4-40s_{2}^{(2)}+49s_{2}^{\left(3\right)}}\right),\\
\tilde{s}^{(3)} & =\frac{7}{2}s_{1}^{\left(3\right)}s_{3}^{\left(3\right)}\left(\frac{2-30s_{2}^{(2)}+147s_{2}^{\left(3\right)}}{8-20s_{2}^{(2)}+63s_{2}^{\left(3\right)}}\right).\label{eq:angularFPs3}
\end{align}
Since the fixed points are such that $\mathbf{s}_{i}=\tilde{\mathbf{s}}_{i}=\mathbf{s}^{*}$,
we find the following physical solutions besides the three stable
ones already obtained $K_{i}^{(J)}=K_{i}\delta_{J,K}$, with $J=1,2,3$:
(i) the semi-stable AF fixed point on the $K^{\left(3\right)}=0$
plane $\mathbf{s}_{4}^{*}=-\frac{1}{30}\left(9+\sqrt{141}\right)\left(1,0\right)\approx\left(-0.70,0\right)$,
which corresponds to $\left(\theta_{4}^{*},\phi_{4}^{*}\right)\approx\left(\frac{\pi}{2},-35^{\circ}\right)$;
(ii) the semi-stable AF fixed point on the $K^{(2)}=0$ plane $\mathbf{s}_{5}^{*}=\left(0,\frac{4}{21}\right)\approx\left(0,0.19\right)$,
or $\left(\theta_{5}^{*},\phi_{5}^{*}\right)\approx\left(79^{\circ},0\right)$;
(iii) the semi-stable AF fixed point on the $K^{(1)}=0$ plane at
which $\left|\mathbf{s}_{6}^{*}\right|$ is infinite but $s_{6}^{(3)*}/s_{6}^{(2)*}=\frac{2}{49}\left(1-\sqrt{141}\right)\approx-0.44$
is finite; this is equivalent to $\left(\theta_{6}^{*},\phi_{6}^{*}\right)\approx\left(66^{\circ},-\frac{\pi}{2}\right)$;
and (iv) the totally unstable SU(4)-symmetric fixed point $\mathbf{s}_{7}^{*}=\left(-\frac{1}{3},\frac{4}{21}\right)$,
which corresponds to $\left(\theta_{7}^{*},\phi_{7}^{*}\right)\approx\left(80^{\circ},-18^{\circ}\right)$.
Notice all these solutions obey the restriction $\left(-1\right)^{J}K_{i}^{(J)*}\leq0$.
Besides these seven solutions, there are additional ones {[}for instance
$\left(\frac{2}{3},\frac{4}{21}\right)${]} which are, however, nonphysical
since the singlet is not the ground state of the local Hamiltonian.

As anticipated in Section~\ref{sub:PairwiseRSP}, we note that the
unstable AF SU(4)-symmetric point can be adiabatically connected to
\emph{all} the other AF stable fixed points. In other words, the ground
state \emph{throughout the AF region }is a collection of the same
singlets as at the SU(4) symmetric point. These singlets are, therefore,
SU(4) singlets, and the ground state is SU(4) invariant. This is one
of our main findings: the ground state singlets inherit the symmetry
of the SU(4) fixed points. The same is true of the lowest excitations,
essentially free spins, which can be seen as transforming according
to either the fundamental of the anti-fundamental representations
of SU(4). The low-energy sector has, therefore, an emergent SU(4)
symmetry, analogous to the mesonic SU(3) RSP found before in spin-1
systems.~\cite{QuitoHoyosMiranda}

As for the other regions of the phase space, we find generically that
the flow is towards a LSP, with only $K_{i}^{\left(1\right)}\ne0$,
except in the region where $\tilde{S}=3$ (not visible in Fig.~\ref{fig:AFoctant}),
where the phase is ferromagnetic.

\subsection{SO(5) line and SU(4) points}

As shown in Ref.~\onlinecite{Wumodphys06}, the Hamiltonian of isotropic
spin-$\frac{3}{2}$ chains has an enlarged SO(5) symmetry in a certain
region of parameter space. Even more interesting, this region happens
to be the one accessible to cold-atom experiments. In such experiments,
for spin-$\frac{3}{2}$ particles at quarter filling and in the limit
of strong interactions, selection rules for the scattering of two
atoms at low energies impose that only channels of even total angular
momentum are allowed. In this case, putting $\epsilon_{i}^{(1)}=\epsilon_{i}^{(3)}=0$
in the Hamiltonian of Eq.~\eqref{eq:Hamilt_projectors} we get

\begin{equation}
H_{{\rm SO\left(5\right)}}=\sum_{i=1}^{N_{{\rm sites}}}\epsilon_{i}^{(0)}P_{0}\left(\mathbf{S}_{i},\mathbf{S}_{i+1}\right)+\epsilon_{i}^{(2)}P_{2}\left(\mathbf{S}_{i},\mathbf{S}_{i+1}\right),
\end{equation}
which was shown to possess an exact SO(5) symmetry.\cite{Wumodphys06}
The IST coupling constants then are, neglecting a constant contribution
(see Appendix~\ref{sec:Appendix:notation})

\begin{eqnarray}
K_{i}^{\left(1\right)} & = & -\frac{\pi}{15}\left(\epsilon_{i}^{(0)}+\epsilon_{i}^{(2)}\right),\\
K_{i}^{\left(2\right)} & = & \frac{\pi}{45}\left(\epsilon_{i}^{(0)}-3\epsilon_{i}^{(2)}\right),\\
K_{i}^{\left(3\right)} & = & -\frac{4\pi}{315}\left(\epsilon_{i}^{(0)}+\epsilon_{i}^{(2)}\right).
\end{eqnarray}
The vector $\mathbf{s}_{i}=\left(K_{i}^{(2)},K_{i}^{(3)}\right)/K_{i}^{(1)}=\left(\varepsilon_{i},\frac{4}{21}\right)$,
where $\varepsilon_{i}=\left[\epsilon_{i}^{(2)}-\epsilon_{i}^{(0)}/3\right]/\left[\epsilon_{i}^{(2)}+\epsilon_{i}^{(0)}\right]$,
with $-\infty<\varepsilon_{i}<\infty$ a free parameter. Thus, the
parameter space in which the SO(5) symmetry is realized is a line
on the surface of our 2-sphere. This is shown as the red line in Fig.~\ref{fig:AFoctant}.
Interestingly, this line contains three of the AF fixed points found:
$\mathbf{s}_{2}^{*}=\left(-\infty,x\right)$, with $x$ being any
finite number (this corresponds to the totally stable fixed point
$K_{i}^{(2)}<0$ and $K_{i}^{(1)}=K_{i}^{(3)}=0$), $\mathbf{s}_{5}^{*}$
(the semi-stable fixed point on the $K_{i}^{(2)}=0$ plane) and $\mathbf{s}_{7}^{*}$
{[}the totally unstable SU(4)-symmetric fixed point{]}.

As shown in Appendix\ \ref{sec:appendix-SU(N)-invariant-Hamil} (see
also \cite{Jonesbook,Wumodphys06}), the Hamiltonian 

\begin{eqnarray}
H_{4-\bar{4}}^{{\rm SU(4)}} & = & \sum_{i}K_{i}\left(\hat{O}_{1}-\frac{1}{3}\hat{O}_{2}+\frac{4}{21}\hat{O}_{3}\right),\label{eq:su4-qaq}
\end{eqnarray}
with generic $K_{i}$, which corresponds to $\mathbf{s}_{i}=\left(-\frac{1}{3},\frac{4}{21}\right)$,
is SU(4)-symmetric. This is indeed the totally unstable fixed point
$\mathbf{s}_{7}^{*}$ in the AF case ($K_{i}>0$). The notation $4-\bar{4}$
indicates that the Hamiltonian \eqref{eq:su4-qaq} corresponds to
a Heisenberg SU(4) spin chain with ``spin'' operators on odd (even)
sites which are the generators of the fundamental (anti-fundamental)
representation of the SU(4) group. In addition, there is another AF
SU(4)-symmetric Hamiltonian given by 

\begin{eqnarray}
H_{4-4}^{{\rm SU(4)}} & = & \sum_{i}K_{i}\left(\hat{O}_{1}+\frac{1}{3}\hat{O}_{2}+\frac{4}{21}\hat{O}_{3}\right).
\end{eqnarray}
The difference with respect to the Hamiltonian \eqref{eq:su4-qaq}
is that all the ``spin'' operators are generators of the fundamental
representation of the SU(4) group regardless of the lattice site.
Although our SDRG scheme is not suitable for treating this case (since
the multiplets $\tilde{S}=0$ and $\tilde{S}=2$ become degenerate
in the ground state of the local Hamiltonian), we know this must be
a fixed point of the RG, since it has to preserve the symmetry. We
then denote this fixed point by $\mathbf{s}_{8}^{*}=\left(\frac{1}{3},\frac{4}{21}\right)$,
which corresponds to $\left(\theta_{8}^{*},\phi_{8}^{*}\right)\approx\left(80^{\circ},18^{\circ}\right)$,
see the pink circle in Fig.\ \ref{fig:AFoctant}. Indeed, in Ref.\ \onlinecite{HoyosMiranda}
it was shown that the low-energy physics of this SU(4) $4-4$ Hamiltonian
is governed by an infinite-randomness fixed point, with the local
energy scales being distributed according to Eq.\ \eqref{eq:P(x)-IRFP}
with a universal tunneling exponent $\psi=\frac{1}{4}$. In addition,
the corresponding ground state is a random singlet state whose singlets
are formed by groups of spins which are multiples of 4: a 4-fold random
singlet state analogous to the triplewise random singlet state depicted
in Fig.\ \hyperref[fig:random-singlets]{\ref{fig:random-singlets}(b)}. 

Therefore, the SO(5) line contains both AF SU(4)-symmetric fixed points
$\mathbf{s}_{7}^{*}$ and $\mathbf{s}_{8}^{*}$. In addition, the
fixed points $\mathbf{s}_{2,5}^{*}$ are also exactly SO(5) symmetric.
Nonetheless, recall they have an emergent SU(4) symmetry.

We have also verified that the SO(5) line is a constant of the flow.
Let us explain this a little further. Starting from any point ($\theta,\phi$)
on the surface of the 2-sphere in Fig.\ \ref{fig:AFoctant}, the
angles change along the RG flow in such a way that one cannot represent
the renormalized Hamiltonian by a single point ($\tilde{\theta},\tilde{\phi}$),
but rather by a distribution of angles. Generically, this distribution
has support on two-dimensional manifolds on the 2-sphere. If one starts
at any point on the SO(5) line, however, all the renormalized angles
($\tilde{\theta}_{i},\tilde{\phi_{i}}$) will remain in the SO(5)
line (a one-dimensional manifold) along the RG flow. This means that
our SDRG scheme preserves the symmetry, as it should.

It is interesting to see that the unstable fixed points on this line
are the points with a larger SU(4) symmetry. Between them there is
the semi-stable fixed point $\mathbf{s}_{5}^{*}$. As we will see
below, the SU(4)-symmetric $4-4$ fixed point $\mathbf{s}_{8}^{*}$
governs the transition between the AF phase and the Large Spin phase.
The topology of the flow requires that there be another fixed point
towards the southwest of the totally stable fixed point $\mathbf{s}_{2}^{*}$.
Very likely, this fixed point delimits the transition between the
FM and AF phases. Like in the spin-1 case,\cite{QuitoHoyosMiranda}
this fixed point is a FM SU(4) symmetric fixed point.

\begin{figure}
\begin{centering}
\includegraphics[clip,width=0.8\columnwidth]{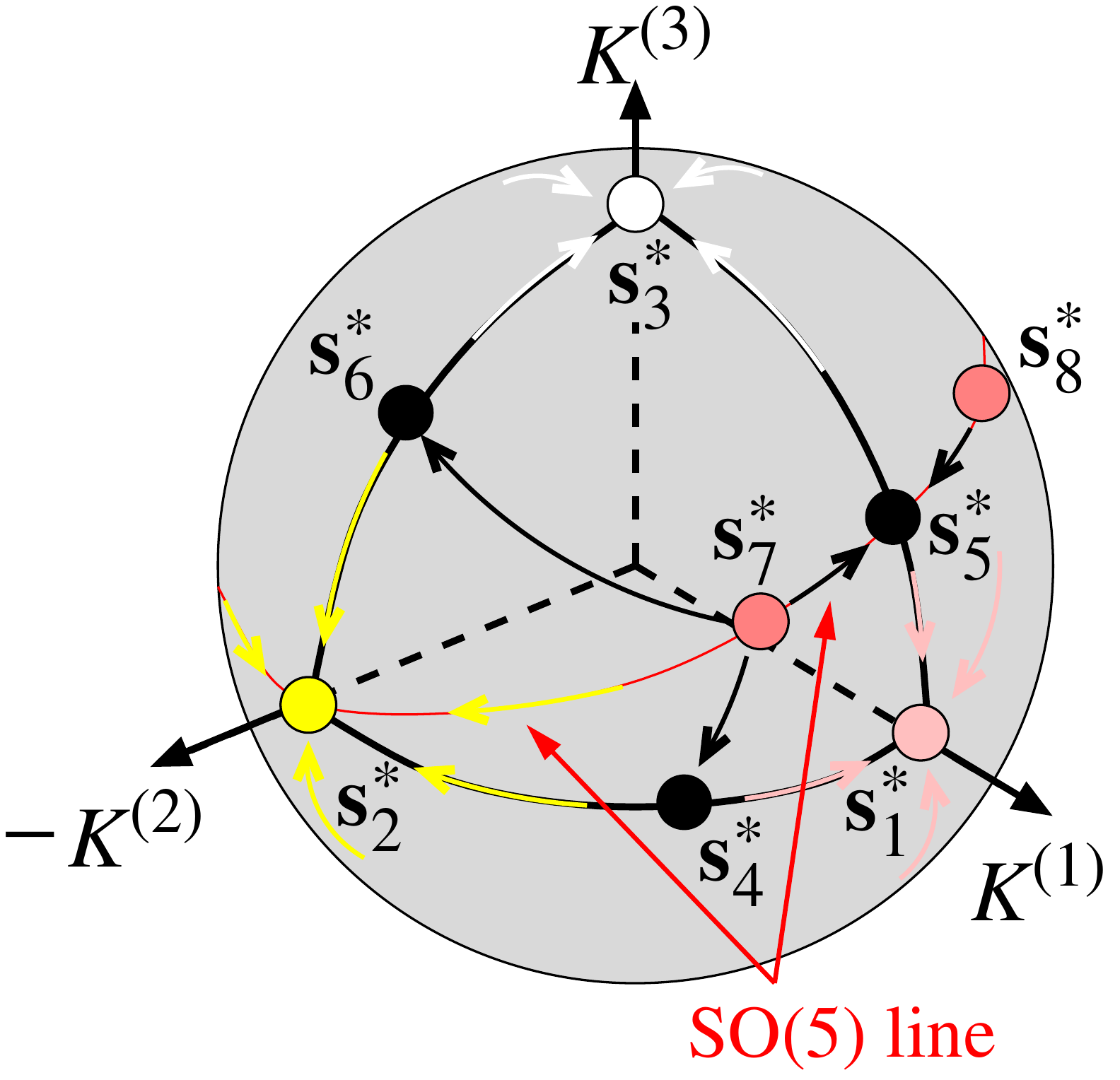}
\par\end{centering}

\caption{Schematic flow diagram of the \emph{average} angular variables in
the AF octant of random spin-$\frac{3}{2}$ chains (based on the phase
diagram of Fig~\ref{fig:AFoctant}). See text for details about the
fixed points $\mathbf{s}_{i}^{*}$. \label{fig:angle-streamlines}}
\end{figure}

It is visually instructive to follow the SDRG flow of the angular
variables $\mathbf{s}_{i}$. It is useful to normalize them so the
flow is confined to the surface of a unit sphere, as in Fig.\ \ref{fig:AFoctant}.
Several trajectories of the \emph{average} values of $\mathbf{s}_{i}$
are shown in Fig.~\ref{fig:angle-streamlines}. This gives some intuition
about the way these variables approach the fixed points. We point
out, however, that the flow of the \emph{distribution} of $\mathbf{s}_{i}$
is not captured by this figure. In particular, their widths start
at zero, become non-zero at intermediate stages of the flow, and tend
to zero again as the fixed points are approached.

\subsection{RG flow on the other semi-axes}

\label{sub:otheraxes}

As mentioned in Section~\ref{sub:First-Order-Perturbation}, the
first-order perturbation theory steps yield a vanishing renormalized
coupling constant in some special cases, even if the ground state
multiplet is degenerate. Curiously, this happens for the spin-3/2
chain on the semi-axes $K^{(2)}>0$ and $K^{(3)}<0$, i.e., on all
other semi-axes except the ones of the AF octant and the purely FM
Heisenberg axis $K^{(1)}<0$. We show how to partially handle these
cases, by finding the first non-zero contributions. 

Before addressing the $J>1$ cases, let us start by reviewing the
first rank case. If $K_{2}^{\left(1\right)}<0$, the ground state
is a spin $\tilde{S}=3$ and the RG rule is the one given in Sec.~\ref{sub:Large-Spin-Phase}.
The mixing of random $K_{i}^{\left(1\right)}$ signs leads to a LSP\cite{westerbergetal},
whereas exclusively negative couplings lead to a FM phase. 

Now we focus on higher rank tensors. On the semi-axis $K^{\left(2\right)}<0$,
first-order perturbation theory generically gives (see Fig.\ \ref{fig:decimation}
for a guide) 

\begin{equation}
\tilde{K}_{1}^{\left(2\right)}=\frac{3\tilde{x}^{2}+\tilde{x}\left(2x_{2}-6x_{3}-3\right)+3\left(x_{2}-x_{3}-1\right)\left(x_{2}-x_{3}\right)}{2\tilde{x}\left(4\tilde{x}-3\right)}K_{1}^{\left(2\right)}\label{eq:renor_sec_rank}
\end{equation}
where $x_{i}=S_{i}\left(S_{i}+1\right)$ and $\tilde{x}=\tilde{S}\left(\tilde{S}+1\right)$.
The $\tilde{K}_{3}^{\left(2\right)}$ renormalization follows analogously,
by exchanging $x_{2}\rightleftharpoons x_{3}$. Starting with a chain
in which all spins are of size $S_{i}=\frac{3}{2}$, then already
in the first decimation step $x_{2}=x_{3}=\frac{15}{4}$. As the ground
state of the local Hamiltonian is $\tilde{S}=2$, then first-order
perturbation theory yields $\tilde{K}_{1,3}^{\left(2\right)}=0$.
In Appendix~\ref{sec:appendix-zeros}, we have shown how to calculate
second-order perturbative corrections. The steps are analogous to
those shown in the second order calculations when the ground state
is a singlet, except that the projector onto the ground state $P_{0}$
has to be replaced by $P_{\tilde{S}}$. The net result, when three-body
non-frustrating terms are neglected, is the appearance of a non-zero
coupling between spin $\mathbf{S}_{1}$ and the effective spin $\tilde{\mathbf{S}}$

\begin{equation}
\Delta H_{1,2}^{\left(2\right)}=\frac{\left(K_{1}^{\left(2\right)}\right)^{2}}{K_{2}^{\left(2\right)}}\left(\frac{9}{16}\hat{O}_{1}-\frac{3}{56}\hat{O}_{3}\right).\label{eq:low-energy-k2-axis-1}
\end{equation}
where $\hat{O}_{i}=\hat{O}_{i}\left(\mathbf{S}_{1}=\frac{3}{2},\mathbf{\tilde{S}}=2\right)$.
By symmetry, we obtain the coupling connecting the spin $\tilde{\mathbf{S}}$
to site 4 by replacing $1\leftrightarrow4$. The RG rules are schematically
shown in Fig.~\ref{fig:RG-step-correc} of Appendix~\ref{sec:appendix-zeros}.
This prescription is enough to fix the first RG decimations, but the
zeros proliferate again in later decimations. Up to the point at which
the flow is not dominated by theses zeros, we do not find indications
of a Large Spin phase. The phase appears to be antiferromagnetic,
even though its full characterization would require going to higher
orders in perturbation theory.

Starting with a negative $K_{2}^{\left(3\right)}$ gives us a spin
$\tilde{S}=1$ as the ground state manifold. But a spin-1 Hamiltonian
does not support third-rank ISTs (as a rule, remember that spin operators
with $2S<J$ do no form rank-$J$ ISTs). Note that this corresponds
to case (a) discussed in Section~\ref{sub:First-Order-Perturbation},
whereas on the $K^{\left(2\right)}<0$ axis it was related to case
(b) of that Section. In Appendix~\ref{sec:appendix-zeros}, we show
in detail how to compute second-order corrections for this case as
well. The effective Hamiltonian that connects a spin $S_{1}=\frac{3}{2}$
with a spin $\tilde{S}=1$ is, neglecting three-body non-frustrating
interactions, is

\begin{equation}
\Delta H_{1,2}=\frac{\left(K_{1}^{\left(3\right)}\right)^{2}}{\left|K_{2}^{\left(3\right)}\right|}\left(\frac{63}{20}\hat{O}_{1}+\frac{189}{100}\hat{O}_{2}\right).
\end{equation}
Here, $\hat{O}_{i}=\hat{O}_{i}\left(\mathbf{S}_{1}=\frac{3}{2},\mathbf{\tilde{S}}=1\right)$
and the analogous term for site 4 follows by symmetry. Again, this
prescription is enough to fix only the first decimation steps. As
before, there are no clear indications of a Large Spin phase being
generated.

Clearly, further investigation is needed to characterize this phase.
Either the inclusion of the three-body interactions remedy the vanishing
renormalized interactions, or the proliferation of these zeroes are
indeed part of the physics, indicating a breakdown of the perturbative
treatment and that a new approach is necessary. We leave as an open
question the elucidation of this problem. Here, we argue, however,
that these problems have little effect on the generic RG flow. In
general, the $K_{i}^{(1)}$ couplings are nonzero and never yield
any vanishing renormalizations. Therefore, they are dominant over
the other higher-rank interactions and the generic RG flow will be
towards the $K_{i}^{(1)}\neq0$ fixed points.

\subsection{RG flow on planes}

\label{sub:flowonplanes}

\begin{figure}
\begin{centering}
\includegraphics[clip,width=1\columnwidth]{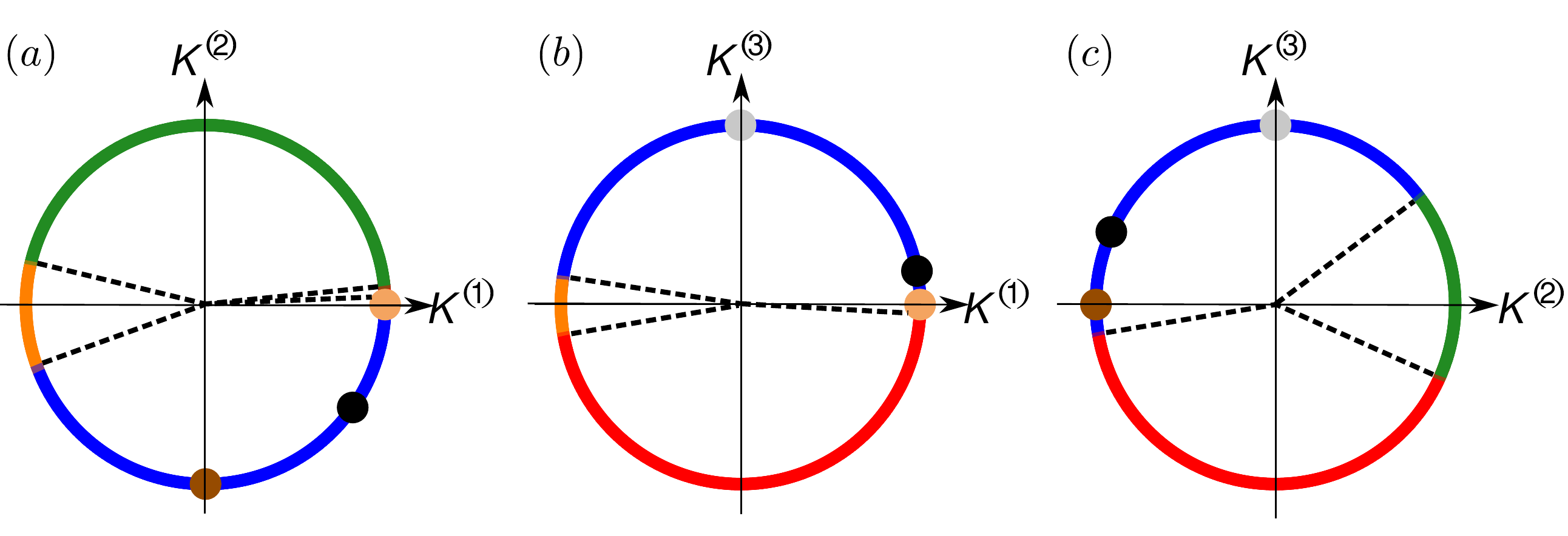}
\par\end{centering}

\caption{(Color online) Diagrams representing the 2-spin ground multiplet for
two spins $3/2$ and the AF RG fixed points of the disordered spin-3/2
chain. (a), (b) and (c) represent the $K^{\left(3\right)}=0$, $K^{\left(2\right)}=0$
and $K^{\left(1\right)}=0$ planes, respectively. The total angular
momentum of the ground multiplets can be $\tilde{S}=0,~1,~2$ or $3$,
which are identified by the color scheme of Table~\ref{tab:Color-scheme}.
The small circles represent the AF angular fixed points. The stable
AF ones lie on the semi-axes with $K^{\left(1\right)}>0$, $K^{\left(2\right)}<0$,
and $K^{\left(3\right)}>0$. The unstable fixed points can also be
found analytically (see the main text for details). \label{fig:planediagrams} }
\end{figure}

Having analyzed the behavior on the different semi-axes, we now explore
planes on which two of the tensor couplings are non-zero. The 2-spin
ground state structure is shown in Fig.~\ref{fig:planediagrams}.
In Table~\ref{tab:Color-scheme}, we list the colors we are going
to use to identify the ground multiplets in the study of both spin-$\frac{3}{2}$
and spin-$2$ chains.

\begin{table}
\begin{centering}
\begin{tabular}{|c|c|}
\hline 
2-spin ground state  & Color\tabularnewline
\hline 
\hline 
$\tilde{S}=0$ & blue\tabularnewline
\hline 
$\tilde{S}=1$ & red\tabularnewline
\hline 
$\tilde{S}=2$ & green\tabularnewline
\hline 
$\tilde{S}=3$ & orange\tabularnewline
\hline 
$\tilde{S}=4$ & purple\tabularnewline
\hline 
\end{tabular}
\par\end{centering}

\caption{Color scheme we are going to use for the identification of ground
multiplets of the local Hamiltonian in the analysis of the spin-$\frac{3}{2}$
and spin-$2$ chains. The same applies to Figs.\ \ref{fig:AFoctant},
\ref{fig:planediagrams}, \ref{fig:Spin-2-AF-circles} and \ref{fig:Spin-2-phase-diagram-ball}.\label{tab:Color-scheme}}
\end{table}

The RG flow in the singlet (blue) region is towards the fully stable
fixed points on the semi-axes discussed in Section~\ref{sub:afoctant}.
We now address the flow in the other non-blue regions of the plane,
where the ground state is not a singlet. We always assume the initial
angles are uniform and the disorder is in the radial direction. Let
us assume additionally that we do not start right on the axes, since
these cases have been discussed previously.

Let us start by analyzing the $K^{\left(1\right)}\times K^{\left(2\right)}$
and $K^{\left(1\right)}\times K^{\left(3\right)}$ planes. In the
orange region, the flow is towards the FM phase, with only $K_{i}^{\left(1\right)}<0$
remaining. Starting in both the red $\left(\tilde{S}=1\right)$ or
the green $\left(\tilde{S}=2\right)$ regions, effective spins that
are not equal to the original spin $\frac{3}{2}$ are generated. Notice
that, in both regions, only the $K_{i}^{\left(1\right)}\ne0$ couplings
remain since the other ones are automatically renormalized to zero,
for the very same reasons discussed in the previous section. After
some initial steps, we end up with a ``soup'' of spin $\frac{3}{2}$
and spins $\tilde{S}$ (equal to 2 or 1, depending on the case) coupled
by $K_{i}^{\left(1\right)}$ couplings. By numerically following the
flow, we find out that after an initial transient, the spins start
to grow and the flow is towards a LSP.

We now focus on the $K^{\left(2\right)}\times K^{\left(3\right)}$
plane. In the green region of the first quadrant, the $K_{i}^{\left(2\right)}$
couplings are renormalized to zero (due to the zeroes discussed in
the previous section), effective spins $\tilde{S}=2$ are generated,
and the corresponding renormalized couplings $K_{i}^{\left(3\right)}$
are negative. The presence of both spins 2 and 3/2 gives rise to the
RG rules shown in Table~\ref{tab:RG-rules-green-plane} (which can
be found with the formulas derived in Section~\ref{sec:Method} and
Appendix~\ref{sec:Appendix:RGstep}). Interestingly, these rules
are closed under the RG transformations. We find numerically that,
at low energies, only rules of type 4 survive, and the phase is again
a RSP on the $K^{\left(3\right)}>0$ semi-axes, with exponent $\psi=\frac{1}{2}$.
Notice that, unlike the RSP found previously (in the blue region),
this one depends strongly on the generated spins $\tilde{S}\ne0$
and the possible combinations of $\tilde{S}$ and $S=\frac{3}{2}$
on later RG steps. In the red region of the third quadrant, the RG
flow is very similar to the one just discussed, with the changes $K^{\left(2\right)}\rightleftarrows-K^{\left(3\right)}$
and $\tilde{S}=1$. Again, the low energy physics is found to be described
by a RSP with exponent $\psi=\frac{1}{2}$, now on the $K^{\left(2\right)}<0$
semi-axis. In the fourth quadrant of this plane, the first RG decimations
make one of the couplings, either $K_{i}^{\left(2\right)}$ or $K_{i}^{\left(3\right)}$,
vanish, depending on the region where the bond is located (red or
green). At later RG steps, however, the remaining non-zero couplings
also vanish. Therefore, we do not have the full low-energy description
in this quadrant, as in the situation discussed in Section~\ref{sub:otheraxes}.
Note that a small perturbation in the perpendicular $K^{\left(1\right)}$
direction already fixes this problem and drives the system towards
a LSP.

\begin{center}
\begin{table*}
\begin{centering}
\begin{tabular}{|c|c|c|c|c|c|}
\hline 
Rule & $S_{2}$  & $S_{3}$  & Couplings & $\tilde{S}$  & RG rules\tabularnewline
\hline 
\hline 
1 & $\frac{3}{2}$  & $\frac{3}{2}$  & $K_{2}^{\left(2\right)}>0,\,\,K_{2}^{\left(3\right)}>0$  & $2$  & $\tilde{K}_{\left\{ 1,3\right\} }^{\left(3\right)}=-\frac{1}{4}K_{\left\{ 1,3\right\} }^{\left(3\right)}$
and $\tilde{K}_{\left\{ 1,3\right\} }^{\left(2\right)}=0$\tabularnewline
\hline 
2 & $2$  & $\frac{3}{2}$  & $K_{2}^{\left(3\right)}<0,\,\,K_{2}^{\left(2\right)}=0$  & $\frac{3}{2}$  & $\tilde{K}_{1}^{\left(3\right)}=-\frac{8}{5}K_{1}^{\left(3\right)}$
and $\tilde{K}_{3}^{\left(3\right)}=\frac{1}{5}K_{3}^{\left(3\right)}$\tabularnewline
\hline 
3 & $2$  & $2$  & $K_{2}^{\left(3\right)}>0,\,\,K_{2}^{\left(2\right)}=0$  & $0$  & 2nd order; $\tilde{K}_{1,3}^{\left(3\right)}=\frac{4}{9}\frac{K_{1}^{\left(3\right)}K_{3}^{\left(3\right)}}{K_{2}^{\left(3\right)}}$\tabularnewline
\hline 
4 & $\frac{3}{2}$  & $\frac{3}{2}$  & $K_{2}^{\left(3\right)}>0,\,\,K_{2}^{\left(2\right)}=0$  & $0$ & 2nd order; $\tilde{K}_{1,3}^{\left(3\right)}=\frac{5}{18}\frac{K_{1}^{\left(3\right)}K_{3}^{\left(3\right)}}{K_{2}^{\left(3\right)}}$\tabularnewline
\hline 
\end{tabular}
\par\end{centering}

\caption{RG rules for the flow that starts in the green region $\left(\tilde{S}=2\right)$
of the $K^{\left(2\right)}\times K^{\left(3\right)}$ plane {[}see
Fig.~\hyperref[fig:planediagrams]{\ref{fig:planediagrams}(c)}{]}.
\label{tab:RG-rules-green-plane}}
\end{table*}

\par\end{center}

\section{Spin-2 Chain \label{sec:Spin-2-Chain}}

We now study the disordered spin-2 chain. The parameter space is spanned
by $J_{\mathrm{max}}=4$ axes $K_{i}^{\left(J\right)}$, with $J=1,2,3,4$.
The RG steps can be found in Appendix~\ref{sec:Appendix-RG-equations-spins}.
We will be brief on features that are analogous to the spin-3/2 case
and will focus on the features which are new.

Let us start by focusing on the flow on the axes. As discussed before,
the AF stable fixed points lie along the semi-axes $K^{\left(1\right)}>0$,
$K^{\left(2\right)}<0$, $K^{\left(3\right)}>0$, and $K^{\left(4\right)}<0$.
Each one of these cases leads to a pairwise RSP (see Section~\ref{sub:PairwiseRSP}).
This conclusion holds throughout the AF hyper-octant, in the unit
3-sphere, defined by the above four semi-axes. Like in the spin-$3/2$
case, the negative $K^{\left(1\right)}$ semi-axis gives rise to a
FM phase and to a LSP when $K_{i}^{\left(1\right)}$ is both negative
and positive. On the other semi-axes, with the exception of the negative
$K^{\left(3\right)}$ axis to be discussed later, our RG scheme suffers
from the presence of zero renormalized couplings in first order of
perturbation theory (see Sec.\ \ref{sub:First-Order-Perturbation}),
in close analogy to the spin-$3/2$ case (see Sec.~\ref{sub:otheraxes}).

\begin{figure}[b]
\begin{centering}
\includegraphics[clip,width=0.8\columnwidth]{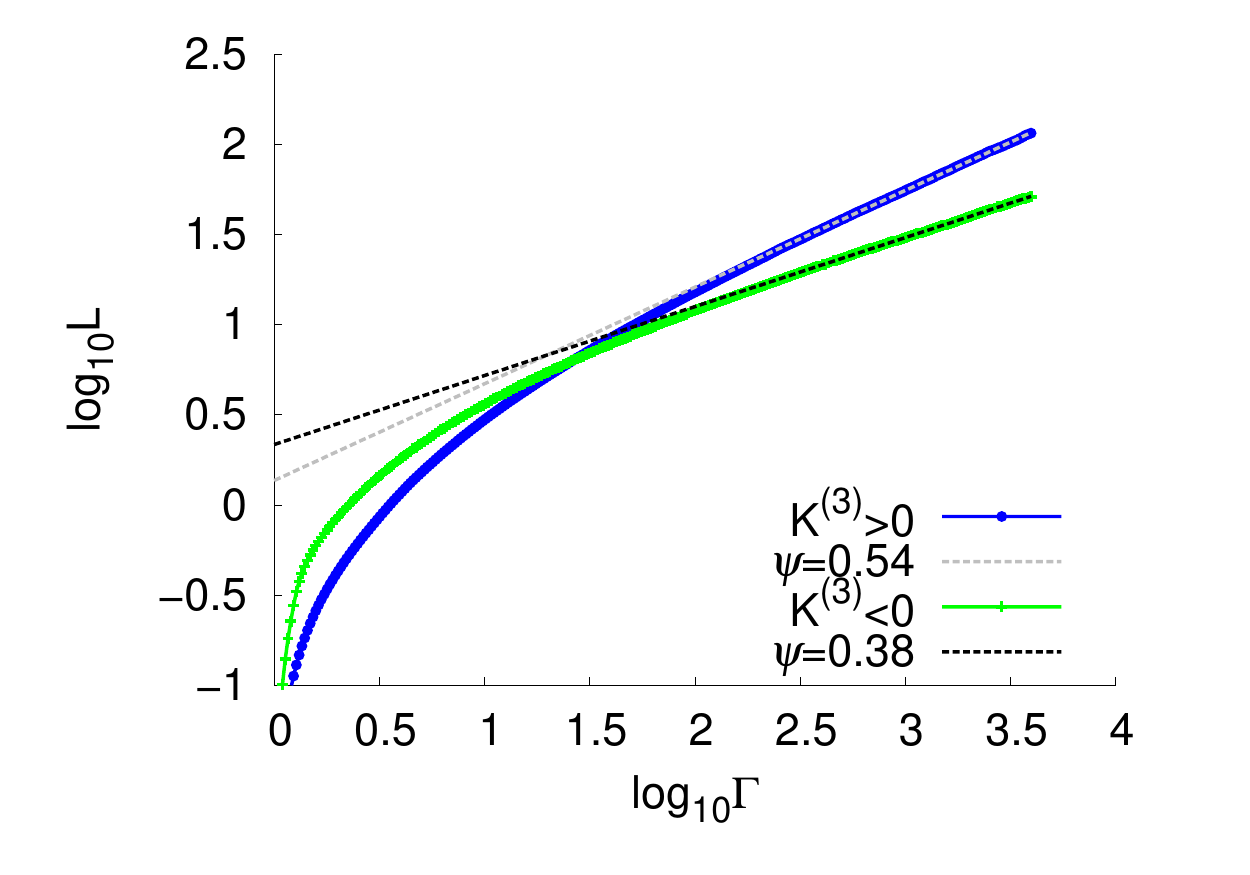}
\par\end{centering}

\caption{(Color online) Energy-length relation for couplings on the $K^{\left(3\right)}$
axes, starting with all spins equal to $S=2$. For negative initial
values, the numerical value of the tunneling exponent is $\psi\approx0.38$,
while for positive initial values $\psi=0.54$, compatible with the
predicted values of $\psi=\frac{1}{3}$ and $\psi=\frac{1}{2}$, respectively.
We have decimated chains of size $N_{{\rm sites}}=8\times10^{6}$.
\label{fig:Energy-length-s2-SU(3)}}
\end{figure}

The $K^{\left(3\right)}$ axis is special because it can also support
a triplewise RSP as discussed in Sec.\ \ref{sub:TriplewiseRSP}.
Whenever $K^{(3)}<0$, the corresponding first order decimation (see
Fig.\ \ref{fig:decimation}) yields an effective spin $\tilde{S}=2$.
Thus, the spin $S=2$ is a constant along the RG flow. In addition,
the effective couplings $\tilde{K}_{1,3}^{(3)}$ change signs meaning
this new effective spin can be later second-order decimated with third
spin into a singlet state. At the fixed point, the distribution of
coupling constants is of the infinite-randomness type given by Eq.~(\ref{eq:P(x)-IRFP})
with $\psi=\frac{1}{3}$. This universal fixed point attracts all
initial conditions in which the $K_{i}^{(3)}$ are either all negative
or have mixed signs. Only in the case of positive signs $K_{i}^{(3)}>0$
for any $i$ does the system flow towards the pairwise random singlet
case. This is illustrated in Fig.\ \ref{fig:Energy-length-s2-SU(3)}
where the length scale $L$ (defined as the mean distance between
effective spin clusters) is plotted along the RG flow parameterized
by the cutoff energy scale $\Omega$. From this plot, the tunneling
exponent $\psi$ can be extracted by fitting the activated dynamical
scaling law $\ln\Omega\sim-L^{\psi}$.

As discussed in Sec.\ \ref{sub:TriplewiseRSP}, this triplewise random
singlet phase is characteristic of integer spin-$S$ chains. In the
case of $S=1$, it was showed that the ground state exhibits an emergent
SU(3) symmetry.~\cite{QuitoHoyosMiranda} This is not the case, however,
for the spin-2 chain. This can be explicitly verified by diagonalizing
a system of 3 spins with $K_{i}^{\left(3\right)}<0$ and then computing
the correlation function of the 24 SU(5) generators $\Lambda_{i}^{\left(a\right)}$,
$i=1,2,3$ and $a=1,...,24$, in the correspond singlet state. Choosing
the normalization such that such that $\mbox{Tr}\left(\Lambda^{\left(a\right)}\Lambda^{\left(b\right)}\right)=2\delta^{a,b}$,
we find that

\begin{equation}
\left|\left\langle \Lambda_{1}^{\left(a\right)}\Lambda_{2}^{\left(a\right)}\right\rangle \right|=\frac{1}{35}\times\begin{cases}
7, & a=1,2,3,\\
3, & a=4,\ldots,8,\\
8, & a=10,\ldots,15,\\
4, & a=16,\ldots,24.
\end{cases}\label{eq:corr-prefactors}
\end{equation}
The number of equal values follows exactly the degeneracies of the
SU(2) multiplets. This implies that there is \textit{no symmetry enhancement}:
there is no symmetry higher than the obvious SU(2). This should be
contrasted with the situation in the AF hyper-octant. In that region,
the expectation value of all the SU(5) generators in the pairwise
singlets is the same as at the SU(5) point

\begin{equation}
\left|\left\langle \Lambda_{1}^{\left(a\right)}\Lambda_{2}^{\left(a\right)}\right\rangle \right|=\frac{2}{5},\,\,a=1,\ldots,24.
\end{equation}

This should also be contrasted with the spin-1 case. An analogous
computation {[}for the SU(3) group{]} of Eq.~\eqref{eq:corr-prefactors}
shows that all expectation values are the same.~\cite{QuitoHoyosMiranda}
For the random spin-2 chain, by contrast, in the triplewise RSP where
the singlets are mostly formed by spin trios, the symmetry remains
SU(2) as in the bare Hamiltonian.

\begin{figure}
\begin{centering}
\includegraphics[clip,width=1\columnwidth]{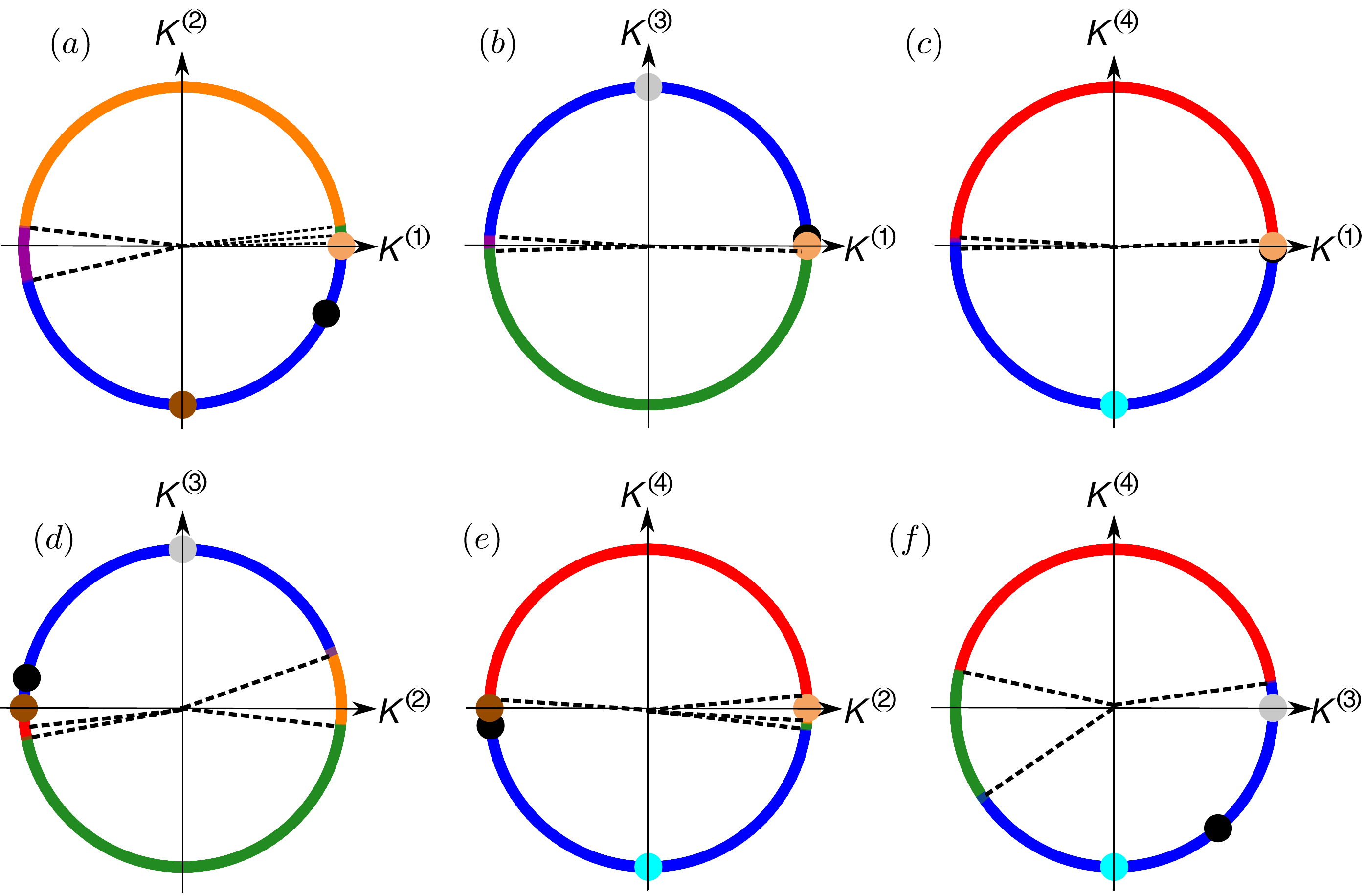}
\par\end{centering}

\caption{(Color online) Diagrams representing the 2-spin ground multiplet for
two spins $2$ and the AF RG fixed points of the disordered spin-2
chain in the various two-dimensional planes in $\mathbf{K}$ space.
The total angular momentum of the ground multiplets vary from $\tilde{S}=0$
to $\tilde{S}=4$ and it is identified by the color scheme of Table~\ref{tab:Color-scheme}.
The small circles represent the AF angular fixed points. The stable
AF ones lie on the semi-axes with $K^{\left(1\right)}>0$, $K^{\left(2\right)}<0$,
$K^{\left(3\right)}>0$, and $K^{\left(3\right)}<0$. The unstable
fixed points can also be found analytically (see the Appendix~\ref{sec:Appendix-RG-equations-spins}).
\label{fig:Spin-2-AF-circles}}
\end{figure}

\begin{figure}
\begin{centering}
\includegraphics[clip,width=0.9\columnwidth]{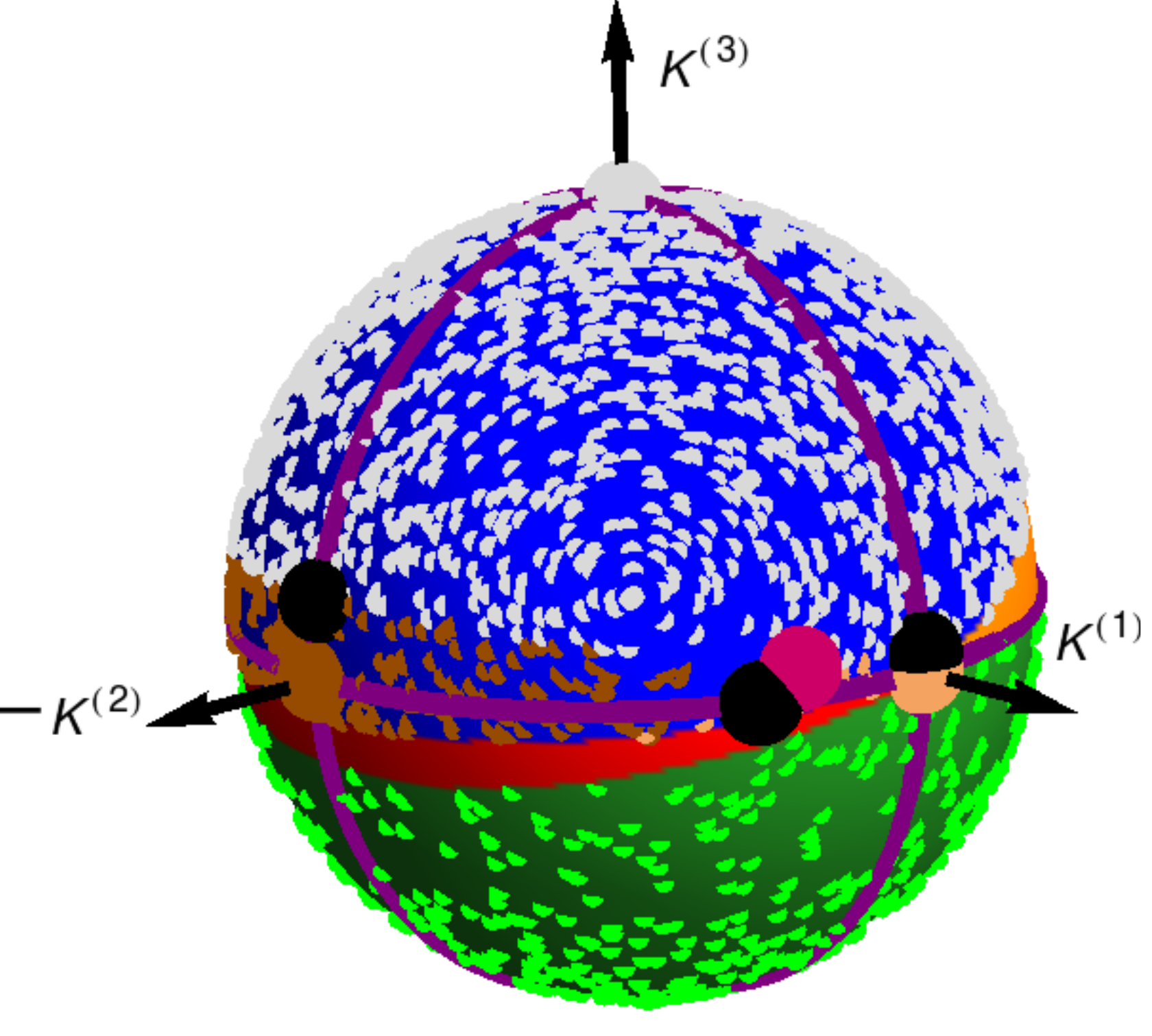}
\par\end{centering}

\caption{(Color online) Schematic phase/flow diagram of the disordered spin-$2$
chain on the unit 2-sphere of the hyper-space with $K^{\left(4\right)}=0$
and axes $K^{\left(1\right)}\times K^{\left(2\right)}\times K^{\left(3\right)}$.
The regions on the sphere are colored according to the spin $\tilde{S}$
of the ground multiplet of the 2-spin Hamiltonian, see Table \ref{tab:Color-scheme}
for the color scheme (the region with $\tilde{S}=4$ is not visible
from this viewing angle). The thick dots on the sphere's surface represent
initial conditions of the numerical RG flow, keeping always $K_{i}^{\left(4\right)}=0$.
A generic flow starting at a point in the blue region ends up at one
of the stable fixed points on the semi-axes $K^{\left(1\right)}>0$,
$K^{\left(2\right)}<0$, and $K^{\left(3\right)}>0$ (dots have the
same color as their final stable fixed points, which are represented
by large circles on the semi-axes). Also in the blue region, semi-stable
planar fixed points belonging to 2-planes (where only two $K^{\left(J\right)}$
are non-zero) are represented in black. A single planar fixed point
(belonging to the hyper-plane $K^{\left(4\right)}=0$) that is fully
unstable on this 2-sphere is shown in pink. RG flows starting in the
green region end up on a fixed point on the $K^{\left(3\right)}$
axis with random signs, corresponding to the triplewise RSP with exponent
$\psi=\frac{1}{3}$. The red and orange regions (not marked by any
dot for clarity) are attracted to LSP fixed points. \label{fig:Spin-2-phase-diagram-ball}}
\end{figure}

We now move to the analysis of the phases when ISTs of two different
ranks are present in the initial Hamiltonian. In analogy with the
spin-3/2 case (see Section~\ref{sub:flowonplanes}), besides the
stable fixed points on the semi-axes, some unstable planar fixed points
exist in the AF hyper-octant, as discussed in Section~\ref{sub:PairwiseRSP}.
Both types of fixed points are shown in Fig.~\ref{fig:Spin-2-phase-diagram-ball}
and their precise locations are given in Table \ref{fig:Fixed-point-conditions-spin-2}
of Appendix~\ref{sec:Appendix-RG-equations-spins}. The RG flow can
be analyzed similarly to the spin $\frac{3}{2}$ chain, with the additional
presence of the $\psi=\frac{1}{3}$ triplewise RSP. We outline some
of the general results. Starting in the blue region leads to a $\psi=\frac{1}{2}$
pairwise RSP, whereas starting in the green region leads to the $\psi=\frac{1}{3}$
triplewise RSP. The purple region is purely FM, as indicated by the
two-spin problem having $\tilde{S}=4$ as the local ground state.
The red region, where $\tilde{S}=1$, has to be analyzed in a case-by-case
manner, leading, for instance, to a LSP, if $K_{i}^{\left(1\right)}$
couplings are non-zero, as in the $K^{\left(1\right)}\times K^{\left(3\right)}$
plane {[}see Fig.~\hyperref[fig:Spin-2-AF-circles]{\ref{fig:Spin-2-AF-circles}(b)}{]},
or to the breakdown of our RG scheme, as in the case of the $K^{\left(3\right)}\times K^{\left(4\right)}$
plane {[}Fig.~\hyperref[fig:Spin-2-AF-circles]{\ref{fig:Spin-2-AF-circles}(f)}{]}.
In the latter, the RG fails because $2\tilde{S}<3$ and, therefore,
both couplings are renormalized to zero at all RG steps. The orange
region behaves similarly, leading also to vanishing coupling constants
{[}as in Fig.~\hyperref[fig:Spin-2-AF-circles]{\ref{fig:Spin-2-AF-circles}(d)}{]}
or to a LSP {[}as in Fig.~\hyperref[fig:Spin-2-AF-circles]{\ref{fig:Spin-2-AF-circles}(a)}{]}. 

Finally, we focus on the 3-dimensional region spanned by the $K^{\left(1\right)},~K^{\left(2\right)}$
and $K^{\left(3\right)}$ axes with $K^{(4)}=0$, which is a 3-dimensional
hyper-plane in $\mathbb{R}^{4}$. Other hyper-planes with $K_{i}^{\left(4\right)}\neq0$
yield qualitatively similar phase diagrams. We also remind the reader
that, according to the analysis of Sec.\ \ref{sub:PairwiseRSP},
small $K_{i}^{(4)}$ perturbations are irrelevant. Our results for
the full RG flow, in analogy to the previous spin-$\frac{3}{2}$ case,
are shown on a unit 2-sphere in the $K^{\left(1\right)}\times K^{\left(2\right)}\times K^{\left(3\right)}$
space in Fig.~\ref{fig:Spin-2-phase-diagram-ball}. The fully stable
AF fixed points on the semi-axes are shown as beige, brown and white
circles. They define the AF hyper-octant (actually, the $K^{\left(4\right)}=0$
section of the hyper-octant). The semi-stable fixed points on the
2-planes {[}see panels (a), (b) and (d) of Fig.~\ref{fig:Spin-2-phase-diagram-ball}{]}
are shown as black circles. The topology of the flow between these
fixed points requires the existence of a third fixed point in the
AF octant. It is shown as a pink circle, that is fully unstable on
the surface of the unit 2-sphere. Note however that, unlike in the
previous spin-$\frac{3}{2}$ case, this is not the totally unstable
FP with enlarged SU(5) symmetry. The latter has $K_{i}^{(4)}\neq0$,
the precise location of which {[}see Appendix~\ref{sec:appendix-SU(N)-invariant-Hamil}
for details about the SU($N$)-symmetric points{]} is given by

\begin{eqnarray}
H_{5-\bar{5}}^{{\rm SU}\left(5\right)} & = & \sum_{i}K_{i}\left(\hat{O}_{1}-\frac{4}{21}\hat{O}_{2}+\frac{1}{21}\hat{O}_{3}-\frac{4}{189}\hat{O}_{4}\right).
\end{eqnarray}

The basins of attraction of the stable AF fixed points are found by
a numerical analysis of the RG flow, following the same protocol as
in the spin-$\frac{3}{2}$ chain. The results are summarized by the
color-coded thick dots in the blue region of Fig.~\ref{fig:Spin-2-phase-diagram-ball},
where the colors used are the same as in the spin $\frac{3}{2}$ chain
in Fig.~\ref{fig:AFoctant}. Furthermore, the green dots map out
the basin of attraction of the $\psi=\frac{1}{3}$ fixed point along
the $K^{\left(3\right)}$ axis, which correspond to the triplewise
RSP already discussed.

To conclude, we see that the phase diagram of the spin-2 chain differs
\emph{qualitatively} from the spin-$\frac{3}{2}$ case only due to
the presence of the triplewise RSP with $\psi=\frac{1}{3}$. We expect
the same trend to hold for disordered chains with higher spin values,
with triplewise RSP appearing only in the cases of integer spins.

\section{Summary of results\label{sec:summary}}

In this Section, we summarize our main findings, focusing mostly on
the different phases we have found as well as the conditions for their
realization in the initial model but leaving out some of the technical
details.

Our focus is on the SU(2)-symmetric, strongly disordered spin-$S$
chains. We have obtained general results for all values of $S$ and
have illustrated in full detail the cases of $S=\frac{3}{2}$ and
$2$. Generic SU(2) symmetry (with only nearest-neighbor interactions,
as we assume here) is usually identified by terms which only involve
powers of the scalar product of spin operators on adjacent sites;
the largest power being $J_{\mathrm{max}}=2S$. There are $J_{\mathrm{max}}$
coupling constants per bond $\alpha_{i}^{\left(J\right)}$ (with $J=1,2,\ldots,J_{\mathrm{max}}$)
between sites $i$ and $i+1$. This form of the Hamiltonian, however,
is not the most suitable for an SDRG treatment. Instead, the flow
becomes simpler if we rewrite the Hamiltonian in terms of scalars
built by contractions of irreducible spherical tensors of spin operators
of a given rank. We called them $\hat{O}_{J}\left(\mathbf{S}_{i},\mathbf{S}_{i+1}\right)$,
where $J$ is the rank of the tensors being contracted and again $J=1,2,\ldots,J_{\mathrm{max}}$.
Each contraction has a coefficient $K_{i}^{\left(J\right)}$. The
sets $\alpha_{i}^{\left(J\right)}$ and $K_{i}^{\left(J\right)}$
are linearly related, as shown for example in Eqs.~\eqref{eq:alpha1inKs}-\eqref{eq:alpha4inKs}
and \eqref{eq:k1inalphas}-\eqref{eq:k4inalphas}.

If we think of the coupling constants as a vector $\mathbf{K}_{i}=\left(K_{i}^{\left(1\right)},...,K_{i}^{\left(J_{{\rm max}}\right)}\right)$,
the SDRG transformations are then a set of rules that change the directions
and magnitudes of these vectors, as well as the spin magnitude on
each site. In general, these three quantities are coupled in the earlier
stages of the SDRG flow. Our results show, however, that the vector
directions decouple from their magnitudes in the final stages of the
flow (near the stable fixed points). This is the main reason why it
is more convenient to rewrite the Hamiltonian in terms of the irreducible
spherical tensors, namely, under the SDRG transformations, tensors
of a given rank do not generate tensors of a different rank. This
is a direct consequence of the SU(2) symmetry. We now summarize the
SDRG flow.

Stable fixed points define stable extended phases. An important set
of stable fixed points are the semi-axes $\left(-1\right)^{J}K^{\left(J\right)}<0$
(see Section~\ref{sub:PairwiseRSP} and Fig.~\ref{fig:Hyper-octant}).
The hyper-octant delimited by $\left(-1\right)^{J}K^{\left(J\right)}<0$
spans most (but not all) of its basin of attraction (see the blue
regions of Figs.~\ref{fig:AFoctant} and \ref{fig:Spin-2-phase-diagram-ball}).
If the system is such that all the vector directions are inside this
basin of attraction and the spin sizes are uniform (equal to $S$),
then the RG flow is such that the spin sizes remain fixed and the
vector directions flow towards one of the semi-axes $\left(-1\right)^{J}K^{\left(J\right)}<0$.
The main feature of these $2S$ fixed points is that they define the
\emph{same} conventional pairwise random singlet state, which is a
generalization of random singlet state of the spin-$1/2$ Heisenberg
chain {[}see Fig. \foreignlanguage{english}{\hyperref[fig:random-singlets]{\ref{fig:random-singlets}(a)}}{]}.
The distribution of vector magnitudes tends towards a universal form
with an asymptotically infinite relative width: a so-called infinite-randomness
fixed point {[}see Eq.~\eqref{eq:P(x)-IRFP} with $\psi=1/2${]}.
The excitations correspond to the breaking of the singlet pairs in
a hierarchy such that the relation between the energy $E$ of a bond
and its size $\xi$ is given by $E\sim\exp\left(-\xi^{\psi}\right)$.
Other physical properties follow from this structure. The magnetic
susceptibility and specific heat behave as $\chi^{-1}\sim T\left|\ln T\right|^{1/\psi}$
and $C\sim\left|\ln T\right|^{1+1/\psi},$ respectively. The average
spin-spin correlation function decays as a power-law $\overline{\left\langle \mathbf{S}_{i}\cdot\mathbf{S}_{i+r}\right\rangle }\sim r^{-4\psi}$.
An important aspect of this set of fixed points is the emergence of
an SU($2S+1$) symmetry: correlation functions and susceptibilities
of combinations of spin operators (dipolar, quadrupolar and other
higher-multipole moments), which transform as generators of the SU($2S+1$)
group, are all equal in the limit of strong disorder. Finally, there
must be unstable fixed points inside this basin of attraction. In
particular, there is a completely unstable fixed point in a particular
direction in $\mathbf{K}_{i}$-space lying within the hyper-octant
with \emph{exact} SU$\left(2S+1\right)$ symmetry (see Fig. \ref{fig:angle-streamlines}).
At all fixed points within this basin of attraction, the ground state
and the low-temperature thermodynamics are the same. 

The second set of fixed points we have found happens for integer spins
up to $S=9$. As in the previous case, the spin size $S$ remains
constant along the RG flow. The vectors $\mathbf{K}_{i}$'s, however,
do not point towards a single direction. Instead, they point towards
the positive and negative directions of the axis $J=I_{S}$, with
equal probability (where $I_{1}=2$, and $I_{S}=3$ for $2\leq S\leq9$,
see Section~\ref{sub:TriplewiseRSP}). The corresponding phase is
also a random-singlet phase. The singlets in this phase are formed
by a number of spins which is a multiplet of 3, with trios being the
most abundant {[}see Fig.\foreignlanguage{english}{\ \hyperref[fig:random-singlets]{\ref{fig:random-singlets}(b)}}{]}.
As in the previous set of fixed points, it is also of the infinite-randomness
type, sharing the same features (correlation functions and thermodynamics)
but with an exponent $\psi=1/3$. Its basin of attraction is illustrated
by the green region in Fig.~\ref{fig:Spin-2-phase-diagram-ball}.
Finally, there is no emergent symmetry at this fixed point larger
than the original SU(2), with the exception of the spin-1 case, for
which there is an emergent SU(3) symmetry. \cite{QuitoHoyosMiranda}

A third important fixed point is found on the $K_{i}^{(1)}$-axis
(the Heisenberg axis). As in the previous case, the vectors point
(with approximately equal probability) towards both the positive (antiferromagnetic)
and the negative (ferromagnetic) $K^{(1)}$-direction (see Section~\ref{sub:Large-Spin-Phase}).
Unlike the previous fixed points, the spin magnitude is not constant
throughout the chain and the distribution of the vector magnitudes
is of \emph{finite}-randomness type {[}see Eq.~\eqref{eq:P(x)-LSP}{]}.
This implies a conventional scaling of energy and length scales $\omega\sim\xi^{-z},$
where $z$ is the dynamical critical exponent. The coarse-grained
degrees of freedom are spins of all sizes and the corresponding phase
is named a Large Spin Phase.\cite{westerbergetal} The magnetic susceptibility
is universal and Curie like $\chi\sim T^{-1}$ while the specific
heat behaves as $C\sim T^{z}$. For weakly disordered systems, $z$
is universal and $\approx2.2$. For strongly disordered systems, $z$
is non-universal and depends on the \textcolor{black}{disorder} strength
of the initial vector magnitude distribution.

The basins of attraction of these three classes of fixed points above
exhaust the parameter space except for the regions with FM long-range
order and other regions of measure zero. Therefore, in the strong
disorder limit, the phase diagram displays four phases whenever $S\leq9$
and integer: the pairwise and triplewise random singlet phases, the
FM phase, and the Large Spin phase (see Fig. \ref{fig:Spin-2-phase-diagram-ball}
for the $S=2$ case, and Fig. 1 of Ref.\ \onlinecite{QuitoHoyosMiranda}
for the $S=1$ case). For the remaining cases (half-integer spins
or $S>9$), the phase diagram shows the same phases with the exception
of the triplewise random singlet phase (see Fig. \ref{fig:AFoctant}).

\section{Outlook\label{sec:Conclusions}}

In this section, we comment on some open questions and give some perspective
for future directions of research. 

Let us start with a technical open issue. Our method is able to produce
an RG flow in the whole parameter space of the Hamiltonian (\ref{eq:Hamilt_tensors})
except for a zero-measure set as discussed in Sec.~\ref{sub:otheraxes}.
This peculiar set of parameters is a subset of the $K_{i}^{\left(1\right)}=0$
hyper-plane (see details in Sec.~\ref{sub:otheraxes}) in which first-order
perturbation theory fails to produce non-vanishing corrections in
the SDRG decimation procedure. Although the RG flow is well-behaved
outside this hyper-plane, it would be desirable to know the real fate
of the flow in this hyper-plane. At the moment, it is unclear whether
including higher-order perturbative terms in the decimation procedure
will suffice to produce a consistent RG flow or whether there is new
physics in this set of parameters.

Our approach is based on a strong-disorder RG method which becomes
asymptotically exact near infinite-randomness fixed points. We have
found two sets of such fixed points (corresponding to the pairwise
and triplewise RSP) and a set of finite-disorder fixed points (corresponding
to the Large Spin phase). Naturally, our approach cannot give exact
results in the Large Spin phase. Unfortunately, there are no studies
on the precision of the SDRG method at finite-disorder fixed points
except for one study in which the accuracy of the SDRG method is shown
to be within $1\%$ (for the values of some critical exponents) when
the dynamical critical exponent is greater than modest values $\approx1.5$.\cite{getelina2015}
We have not performed an analysis of the dynamical critical exponent
dependence on the initial conditions and leave this task for the future.
It is known to be different from the one reported in Ref.\ \onlinecite{westerbergetal}
when other terms beyond the usual nearest-neighbor Heisenberg interactions
are present.\cite{Hoyosladders} In addition, as pointed out in Ref.\ \onlinecite{QuitoHoyosMiranda},
there are regions in which three phases meet. A complete analysis
of the corresponding unstable critical points is also left as an open
question. Finally, we remind that our method can neither capture the
phases of the clean system nor predict if they are stable against
weak disorder. For the latter one, other methods are necessary (see
discussion in Ref.\ \onlinecite{QuitoHoyosMiranda} for the $S=1$
case). Thus, there is the possibility of a much richer phase diagram
in the intermediate disorder regime.

One particular outcome of our results concerns the so-called permutation-symmetric
multicritical points.\cite{PhysRevLett.89.277203} In spin-$S$ random
chains, it was shown that whenever $N_{1}=2S+1$ different dimerized
phases meet at a single multicritical point, this point is of infinite-randomness
type with tunneling exponent $\psi=1/N_{1}$. At that point, it was
left as an open question what is the (fine-tuned) condition necessary
for these phases to meet at a single point. As we have shown here,
there are no phases with arbitrarily small tunneling exponent. Such
feature only happens at special unstable fixed points which possess
explicit SU($N$) symmetry (with $N=N_{1}$) and which we have shown
how to precisely define. Therefore, we have now discovered the exact
location of the permutation-symmetric multicritical points in SU(2)-symmetric
spin-$S$ random chains: they occur at the SU($2S+1$)-symmetric point.

As mentioned in the Introduction, the main motivation for this work
is the search for phases displaying emergent enlarged symmetries and
the understanding of the corresponding physical mechanism. Indeed,
we have found an emergent SU($2S+1$) symmetry in the entire pairwise
RSP. In addition, this phase is pairwise, and thus, $\psi=1/2$ (a
``mesonic'' phase). As shown in Ref.\ \onlinecite{QuitoHoyosMiranda},
the triplewise RSP of the spin-1 chain also possesses an emergent
SU(3) symmetry, with $\psi=1/3$ (a ``baryonic'' phase). Unfortunately,
we have not found a generalization of the baryonic RSP for higher
spins. This would be interesting because it would imply in the existence
of entire phases with $\psi=\left(2S+1\right)^{-1}$. We have found
instead some triplewise RSPs for integer spins $S$ with $2\leq S\leq9$,
which, however, do not possess any enhanced symmetry. In some sense,
this supports the conventional wisdom that emergent enhanced symmetries
are indeed more the exception than the rule. Additionally, a line
with SO(5) symmetry was found in the spin-$3/2$ random chains which
contains the corresponding SU(4) baryonic and mesonic points. Since
it is confined to a lower dimensional manifold of the full phase space,
it suggests that an emergent SU($2S+1$) baryonic RSP may be realized
in a different symmetry group. Exploring spin chain Hamiltonians with
SO($N$) symmetry is the next step of our research. This is not of
just academic curiosity since experimental realizations of these groups
have been proposed in cold-atomic systems.~\cite{wuetal2003,Wumodphys06}

Finally, we point out that our method has exciting applications to
the Hamiltonian in Eq.~\eqref{eq:Hamiltonian_SU2} generalized to
any dimension, geometry (as in ladders) or with long-range interactions.
For instance, in higher dimensions there are quantum phase transitions
between the Néel AFM state to other phases (such as a valence bond
crystal phase) upon increasing the value of the terms besides the
bilinear one. Our method can thus be directly used to study the disorder
effects on such quantum phase transitions.

\section{Acknowledgments}

We gratefully acknowledge P. L. S. Lopes for useful discussions. This
work has been supported by FAPESP through grants 2009/17531-3 (VLQ),
07/57630-5 (EM)\textbf{ }and 2015/23849-7 (JAH), and by CNPq through
grants 307548/2015-5 (JHA), 304311/2010-3 (EM) and 590093/2011-8 (JHA
and EM).

\appendix
\begin{widetext}

\section{Dictionary of conversion of different notations \label{sec:Appendix:notation}}

In this Appendix, we list several conversions between different forms
of the relevant operators and Hamiltonians used in the paper. The
calculations are tedious but straightforward. Alternatively, the use
of a software like MATHEMATICA expedites the procedure. 

We first explicitly show the decomposition of $\left(\mathbf{S}_{1}\cdot\mathbf{S}_{2}\right)^{J}$,
for $J=1,\ldots,4$, in terms of the $\hat{O}_{J}$ operators
\begin{eqnarray}
\left(\mathbf{S}_{1}\cdot\mathbf{S}_{2}\right) & = & \frac{4\pi}{3}\hat{O}_{1},\label{eq:S1S2}\\
\left(\mathbf{S}_{1}\cdot\mathbf{S}_{2}\right)^{2} & = & -\frac{2\pi}{3}\hat{O}_{1}+\frac{8\pi}{15}\hat{O}_{2}+\frac{1}{3}\mathbf{S}_{1}^{2}\mathbf{S}_{2}^{2},\label{eq:S1S2squared}\\
\left(\mathbf{S}_{1}\cdot\mathbf{S}_{2}\right)^{3} & = & \left(-\frac{8\pi}{30}\left(\mathbf{S}_{1}^{2}+\mathbf{S}_{2}^{2}-3\mathbf{S}_{1}^{2}\mathbf{S}_{2}^{2}\right)+\frac{8\pi}{15}\right)\hat{O}_{1}-\frac{16\pi}{15}\hat{O}_{2}+\frac{8\pi}{35}\hat{O}_{3}-\frac{1}{6}\mathbf{S}_{1}^{2}\mathbf{S}_{2}^{2},\label{eq:S1S2cube}\\
\left(\mathbf{S}_{1}\cdot\mathbf{S}_{2}\right)^{4} & = & \frac{2\pi}{3}\left(\mathbf{S}_{1}^{2}+\mathbf{S}_{2}^{2}-2\mathbf{S}_{1}^{2}\mathbf{S}_{2}^{2}-1\right)\hat{O}_{1}+\frac{8\pi}{105}\left(31-5\mathbf{S}_{1}^{2}-5\mathbf{S}_{2}^{2}+6\mathbf{S}_{1}^{2}\mathbf{S}_{2}^{2}\right)\hat{O}_{2}-\frac{8\pi}{7}\hat{O}_{3}+\frac{32\pi}{315}\hat{O}_{4}\nonumber \\
 &  & +\frac{2}{15}\mathbf{S}_{1}^{2}\mathbf{S}_{2}^{2}-\frac{1}{15}\left(\mathbf{S}_{1}^{4}\mathbf{S}_{2}^{2}+\mathbf{S}_{1}^{2}\mathbf{S}_{2}^{4}\right)+\frac{1}{5}\mathbf{S}_{1}^{4}\mathbf{S}_{2}^{4}.\label{eq:S1S2quartic}
\end{eqnarray}

Conversely, 

\begin{eqnarray}
\hat{O}_{1} & = & \frac{3}{4\pi}\left(\mathbf{S}_{1}\cdot\mathbf{S}_{2}\right),\label{eq:S1S2-1}\\
\hat{O}_{2} & = & \frac{15}{16\pi}\left(\mathbf{S}_{1}\cdot\mathbf{S}_{2}\right)+\frac{15}{8\pi}\left(\mathbf{S}_{1}\cdot\mathbf{S}_{2}\right)^{2}-\frac{5}{8\pi}\mathbf{S}_{1}^{2}\mathbf{S}_{2}^{2},\label{eq:S1S2squared-1}\\
\hat{O}_{3} & = & \frac{7}{8\pi}\left(\mathbf{S}_{1}^{2}+\mathbf{S}_{2}^{2}-3\mathbf{S}_{1}^{2}\mathbf{S}_{2}^{2}+3\right)\left(\mathbf{S}_{1}\cdot\mathbf{S}_{2}\right)+\frac{35}{4\pi}\left(\mathbf{S}_{1}\cdot\mathbf{S}_{2}\right)^{2}+\frac{35}{8\pi}\left(\mathbf{S}_{1}\cdot\mathbf{S}_{2}\right)^{3}-\frac{35}{16\pi}\mathbf{S}_{1}^{2}\mathbf{S}_{2}^{2},\label{eq:S1S2cube-1}\\
\hat{O}_{4} & = & -\frac{45}{32\pi}\left(17\mathbf{S}_{1}^{2}\mathbf{S}_{2}^{2}-6\left(\mathbf{S}_{1}^{2}+\mathbf{S}_{2}^{2}\right)-9\right)\left(\mathbf{S}_{1}\cdot\mathbf{S}_{2}\right)-\frac{45}{32\pi}\left(6\mathbf{S}_{1}^{2}\mathbf{S}_{2}^{2}-5\left(\mathbf{S}_{1}^{2}+\mathbf{S}_{2}^{2}\right)-39\right)\left(\mathbf{S}_{1}\cdot\mathbf{S}_{2}\right)^{2}\nonumber \\
 & + & \frac{1575}{32\pi}\left(\mathbf{S}_{1}\cdot\mathbf{S}_{2}\right)^{3}+\frac{315}{32\pi}\left(\mathbf{S}_{1}\cdot\mathbf{S}_{2}\right)^{4}-\frac{729}{64\pi}\mathbf{S}_{1}^{2}\mathbf{S}_{2}^{2}-\frac{27}{16\pi}\left(\mathbf{S}_{1}^{2}\mathbf{S}_{2}^{4}+\mathbf{S}_{1}^{4}\mathbf{S}_{2}^{2}\right)+\frac{27}{32\pi}\mathbf{S}_{1}^{4}\mathbf{S}_{2}^{4}.\label{eq:S1S2quartic-1}
\end{eqnarray}

The coupling constants can be mapped according to

\begin{eqnarray}
\alpha^{\left(1\right)} & = & \frac{1}{16\pi}\left(12K^{\left(1\right)}+15K^{\left(2\right)}+14\left(3+\mathbf{S}_{1}^{2}+\mathbf{S}_{2}^{2}-3\mathbf{S}_{1}^{2}\mathbf{S}_{2}^{2}\right)K^{\left(3\right)}\right)+\left(\frac{405}{32\pi}+\frac{135}{16\pi}\left(\mathbf{S}_{1}^{2}+\mathbf{S}_{2}^{2}\right)-\frac{765}{32\pi}\mathbf{S}_{1}^{2}\mathbf{S}_{2}^{2}\right)K^{\left(4\right)},\label{eq:alpha1inKs}\\
\alpha^{\left(2\right)} & = & \frac{5}{8\pi}\left(3K^{\left(2\right)}+14K^{\left(3\right)}\right)+\left(\frac{1755}{32\pi}+\frac{225}{32\pi}\left(\mathbf{S}_{1}^{2}+\mathbf{S}_{2}^{2}\right)-\frac{135}{16\pi}\mathbf{S}_{1}^{2}\mathbf{S}_{2}^{2}\right)K^{\left(4\right)},\label{eq:alpha2inKs}\\
\alpha^{\left(3\right)} & = & \frac{35}{8\pi}K^{\left(3\right)}+\frac{1575}{32\pi}K^{\left(4\right)},\label{eq:alpha3inKs}\\
\alpha^{\left(4\right)} & = & \frac{315}{32\pi}K^{\left(4\right)},\label{eq:alpha4inKs}
\end{eqnarray}
and

\begin{eqnarray}
K^{\left(1\right)} & = & \frac{4\pi}{3}\alpha^{\left(1\right)}-\frac{2\pi}{3}\alpha^{\left(2\right)}-\frac{8\pi}{30}\left(\mathbf{S}_{1}^{2}+\mathbf{S}_{2}^{2}-3\mathbf{S}_{1}^{2}\mathbf{S}_{2}^{2}-2\right)\alpha^{\left(3\right)}+\frac{2\pi}{3}\left(\mathbf{S}_{1}^{2}+\mathbf{S}_{2}^{2}-2\mathbf{S}_{1}^{2}\mathbf{S}_{2}^{2}-1\right)\alpha^{\left(4\right)},\label{eq:k1inalphas}\\
K^{\left(2\right)} & = & \frac{8\pi}{15}\alpha^{\left(2\right)}-\frac{16\pi}{15}\alpha^{\left(3\right)}+\frac{8\pi}{105}\left(31-5\mathbf{S}_{1}^{2}-5\mathbf{S}_{2}^{2}+6\mathbf{S}_{1}^{2}\mathbf{S}_{2}^{2}\right)\alpha^{\left(4\right)},\label{eq:k2inalphas}\\
K^{\left(3\right)} & = & \frac{8\pi}{35}\alpha^{\left(3\right)}-\frac{8\pi}{7}\alpha^{\left(4\right)},\label{eq:k3inalphas}\\
K^{\left(4\right)} & = & \frac{32\pi}{315}\alpha^{\left(4\right)}.\label{eq:k4inalphas}
\end{eqnarray}

Also relevant is the decomposition of projection operators into spin
objects. Unlike the previous equations, which can be found for any
values of $\mathbf{S}_{2}$ and $\mathbf{S}_{3}$, this has to be
done in a case-by-case manner. For two coupled spin $\frac{3}{2}$,
we find the following correspondence 
\begin{eqnarray}
\epsilon^{\left(0\right)} & = & \alpha^{\left(0\right)}-\frac{15}{4}\alpha^{\left(1\right)}+\frac{225}{16}\alpha^{\left(2\right)}-\frac{3375}{64}\alpha^{\left(3\right)},\\
\epsilon^{\left(1\right)} & = & \alpha^{\left(0\right)}-\frac{11}{4}\alpha^{\left(1\right)}+\frac{121}{16}\alpha^{\left(2\right)}-\frac{1331}{64}\alpha^{\left(3\right)},\\
\epsilon^{\left(2\right)} & = & \alpha^{\left(0\right)}-\frac{3}{4}\alpha^{\left(1\right)}+\frac{9}{16}\alpha^{\left(2\right)}-\frac{25}{64}\alpha^{\left(3\right)},\\
\epsilon^{\left(3\right)} & = & \alpha^{\left(0\right)}+\frac{9}{4}\alpha^{\left(1\right)}+\frac{81}{16}\alpha^{\left(2\right)}+\frac{729}{64}\alpha^{\left(3\right)},
\end{eqnarray}
and, conversely,

\begin{eqnarray}
K^{\left(0\right)} & = & \frac{\pi}{4}\epsilon^{\left(0\right)}-\frac{3\pi}{4}\epsilon^{\left(1\right)}+\frac{5\pi}{4}\epsilon^{\left(2\right)}+\frac{7\pi}{4}\epsilon^{\left(3\right)},\\
K^{\left(1\right)} & = & -\frac{\pi}{15}\epsilon^{\left(0\right)}-\frac{11\pi}{75}\epsilon^{\left(1\right)}-\frac{\pi}{15}\epsilon^{\left(2\right)}+\frac{7\pi}{25}\epsilon^{\left(3\right)},\\
K^{\left(2\right)} & = & \frac{\pi}{45}\epsilon^{\left(0\right)}+\frac{\pi}{75}\epsilon^{\left(1\right)}-\frac{\pi}{15}\epsilon^{\left(2\right)}+\frac{7\pi}{225}\epsilon^{\left(3\right)},\\
K^{\left(3\right)} & = & -\frac{4\pi}{315}\epsilon^{\left(0\right)}+\frac{4\pi}{175}\epsilon^{\left(1\right)}-\frac{4\pi}{315}\epsilon^{\left(2\right)}+\frac{4\pi}{1575}\epsilon^{\left(3\right)}.
\end{eqnarray}

\section{Derivation of the RG step \label{sec:Appendix:RGstep}}

In this Appendix, we give details about the perturbative calculations
that allow one to make one RG decimation step. We divide this Appendix
into three subsections. In the first and second subsections, we derive
how the first- and second-order perturbation theories are applied
to this problem, while the details about the calculation of coefficients
that appear in the first two subsections are left for the final subsection.

Throughout this derivation, we will need to compute matrix elements
of ISTs in a two-site problem. Let us then already set the notation
we are going to follow. The notation is the same as in Edmonds' book\cite{Edmondsbook}
(see, for instance, page 74). Assume that the largest energy gap is
due to a bond connecting sites 2 and 3 and let us call the spins of
this 2-site problem $\mathbf{S}_{2}$ and $\mathbf{S}_{3}$. The Wigner-Eckart
theorem, that gives the matrix elements of ISTs of rank $J$ and component
$M$, $Y_{JM}\left(\mathbf{S}_{i}\right)$ ($i=2,3$) in the total
angular momentum basis, is given by

\begin{eqnarray}
\left\langle S_{2}S_{3},J'M'\left|Y_{JM}\left(\mathbf{S}_{i}\right)\right|S_{2}S_{3},J''M''\right\rangle  & = & \left(-1\right)^{J-J'+J''}\frac{\left\langle JJ'';MM''\left|JJ'';J'M'\right.\right\rangle }{\sqrt{2J'+1}}\left\langle S_{2}S_{3},J'\left|\left|Y_{J}\left(\mathbf{S}_{i}\right)\right|\right|S_{2}S_{3},J''\right\rangle ,\label{eq:wigner-eckart}
\end{eqnarray}
where $\left\langle J'\left|\left|Y_{k}\left(\mathbf{S}\right)\right|\right|J\right\rangle $
is the reduced matrix element, independent of the IST component $M$
component and the angular momentum projections $M'$ and $M'$'. To
simplify the notation, in this Appendix we write $\left\langle S_{2}S_{3}J'\left|\left|Y_{J}\left(\mathbf{S}_{i}\right)\right|\right|S_{2}S_{3}J''\right\rangle \equiv\left\langle J'\left|\left|Y_{J}\left(\mathbf{S}_{i}\right)\right|\right|J''\right\rangle $.

\subsection{First-order perturbation theory}

The algebraic challenge is to simplify the projection defined in the
main text in Eq.~\eqref{eq:first_order_projection}, i. e., to find
effective couplings between $S_{1}$ and the new spin $\tilde{S}$,
introduced to replace $S_{2}$ and $S_{3}$, at low energies. For
concreteness, we focus on operators of site 2, $Y_{J-M}\left(\mathbf{S}_{2}\right)$.
By using the projection operator onto a multiplet of total angular
momentum $\tilde{S}$, as defined in Eq.~\eqref{eq:projector_multiplets},

\begin{equation}
P_{\tilde{S}}=\sum_{M'=-\tilde{S}}^{\tilde{S}}\left|\tilde{S}M'\right\rangle \left\langle \tilde{S}M'\right|,
\end{equation}
we find the projection to be

\begin{singlespace}
\begin{eqnarray}
P_{\tilde{S}}Y_{J-M}\left(\mathbf{S}_{2}\right)P_{\tilde{S}} & = & \left(\sum_{M'}\left|\tilde{S}M'\right\rangle \left\langle \tilde{S}M'\right|\right)Y_{J-M}\left(\mathbf{S}_{2}\right)\left(\sum_{M''}\left|\tilde{S}M''\right\rangle \left\langle \tilde{S}M''\right|\right),\nonumber \\
 & = & \sum_{M',M''}\left|\tilde{S}M'\right\rangle \left\langle \tilde{S}M''\right|\left\langle \tilde{S}M'\left|Y_{J-M}\left(\mathbf{S}_{2}\right)\right|\tilde{S}M''\right\rangle ,
\end{eqnarray}

\end{singlespace}

We now apply the Wigner-Eckart theorem, Eq.~\eqref{eq:wigner-eckart},
in order to calculate the matrix element $\left\langle \tilde{S}M'\left|Y_{J-M}\left(\mathbf{S}_{2}\right)\right|\tilde{S}M''\right\rangle $.
This matrix element is proportional to $\left\langle \tilde{S}\left|\left|Y_{J}\left(\mathbf{S}_{2}\right)\right|\right|\tilde{S}\right\rangle $
and to the Clebsch-Gordan coefficient $\left\langle J\tilde{S};-MM''\left|J\tilde{S};\tilde{S}M'\right.\right\rangle $.
In order to rewrite the projection as a new IST acting on the ground
state manifold, we use the Wigner-Eckart theorem again to calculate
$\left\langle \tilde{S}M'\right|Y_{J-M}\left(\mathbf{\tilde{S}}\right)\left|\tilde{S}M''\right\rangle $,
where $\mathbf{S}_{2}$ has been replaced by $\tilde{\mathbf{S}}$.
The latter is proportional to $\left\langle \tilde{S}\left|\left|Y_{J}\left(\mathbf{\tilde{S}}\right)\right|\right|\tilde{S}\right\rangle $
and to the same Clebsch-Gordan coefficient, which implies 

\begin{equation}
\frac{\left\langle \tilde{S}M'\left|Y_{J-M}\left(\mathbf{S}_{2}\right)\right|\tilde{S}M''\right\rangle }{\left\langle \tilde{S}M'\right|Y_{J-M}\left(\mathbf{\tilde{S}}\right)\left|\tilde{S}M''\right\rangle }=\frac{\left\langle \tilde{S}\left|\left|Y_{J}\left(\mathbf{S}_{2}\right)\right|\right|\tilde{S}\right\rangle }{\left\langle \tilde{S}\left|\left|Y_{J}\left(\mathbf{\tilde{S}}\right)\right|\right|\tilde{S}\right\rangle }.
\end{equation}

Therefore, we get

\begin{eqnarray}
P_{\tilde{S}}Y_{J-M}\left(\mathbf{S}_{2}\right)P_{\tilde{S}} & = & \frac{\left\langle \tilde{S}\left|\left|Y_{J}\left(\mathbf{S}_{2}\right)\right|\right|\tilde{S}\right\rangle }{\left\langle \tilde{S}\left|\left|Y_{J}\left(\mathbf{\tilde{S}}\right)\right|\right|\tilde{S}\right\rangle }\sum_{M'}\sum_{M''}\left|\tilde{S}M'\right\rangle \left\langle \tilde{S}M'\right|Y_{J-M}\left(\mathbf{\tilde{S}}\right)\left|\tilde{S}M''\right\rangle \left\langle \tilde{S}M''\right|,\\
 &  & =\frac{\left\langle \tilde{S}\left|\left|Y_{J}\left(\mathbf{S}_{2}\right)\right|\right|\tilde{S}\right\rangle }{\left\langle \tilde{S}\left|\left|Y_{J}\left(\mathbf{\tilde{S}}\right)\right|\right|\tilde{S}\right\rangle }P_{\tilde{S}}Y_{J-M}\left(\mathbf{\tilde{S}}\right)P_{\tilde{S}}.\label{eq:ratio-matrix-elem}\\
 & = & f^{\left(J\right)}\left(S_{2},S_{3},\tilde{S}\right)P_{\tilde{S}}Y_{J-M}\left(\mathbf{\tilde{S}}\right)P_{\tilde{S}}.
\end{eqnarray}
This a fundamental part of the process, which guarantees that the
renormalized Hamiltonian has the same functional form as the undecimated
one. We leave the calculation of the ratio of reduced matrix elements
$f^{\left(J\right)}\left(S_{2},S_{3},\tilde{S}\right)$ to the third
part of this Appendix. The new coupling is, therefore,

\begin{equation}
\tilde{K}_{1}^{\left(J\right)}=f^{\left(J\right)}\left(S_{2},S_{3},\tilde{S}\right)K_{1}^{\left(J\right)}.
\end{equation}
For the effective coupling between $\tilde{S}$ and $S_{3}$, the
calculation can be done following the same steps with the replacement
$S_{2}\rightleftarrows S_{3}$. Therefore,

\begin{equation}
\tilde{K}_{3}^{\left(J\right)}=f^{\left(J\right)}\left(S_{3},S_{2},\tilde{S}\right)K_{3}^{\left(J\right)}.
\end{equation}

\subsection{Second-order perturbation theory }

In the main text, we gave some of the steps for the second-order perturbation
theory calculation. Particularly, we showed that selection rules restrict
the values of the angular momentum of virtual states in such a way
that we are left with the task of computing $\left\langle 00\left|Y_{J-M}\left(\mathbf{S}_{2}\right)\right|JM\right\rangle \left\langle JM\left|Y_{JM}\left(\mathbf{S}_{3}\right)\right|00\right\rangle $.
In this section, we want to justify the simplification we made from
Eq~\textbf{\eqref{eq:second-order-eq}} to Eq.~\textbf{\eqref{eq:second-order-effec-coupl},
}that is, the $M$-independence of $g\left(J,S\right)$. From the
Wigner-Eckart theorem, Eq.~\eqref{eq:wigner-eckart}, the $M$-dependence
of the product $\left\langle 00\left|Y_{J-M}\left(\mathbf{S}_{2}\right)\right|JM\right\rangle \left\langle JM\left|Y_{JM}\left(\mathbf{S}_{3}\right)\right|00\right\rangle $
is given by the product of the Clebsch-Gordan coefficients $\left\langle JJ;-MM\left|JJ;00\right.\right\rangle $
and $\left\langle J0;M0\left|J0;JM\right.\right\rangle $, since both
the pre-factor and the reduced matrix elements are $M$-independent.
These Clebsch-Gordan coefficients can be explicitly calculated, and
are equal to 

\begin{eqnarray}
\left\langle JJ;-MM\left|JJ;00\right.\right\rangle  & = & \frac{\left(-1\right)^{J+M}}{\sqrt{1+2J}},\\
\left\langle J0;M0\left|J0;JM\right.\right\rangle  & = & 1.
\end{eqnarray}
The $M$-dependence is, then, $\left(-1\right)^{M}$, and 

\begin{equation}
\left\langle 00\left|Y_{J-M}\left(\mathbf{S}_{2}\right)\right|JM\right\rangle \left\langle JM\left|Y_{JM}\left(\mathbf{S}_{3}\right)\right|00\right\rangle =\left(-1\right)^{M}g\left(J,S\right),
\end{equation}
as defined in the main text. The function $g$ is going to be calculated
in the next part of this Appendix.

\subsection{Reduced matrix element calculations}

\subsubsection{Reduced matrix elements of ISTs}

We now need to find analytic expressions for the reduced matrix elements.
There is a great simplification in the cases we treat in this paper,
in which the operators act only on the degrees of freedom of one of
the sites, as shown in Ref.~\onlinecite{Edmondsbook} (page 111),

\begin{equation}
\left\langle J'\left|\left|Y_{J}\left(\mathbf{S}_{2}\right)\right|\right|J''\right\rangle =\left(-1\right)^{S_{2}+S_{3}+J''+J}\sqrt{\left(2J'+1\right)\left(2J''+1\right)}\left\{ \begin{array}{ccc}
S_{2} & J' & S_{3}\\
J'' & S_{2} & J
\end{array}\right\} \left\langle S_{2}\left|\left|Y_{J}\left(\mathbf{S}_{2}\right)\right|\right|S_{2}\right\rangle ,\label{eq:red_tensor}
\end{equation}
where $\left\{ ...\right\} $ represents the Wigner's \textit{6-j}
symbol, and analogously for the reduced matrix elements of $Y_{J}\left(\mathbf{S}_{3}\right)$.
The reduced matrix element above, $\left\langle S_{2}\left|\left|Y_{J}\left(\mathbf{S}_{2}\right)\right|\right|S_{2}\right\rangle $,
can be easily calculated by going back to the Wigner-Eckart theorem,
Eq.~\eqref{eq:wigner-eckart}, and choosing the state with highest
possible value of $M$, $M=S$, and $J'=J''=S_{2}$,

\begin{align}
\left\langle S_{2}\left|\left|Y_{J}\left(\mathbf{S}_{2}\right)\right|\right|S_{2}\right\rangle  & =\left(-1\right)^{J}\sqrt{2S_{2}+1}\frac{\left\langle S_{2}S_{2}\left|Y_{J0}\right|S_{2}S_{2}\right\rangle }{\left\langle JS_{2};0S_{2}\left|JS_{2};S_{2}S_{2}\right.\right\rangle }.\label{eq:redA}
\end{align}

\subsubsection{First-order perturbation theory}

The ratio of reduced matrix elements is what is left in the first-order
perturbation theory calculation {[}Eq.~\eqref{eq:first_order_ratio}{]}.
The numerator $\left\langle \tilde{S}\left|\left|Y_{J}\left(\mathbf{S}_{2}\right)\right|\right|\tilde{S}\right\rangle $
corresponds to the case when $J''=J'=\tilde{S}$ of Eq.~\eqref{eq:red_tensor}.
The denominator is the reduced matrix element $\left\langle \tilde{S}\left|\left|Y_{J}\left(\tilde{\mathbf{S}}\right)\right|\right|\tilde{S}\right\rangle $,
that can be easily calculated from the Wigner-Eckart theorem, Eq.~\eqref{eq:wigner-eckart},
by again choosing $J''=J'=\tilde{S}$

\begin{equation}
\left\langle \tilde{S}\left|\left|Y_{J}\left(\tilde{\mathbf{S}}\right)\right|\right|\tilde{S}\right\rangle =\frac{\left(-1\right)^{J}\sqrt{2\tilde{S}+1}\left\langle \tilde{S}\tilde{S}\left|Y_{J0}\left(\tilde{\mathbf{S}}\right)\right|\tilde{S}\tilde{S}\right\rangle }{\left\langle J\tilde{S};0\tilde{S}\left|J\tilde{S};\tilde{S}\tilde{S}\right.\right\rangle }.\label{eq:reduced-matel-single-site}
\end{equation}
After some simplifications, we get

\begin{eqnarray}
\frac{\left\langle \tilde{S}\left|\left|Y_{J}\left(\mathbf{S}_{2}\right)\right|\right|\tilde{S}\right\rangle }{\left\langle \tilde{S}\left|\left|Y_{J}\left(\tilde{\mathbf{S}}\right)\right|\right|\tilde{S}\right\rangle } & = & \left(-1\right)^{S_{2}+S_{3}+\tilde{S}+J}\frac{\left(2\tilde{S}+1\right)!}{\left(2S_{2}\right)!}\sqrt{\frac{\left(2S_{2}-J\right)!\left(2S_{2}+J+1\right)!}{\left(2\tilde{S}-J\right)!\left(2\tilde{S}+J+1\right)!}}\left\{ \begin{array}{ccc}
S_{2} & \tilde{S} & S_{3}\\
\tilde{S} & S_{2} & J
\end{array}\right\} \frac{\left\langle S_{2}S_{2}\left|Y_{J0}\left(\mathbf{S}_{2}\right)\right|S_{2}S_{2}\right\rangle }{\left\langle \tilde{S}\tilde{S}\left|Y_{J0}\left(\tilde{\mathbf{S}}\right)\right|\tilde{S}\tilde{S}\right\rangle },\label{eq:first_order_ratio}
\end{eqnarray}
where the last term of the right-hand side is
\begin{equation}
\frac{\left\langle S_{2}S_{2}\left|Y_{J0}\left(\mathbf{S}_{2}\right)\right|S_{2}S_{2}\right\rangle }{\left\langle \tilde{S}\tilde{S}\left|Y_{J0}\left(\tilde{\mathbf{S}}\right)\right|\tilde{S}\tilde{S}\right\rangle }=\frac{\prod_{S=\left\{ 0,\frac{1}{2},...,\frac{J-1}{2}\right\} }\left(S_{2}-S\right)}{\prod_{S=\left\{ 0,\frac{1}{2},...,\frac{J-1}{2}\right\} }\left(\tilde{S}-S\right)}.
\end{equation}

Putting it all together, we find

\begin{singlespace}
\begin{eqnarray}
f^{\left(J\right)}\left(S_{2},S_{3},\tilde{S}\right) & = & \frac{\left\langle \tilde{S}\left|\left|Y_{J}\left(\mathbf{S}_{2}\right)\right|\right|\tilde{S}\right\rangle }{\left\langle \tilde{S}\left|\left|Y_{J}\left(\tilde{\mathbf{S}}\right)\right|\right|\tilde{S}\right\rangle }\label{eq:first_order_general}\\
 & = & \left(-1\right)^{S_{2}+S_{3}+\tilde{S}+J}\frac{\left(2\tilde{S}+1\right)!}{\left(2S_{2}\right)!}\sqrt{\frac{\left(2S_{2}-J\right)!\left(2S_{2}+J+1\right)!}{\left(2\tilde{S}-J\right)!\left(2\tilde{S}+J+1\right)!}}\left\{ \begin{array}{ccc}
S_{2} & \tilde{S} & S_{3}\\
\tilde{S} & S_{2} & J
\end{array}\right\} \frac{\prod_{S<J-1}\left(S_{2}-S\right)}{\prod_{S<J-1}\left(\tilde{S}-S\right)}.
\end{eqnarray}

\end{singlespace}

Using the properties of the \emph{6-j} symbol, we also find (page
94 of Ref.~\onlinecite{Edmondsbook})

\begin{equation}
\left\{ \begin{array}{ccc}
S_{2} & \tilde{S} & S_{3}\\
\tilde{S} & S_{2} & J
\end{array}\right\} =\left\{ \begin{array}{ccc}
\tilde{S} & \tilde{S} & J\\
S_{2} & S_{2} & S_{3}
\end{array}\right\} .\label{eq:6-j-symbol}
\end{equation}

A necessary condition for the above \emph{6-j} symbol above \textit{\emph{to
be}}\textit{ non-zero} is that the so-called triangular conditions
are satisfied by $\left(\tilde{S},\tilde{S},J\right)$ and $\left(S_{2},S_{2},J\right)$.
A triad $\left(l_{1},l_{2},l_{3}\right)$ is said to satisfy a triangular
condition when it is possible to build a triangle with edges of sizes
$l_{1},$ $l_{2}$ and $l_{3}$. The triad $\left(\tilde{S},\tilde{S},J\right)$
satisfies the triangular condition if, and only if, $J<2\tilde{S}$.
In one case discussed in the main text, $S_{2}=S_{3}=\frac{3}{2}$,
$\tilde{S}=1$, and $J=3$, this condition is not satisfied. That
is what we have called case (a) in Section~\ref{sub:First-Order-Perturbation}.
The second triangular condition is equivalent to $S_{2}<2\tilde{S}$,
and is always satisfied. On the other hand, in another case discussed
in the main text, $S_{2}=S_{3}=\frac{3}{2}$ again, but $\tilde{S}=2$
and $J=2$. Even though the triangular conditions are satisfied, the
\emph{6-j} symbol in Eq.~\eqref{eq:6-j-symbol} vanishes. This is
what we have called case (b) in Section~\ref{sub:First-Order-Perturbation}.
Its occurrence cannot be predicted in general.

\subsubsection{Second-order perturbation theory}

Now, we go back to the second-order perturbation theory calculation
and find explicitly the $g\left(J,S\right)$ function. The difference
when compared to first order perturbation theory is that the matrix
elements of Eq.~\eqref{eq:def-g} are calculated between a state
of finite angular momentum and a singlet. The matrix elements are,
using Eqs.~\eqref{eq:wigner-eckart}, \eqref{eq:red_tensor} and
\eqref{eq:redA} 

\begin{eqnarray*}
\left\langle 00\left|Y_{J-M}\left(\mathbf{S}_{2}\right)\right|JM\right\rangle  & = & \left(-1\right)^{J+2S}\frac{\left\langle JJ;-MM\left|JJ;00\right.\right\rangle }{\left\langle JS;0S\left|JS;SS\right.\right\rangle }\sqrt{\left(2J+1\right)\left(2S+1\right)}\left\{ \begin{array}{ccc}
S & 0 & S\\
J & S & J
\end{array}\right\} \left\langle SS\left|Y_{J0}\right|SS\right\rangle ,\\
\left\langle JM\left|Y_{JM}\left(\mathbf{S}_{3}\right)\right|00\right\rangle  & = & \left(-1\right)^{J+2S}\frac{\left\langle J0;M0\left|J0;JM\right.\right\rangle }{\left\langle JS;0S\left|JS;SS\right.\right\rangle }\sqrt{\left(2J+1\right)\left(2S+1\right)}\left\{ \begin{array}{ccc}
S & J & S\\
0 & S & J
\end{array}\right\} \left\langle SS\left|Y_{J0}\right|SS\right\rangle .
\end{eqnarray*}

Multiplying the previous equations and simplifying the \emph{6-j}
symbol, we get

\begin{eqnarray}
g\left(J,S\right) & = & \left(-1\right)^{J}\left(\frac{1+2S}{1+2J}\right)\frac{\left(2S-J\right)!\left(2S+J+1\right)!}{\left(2S+1\right)!\left(2S+1\right)!}\left|\left\langle SS\left|Y_{J0}\right|SS\right\rangle \right|^{2}\label{eq:g-function}
\end{eqnarray}

The matrix elements that are explicitly used in this manuscript are

\begin{eqnarray}
\left\langle S,S\left|Y_{10}\right|S,S\right\rangle  & = & \frac{1}{2}\sqrt{\frac{3}{\pi}}S,\\
\left\langle S,S\left|Y_{20}\right|S,S\right\rangle  & = & \frac{1}{2}\sqrt{\frac{5}{\pi}}S\left(S-\frac{1}{2}\right),\\
\left\langle S,S\left|Y_{30}\right|S,S\right\rangle  & = & \frac{1}{2}\sqrt{\frac{7}{\pi}}S\left(S-\frac{1}{2}\right)\left(S-1\right),\\
\left\langle S,S\left|Y_{40}\right|S,S\right\rangle  & = & \frac{3}{2}\sqrt{\frac{1}{\pi}}S_{2}\left(S_{2}-\frac{1}{2}\right)\left(S_{2}-1\right)\left(S_{2}-\frac{3}{2}\right).
\end{eqnarray}

Note that the matrix elements are zero if $S<\frac{J}{2}$, which
immediately implies a product of $S-S_{i}$, with $S_{i}<S$. The
only task for higher rank tensors is, then, to find the overall prefactors.

Explicitly, for the tensors studied in this paper,
\begin{align}
g\left(1,S\right) & =-\frac{\left(S+1\right)S}{4\pi},\label{eq:g1}\\
g\left(2,S\right) & =\frac{\left(S+\frac{3}{2}\right)\left(S+1\right)S\left(S-\frac{1}{2}\right)}{4\pi},\label{eq:g2}\\
g\left(3,S\right) & =-\frac{\left(S+2\right)\left(S+\frac{3}{2}\right)\left(S+1\right)S\left(S-\frac{1}{2}\right)\left(S-1\right)}{4\pi},\label{eq:g3}\\
g\left(4,S\right) & =\frac{\left(S+\frac{5}{2}\right)\left(S+2\right)\left(S+\frac{3}{2}\right)\left(S+1\right)S\left(S-\frac{1}{2}\right)\left(S-1\right)\left(S-\frac{3}{2}\right)}{4\pi}.\label{eq:g4}
\end{align}
The above equations suggest a general formula for $g\left(J,S\right)$,
which, however, we were not able to prove

\begin{eqnarray}
g\left(J,S\right) & = & \frac{\left(-1\right)^{J}}{4\pi}\frac{\left(2S+1+J\right)!}{2^{2J}\left(2S-J\right)!},\\
 & = & \frac{\left(-1\right)^{J}}{4\pi}\left(S+\frac{J+1}{2}\right)\left(S+\frac{J}{2}\right)\ldots\left(S+1\right)S\left(S-\frac{1}{2}\right)\ldots\left[S-\frac{\left(J-1\right)}{2}\right].
\end{eqnarray}

\end{widetext}

\section{RG decimation rules: Spin $\frac{3}{2}$ and Spin $2$ \label{sec:Appendix-RG-equations-spins}}

In this Appendix, we list the RG decimation rules in the AF region
for both spin-$\frac{3}{2}$ and spin-$2$ decimations. We also comment
on the planar fixed point in both cases.

\subsection{Spin-$\frac{3}{2}$}

The RG decimation rules in the AF region are given by

\begin{align}
\tilde{K}_{14}^{\left(1\right)} & =15\frac{K_{1}^{\left(1\right)}K_{3}^{\left(1\right)}}{6K_{2}^{\left(1\right)}-90K_{2}^{\left(2\right)}+441K_{2}^{\left(3\right)}},\label{eq:K1threehalves}\\
\tilde{K}_{14}^{\left(2\right)} & =-10\frac{K_{1}^{\left(2\right)}K_{3}^{\left(2\right)}}{4K_{2}^{\left(1\right)}-40K_{2}^{\left(2\right)}+49K_{2}^{\left(3\right)}},\label{eq:K2threehalves}\\
\tilde{K}_{14}^{\left(3\right)} & =\frac{35}{2}\frac{K_{1}^{\left(3\right)}K_{3}^{\left(3\right)}}{8K_{2}^{\left(1\right)}-20K_{2}^{\left(2\right)}+63K_{2}^{\left(3\right)}}.\label{eq:K3threehalves}
\end{align}

\subsection{Spin 2}

The RG equations for a decimation of a spin-2 pair at the AF region
are

\begin{eqnarray}
\tilde{K}_{1,4}^{\left(1\right)} & = & \frac{16K_{1}^{\left(1\right)}K_{3}^{\left(1\right)}}{4K^{\left(1\right)}-21\left[5K^{\left(2\right)}-56K^{\left(3\right)}+270K^{\left(4\right)}\right]},\\
\tilde{K}_{1,4}^{\left(2\right)} & = & -\frac{28K_{1}^{\left(2\right)}K_{3}^{\left(2\right)}}{4K^{\left(1\right)}-85K^{\left(2\right)}+616K^{\left(3\right)}-810K^{\left(4\right)}},\\
\tilde{K}_{1,4}^{\left(3\right)} & = & \frac{112K_{1}^{\left(3\right)}K_{3}^{\left(3\right)}}{8K^{\left(1\right)}-110K^{\left(2\right)}+252K^{\left(3\right)}-1215K^{\left(4\right)}},\\
\tilde{K}_{1,4}^{\left(4\right)} & = & -\frac{378K_{1}^{\left(4\right)}K_{3}^{\left(4\right)}}{5\left[8K^{\left(1\right)}-30K^{\left(2\right)}+252K^{\left(3\right)}-675K^{\left(4\right)}\right]}.
\end{eqnarray}

The method to find planar fixed points is similar to the one we have
used for the spin-$\frac{3}{2}$ chain in the main text. The solutions
are shown in Table~\ref{fig:Fixed-point-conditions-spin-2} and represented
as black circles in Fig.~\ref{fig:Spin-2-AF-circles}.

\begin{table}
\begin{centering}
\begin{tabular}{|c|c|}
\hline 
Plane & Planar Fixed Point\tabularnewline
\hline 
\hline 
$K^{\left(1\right)}\times K^{\left(2\right)}$ & $K^{\left(2\right)}=-\frac{4}{735}\left(39+4\sqrt{141}\right)K^{\left(1\right)},~~K^{\left(1\right)}>0$\tabularnewline
\hline 
$K^{\left(1\right)}\times K^{\left(3\right)}$ & $K^{\left(3\right)}=\frac{1}{21}K^{\left(1\right)},~~K^{\left(1\right)}>0$\tabularnewline
\hline 
$K^{\left(1\right)}\times K^{\left(4\right)}$ & $K^{\left(4\right)}=-\frac{2}{19845}\left(59+\sqrt{18181}\right)K^{\left(1\right)},~~K^{\left(1\right)}>0$\tabularnewline
\hline 
$K^{\left(2\right)}\times K^{\left(3\right)}$ & $K^{\left(3\right)}=\frac{1}{56}\left(1-\sqrt{141}\right)K^{\left(2\right)},~~K^{\left(2\right)}<0$\tabularnewline
\hline 
$K^{\left(4\right)}\times K^{\left(2\right)}$ & $K^{\left(4\right)}=\frac{1}{9}K^{\left(2\right)},~~K^{\left(2\right)}<0$\tabularnewline
\hline 
$K^{\left(4\right)}\times K^{\left(3\right)}$ & $K^{\left(4\right)}=-\frac{1175-\sqrt{2328145}}{1080}K^{\left(2\right)},~~K^{\left(4\right)}<0$\tabularnewline
\hline 
\end{tabular}
\par\end{centering}

\caption{Planar fixed points of the disordered spin-2 chain. \label{fig:Fixed-point-conditions-spin-2}}
\end{table}

Finally, defining $s^{\left(i\right)}=\frac{K^{\left(i\right)}}{K^{\left(1\right)}}$,
the globally unstable point in the AF region of the 2-sphere of the
$K^{\left(1\right)}\times K^{\left(2\right)}\times K^{\left(3\right)}$
space is 

\begin{eqnarray}
\tilde{s}^{\left(2\right)*} & = & -\frac{4}{1617}\left(55+2\sqrt{1969}\right),\\
\tilde{s}^{\left(3\right)*} & = & \frac{1}{6468}\left(253+5\sqrt{1969}\right).
\end{eqnarray}
It is represented as a pink circle in Fig.~\ref{fig:Spin-2-AF-circles}.

\section{Generating SU(N)-invariant Hamiltonians using spin operators}

\label{sec:appendix-SU(N)-invariant-Hamil}

In this Appendix, we find which SU(2)-symmetric spin-$S$ Hamiltonians
of the form (\ref{eq:Hamiltonian_SU2}) are also explicitly invariant
under SU($N$) transformations. The idea is to fine-tune the parameters
in Eq.~\eqref{eq:Hamiltonian_SU2} in order to match the spectra
of SU($N$)-symmetric spin Hamiltonians. Notice that this task can
be accomplished by considering just the two-site Hamiltonian.

From the dimension of the Hilbert space of a spin $S$ in Eq.~\eqref{eq:Hamiltonian_SU2}
that $N$ must be equal to $2S+1$. This leaves us with two possibilities:
either all the SU($N$) spin operators are the generators of the fundamental
representation of the SU($N$) group, or the SU($N$) spin operators
at odd (even) sites are the generators of the fundamental (anti-fundamental)
representation of the group.

\subsubsection{Fundamental and antifundamental representation on alternating sites }

Consider the 2-site problem $h_{N-\bar{N}}^{{\rm SU}\left(N\right)}=\boldsymbol{\Gamma}_{2}\cdot\bar{\boldsymbol{\Gamma}}_{3}$,
where $\boldsymbol{\Gamma}=\left(\Gamma^{\left(1\right)},\dots,\Gamma^{\left(N^{2}-1\right)}\right)$
and $\bar{\boldsymbol{\Gamma}}=\left(\bar{\Gamma}^{\left(1\right)},\ldots,\bar{\Gamma}^{\left(N^{2}-1\right)}\right)$,
with $\Gamma^{(a)}$ ($\bar{\Gamma}^{(a)}$) being a generator of
the fundamental (anti-fundamental) representation of the SU($N$)
group. The Clebsch-Gordan series is simply $N\otimes\bar{N}=1\oplus\left(N^{2}-1\right)$.
Notice the spectrum is very simple. It has two states, of which one
is an SU($N$) singlet, which must correspond to the SU(2) singlet,
with zero total spin $\left|\tilde{S}=0\right\rangle $. The energy
difference can be obtained via the Casimir of the corresponding Young
tableau (which can be found, e. g., in Ref.\ \onlinecite{HoyosMiranda})
but this knowledge is of no importance here. We now want to recover
this spectrum with a spin-$S$ Hamiltonian. Naturally, it must read,
up to a constant,

\begin{align}
h_{N-\bar{N}}^{{\rm SU}\left(N\right)} & =0\times P_{0}+\Delta E\left(1-P_{0}\right),\\
 & =-\Delta EP_{0}+{\rm const.,}
\end{align}
where $P_{0}$ is the projector onto the singlet as defined in Eq.~\eqref{eq:projector_multiplets}
and $\Delta E$ is the energy difference between the singlet and all
the other degenerate levels.

How does this translate to the spin-spin couplings in the Hamiltonian
\eqref{eq:Hamiltonian_SU2}? This is given by Eq.~\eqref{eq:projector_multiplets}.
Here, we simply list a few examples. Defining the vector $\boldsymbol{\alpha}_{N,S}=\left(\alpha^{\left(1\right)},\ldots,\alpha^{\left(2S\right)}\right)$,
some spin-$S$ SU($N$)-symmetric cases are $\boldsymbol{\alpha}_{N=3,S=1}=\left(0,-1\right)$,
$\boldsymbol{\alpha}_{N=4,S=\frac{3}{2}}=\left(-93,20,16\right)$,
and $\boldsymbol{\alpha}_{N=5,S=2}=\left(60,17,-4,-1\right)$. Evidently,
$-\boldsymbol{\alpha}_{N,S}$ yields a Hamiltonian which possesses
the same symmetry but represents the FM case.

It is also interesting to recast these Hamiltonians in terms of the
ISTs defined in Eq.~\eqref{eq:irred_spherical_harm}. It can be done
by directly using the dictionary between the $\alpha$-couplings and
the $K$-couplings. For instance, in obvious notation, $\mathbf{K}_{N=3,S=1}=\left(1,-\frac{4}{5}\right)$,
$\mathbf{K}_{N=4,S=\frac{3}{2}}=\left(1,-\frac{1}{3},\frac{4}{21}\right)$,
and $\mathbf{K}_{N=5,S=2}=\left(1,-\frac{4}{21},\frac{1}{21},-\frac{4}{189}\right)$. 

We would like to show now that these Hamiltonians lie inside the AF
hyper-octant, as discussed in Sec.\ \eqref{sub:PairwiseRSP}. For
that, we will derive a more general approach. Let us start by decomposing
the projector $P_{0}$ as 

\begin{equation}
-P_{0}=\sum_{J}\phi_{J}\left(S\right)\hat{O}_{J}\left(\mathbf{S}_{2,}\mathbf{S}_{3}\right),\label{eq:projasoj}
\end{equation}
where $S_{2}=S_{3}=S$, and find the coefficients $\phi_{J}\left(S\right)$.
We are going to compute the matrix elements of the above equation
in states of total angular momentum $\tilde{S}$, that is, the multiplet
coming from the sum of angular momenta $S_{2}$ with $S_{3}$, denoted
by $\left|\tilde{S},\tilde{M}\right\rangle =\left|S_{2}S_{3};\tilde{S},\tilde{M}\right\rangle $.
The matrix element of the $\hat{O}_{J}$ operator is found from Ref.~\onlinecite{Edmondsbook}
(page 111) to be given by

\begin{eqnarray}
\left\langle J'M'\left|\hat{O}_{J}\right|J''M''\right\rangle  & = & \left(-1\right)^{2S+J''}\delta_{J'J''}\delta_{M'M''}\left\{ \begin{array}{ccc}
J'' & S & S\\
J & S & S
\end{array}\right\} \nonumber \\
 &  & \times\left|\left\langle S\left|\left|Y_{J}\left(\mathbf{S}\right)\right|\right|S\right\rangle \right|^{2}.\label{eq:matrix-element-OJ}
\end{eqnarray}
where $\left\{ ...\right\} $ is Wigner's \textit{6-j} symbol. The
reduced matrix element of $Y_{J}\left(\mathbf{S}\right)$, $\left\langle S\left|\left|Y_{J}\left(\mathbf{S}\right)\right|\right|S\right\rangle $,
was calculated in Eq.~\eqref{eq:reduced-matel-single-site}. Since
$\left\langle \tilde{S},\tilde{M}\left|P_{0}\right|\tilde{S},\tilde{M}\right\rangle =\delta_{\tilde{S},0}$,
the matrix elements of Eq.~\eqref{eq:projasoj} in the $\left|\tilde{S},\tilde{M}\right\rangle $
states yield

\begin{equation}
\left(-1\right)^{2S+\tilde{S}+1}\delta_{\tilde{S},0}=\sum_{J}\phi_{J}\left\{ \begin{array}{ccc}
\tilde{S} & S & S\\
J & S & S
\end{array}\right\} \left|\left\langle S\left|\left|Y_{J}\left(\mathbf{S}\right)\right|\right|S\right\rangle \right|^{2}.\label{eq:matrix-elements-sides}
\end{equation}
Multiplying Eq.~\eqref{eq:matrix-elements-sides} by $\left(2\tilde{S}+1\right)\left\{ \begin{array}{ccc}
J'' & S & S\\
\tilde{S} & S & S
\end{array}\right\} $, summing over $\tilde{S}$, and using the orthogonality relation
(page 96 of Ref.~\onlinecite{Edmondsbook})

\begin{equation}
\sum_{\tilde{S}}\left(2\tilde{S}+1\right)\left\{ \begin{array}{ccc}
\tilde{S} & S & S\\
J & S & S
\end{array}\right\} \left\{ \begin{array}{ccc}
J'' & S & S\\
\tilde{S} & S & S
\end{array}\right\} =\frac{\delta_{J,J''}}{\left(2J''+1\right)},
\end{equation}
we find

\begin{eqnarray}
\sum_{\tilde{S}}\left(-1\right)^{2S+\tilde{S}+1}\delta_{\tilde{S},0}\left\{ \begin{array}{ccc}
\tilde{S} & S & S\\
J'' & S & S
\end{array}\right\}  & = & \frac{\left|\left\langle S\left|\left|Y_{J}\left(\mathbf{S}\right)\right|\right|S\right\rangle \right|^{2}}{\left(2J''+1\right)}\phi_{J''}.
\end{eqnarray}
Since $\left\{ \begin{array}{ccc}
0 & S & S\\
J'' & S & S
\end{array}\right\} =\frac{\left(-1\right)^{J''+2S}}{2S+1}$ (page 98 of Ref.~\onlinecite{Edmondsbook}), we obtain finally

\begin{eqnarray}
\phi_{J}\left(S\right) & = & \frac{\left(-1\right)^{J+1}}{\left|\left\langle S\left|\left|Y_{J}\left(\mathbf{S}\right)\right|\right|S\right\rangle \right|^{2}}\frac{\left(2J+1\right)}{\left(2S+1\right)}.\label{eq:phiJ}
\end{eqnarray}
The most important feature of Eq.~\eqref{eq:phiJ} is that it alternates
sign with $J$. This guarantees that the SU($N$)-symmetric Hamiltonian
is in a region in parameter space where the product $\left(-1\right)^{J+1}K_{J}$
is always positive, i.e., in the middle of the AF hyper-octant (see
Fig.\ \eqref{fig:Hyper-octant}). Besides, as this large symmetry
is preserved along the RG flow, it also corresponds to a fixed point
in the middle of the AF hyper-octant, which is totally unstable. Since,
as mentioned in the main text, the random singlets generated by the
RG process are the same in the entire AF hyper-octant, we conclude
that they are all also SU($N$) singlets.

\subsubsection{Fundamental representation on all sites}

We now repeat the same steps above for the 2-site problem $h_{N-N}^{{\rm SU}\left(N\right)}=\boldsymbol{\Gamma}_{2}\cdot\boldsymbol{\Gamma}_{3}$.
As before, the spectrum also has two states but with a different degeneracy.
From the Clebsch-Gordan series, we have 
\begin{align}
N\otimes N & =\frac{N\left(N-1\right)}{2}\oplus\frac{N\left(N+1\right)}{2}\nonumber \\
 & =S\left(2S+1\right)\oplus\left(2S+1\right)\left(S+1\right).
\end{align}
It can be checked that this spectrum can be generated, up to a constant,
by 
\[
h_{N-N}^{{\rm SU}\left(N\right)}=-\sum_{J=0}^{2S}\left(1-\left(-1\right)^{J+2S}\right)P_{J},
\]
where $P_{J}$ is the projector onto the multiplet of total angular
momentum $J$ {[}see Eq.\ (\ref{eq:projector_multiplets}){]}. Notice
that if $S$ is integer (semi-integer), then only the projectors onto
the odd (even) $J$ multiplets are included. Even though there are
other ways of reproducing the spectrum for particular values of $S$
(e. g., $S=2$), this is the only choice that does so for generic
spin values and, indeed, the one that realizes the SU($N$) symmetry.

As in the $N-\bar{N}$ case, we can use Eq.~\eqref{eq:projector_multiplets}
in order to find the corresponding Hamiltonian in the form \eqref{eq:Hamiltonian_SU2}
in terms of spin operators. Some examples are $\boldsymbol{\alpha}_{N=3,S=1}=\left(1,1\right)$,
$\boldsymbol{\alpha}_{N=4,S=\frac{3}{2}}=\left(-81,44,16\right)$,
and $\boldsymbol{\alpha}_{N=5,S=2}=\left(-90,-13,6,1\right)$. They
correspond to $\mathbf{K}_{N=3,S=1}=\left(1,\frac{4}{5}\right)$,
$\mathbf{K}_{N=4,S=\frac{3}{2}}=\left(1,\frac{1}{3},\frac{4}{21}\right)$,
and $\mathbf{K}_{N=5,S=2}=\left(1,\frac{4}{21},\frac{1}{21},\frac{4}{189}\right)$.
Note the similarities with the $N-\bar{N}$ case: only the signs of
the even-rank couplings are reversed.

As in the previous case, by using Eq.~\eqref{eq:irred_spherical_harm},
the Hamiltonian can be rewritten in terms of ISTs. The generators
of the fundamental representation of the SU($N$) group are $N^{2}-1$
$N\times N$ traceless hermitian matrices. We choose them to satisfy
the trace condition

\begin{equation}
\mbox{Tr}\left(\Lambda_{i}\Lambda_{j}\right)\propto\delta_{i,j}.\label{eq:sun-trace-cond}
\end{equation}
There are $2J+1$ linearly independent components for each rank-$J$
IST. By collecting all of them (except for $J=0$), the number of
linear independent ISTs up to order $2S$ is

\begin{eqnarray}
\sum_{J=1}^{2S}\left(2J+1\right) & = & 2S\left(2S+1\right)+2S\\
 & = & 4S\left(S+1\right),
\end{eqnarray}
which is exactly $N^{2}-1$, with $N=2S+1$. The $J=0$ IST was excluded
because its trace is non-zero. The proper choice of the SU($N$) generators
is found by combining the ISTs of same rank and components $M$ and
$-M$, namely, 

\begin{equation}
\Lambda_{J,M}\propto\begin{cases}
Y_{J,M}\left(\mathbf{S}\right)+Y_{J,-M}\left(\mathbf{S}\right), & \,\,M>0,\\
Y_{J,0}\left(\mathbf{S}\right), & \,\,M=0,\\
Y_{J,M}\left(\mathbf{S}\right)-Y_{J,-M}\left(\mathbf{S}\right), & \,\,M<0.
\end{cases}\label{eq:sun-ISTs}
\end{equation}
When the linear combination with minus sign is taken, the overall
constant is an imaginary number. The above $\Lambda_{J,M}$ matrices
are hermitian, traceless and also linearly independent, since, by
construction, the ISTs are linearly independent. This shows that the
set of matrices in Eq.~\eqref{eq:sun-trace-cond} are generators
of the fundamental representation of the SU($N$) group.

The trace orthogonality condition Eq.~\eqref{eq:sun-trace-cond}
is also satisfied. In order to show that, we start by expanding a
product of two ISTs as a linear combination of ISTs (page 69 of Ref.~\onlinecite{Edmondsbook}) 

\begin{align}
Y_{J,M}\left(\mathbf{S}\right)Y_{J',M'}\left(\mathbf{S}\right) & =\sum_{J'',M''}\zeta\left(J,J',J''\right)\times\nonumber \\
\left\langle JJ';JM\right| & \left.\!JJ';MM'\right\rangle Y_{J'',M''}\left(\mathbf{S}\right),\label{eq:2-IST-decomp}
\end{align}
where $\zeta\left(J,J',J''\right)$ does not depend on the tensor
components, and the sum over $J''$ runs from $\left|J-J'\right|$
to $J+J'$. The trace involves the computation of diagonal elements
of the above equation

\begin{align}
\left\langle SM\right|Y_{J,M}\left(\mathbf{S}\right) & Y_{J',M'}\left(\mathbf{S}\right)\left|SM\right\rangle \nonumber \\
= & \sum_{J'',M''}\left\langle JJ';J''M''\left|JJ';MM'\right.\right\rangle \nonumber \\
 & \times\left\langle SM\left|Y_{J'',M''}\left(\mathbf{S}\right)\right|SM\right\rangle .
\end{align}
From the Wigner-Eckart theorem, Eq.~\eqref{eq:wigner-eckart}, the
only value of $M^{\prime\prime}$ that survives the sum is $M''=0$,
and the Clebsch-Gordan coefficient then requires $M'=-M$. Thus, 

\begin{align}
\mbox{Tr}\left(Y_{J,M}\left(\mathbf{S}\right)Y_{J',-M}\left(\mathbf{S}\right)\right) & =\nonumber \\
\sum_{J''=\left|J-J'\right|}^{J+J'}\left\langle JJ';J''0\left|JJ';M-M\right.\right\rangle  & \mbox{Tr}\left(Y_{J'',0}\left(\mathbf{S}\right)\right).
\end{align}
But only $Y_{0,0}\left(\mathbf{S}\right)$ has a non-vanishing trace.
Therefore, the only term that survives in the sum is $J''=0$. But
$J''=0$ requires $J=J'$. It follows that only $Y_{J,M}\left(\mathbf{S}\right)Y_{J,-M}\left(\mathbf{S}\right)$
has a non-zero trace. Let us assume, for concreteness, that $M,M'>0$.
The other cases follow analogously. Then,

\begin{eqnarray}
\mbox{Tr}\left(\Lambda_{J,M}\Lambda_{J',M'}\right) & \propto & \mbox{Tr}\left(Y_{J,M}\left(\mathbf{S}\right)+Y_{J,-M}\left(\mathbf{S}\right)\right)\times\nonumber \\
 &  & \left(Y_{J',M'}\left(\mathbf{S}\right)+Y_{J',-M'}\left(\mathbf{S}\right)\right),\\
 & \propto & 2\delta_{J,J'}\left(\delta_{M,M'}+\delta_{M,-M'}\right).
\end{eqnarray}
Therefore, the trace condition $\mbox{Tr}\left(\Lambda_{J,M}\Lambda_{J',M'}\right)\propto\delta_{J,J'}\delta_{M,M'}$
is satisfied by the operators defined in Eq.~\eqref{eq:sun-ISTs}.
We have thus proved all the conditions that are necessary in order
that the collection of the $N^{2}-1$ traceless ISTs of rank up to
$2S$ (excluding zero rank) can be chosen as generators of the fundamental
representation of the SU(N) group.

\section{When the two-spin Hamiltonian has a singlet ground state\label{sec:S=00003D0-condition}}

Fundamental to our understanding of the $\psi=\frac{1}{2}$ AF phases
is the fact that singlets are formed in the decimations on some of
the semi-axes. In this Appendix, we analyze in detail the conditions
under which the ground state of the two-spin problem $S_{2}=S_{3}=S$
is a singlet. We focus on a given axis and, therefore, only ISTs of
a given rank, say $J$, are non-zero.

The energy of a multiplet of total angular momentum $J'$, $E_{J}\left(J'\right)$,
can be found by using Eq.~\eqref{eq:matrix-element-OJ} in a generic
eigenstate of the Hamiltonian, $\left|SS;J'M'\right\rangle $

\begin{eqnarray}
E_{J}\left(J'\right) & = & \left\langle SS;J'M'\left|\hat{O}_{J}\right|SS;J'M'\right\rangle \label{eq:red_tensor-1-1}\\
 & = & \left(-1\right)^{2S+J'}\left\{ \begin{array}{ccc}
J' & S & S\\
J & S & S
\end{array}\right\} \left|\left\langle S\left|\left|Y_{J}\left(\mathbf{S}\right)\right|\right|S\right\rangle \right|^{2}
\end{eqnarray}
Note that, since the operator is SU(2)-symmetric, the right-hand side
is the energy of the system, independent of $M'$. One can use Eq.~\eqref{eq:reduced-matel-single-site}
to compute the reduced matrix element but, for now, we just note that
it is a function of only $J$ and $S$, assumed to be fixed in this
analysis. We define the ratio

\begin{eqnarray}
\tilde{E}_{J}\left(J'\right) & = & \frac{E_{J}\left(J'\right)}{\left(-1\right)^{2S}\left|\left\langle S\left|\left|Y_{J}\left(\mathbf{S}\right)\right|\right|S\right\rangle \right|^{2}}\nonumber \\
 & = & \left(-1\right)^{J'}\left\{ \begin{array}{ccc}
J' & S & S\\
J & S & S
\end{array}\right\} .
\end{eqnarray}
\textcolor{black}{The task is to find the value of $J'$ that minimizes
(maximizes) $\tilde{E}_{J}\left(J'\right)$ for integer (half-integer)
$S$, with $J$ varying from $0$ to $2S$. We have numerically checked
up to $J=8$ and $S=80$ that this requirement is satisfied for $J'=0$.
}This provides strong evidence that the singlet is the ground state
when $\left(-1\right)^{J}K^{\left(J\right)}>0$, which is the result
quoted in the main text.

\section{Beyond first-order perturbation theory of degenerate multiplets \label{sec:appendix-zeros}}

We show explicitly how to compute second-order corrections to two
concrete cases where the ground multiplet is not a singlet but the
first-order perturbation theory renormalization vanishes. The calculations
to find such corrections are lengthy and have to be done case by case.
We deal with the case where the spins are equal to $\frac{3}{2}$.
We start by showing the steps to derive the RG renormalization when
decimations are performed on the $K^{\left(3\right)}<0$ axis, where
the local ground state is a spin-1 multiplet. The first-order perturbation
theory vanishes due to case (a) discussed in Section~\ref{sub:First-Order-Perturbation}.
After that, we compute second-order effects on spins connected by
$K^{\left(2\right)}>0$ tensors. This is an axis where case (b) of
Section \ref{sub:First-Order-Perturbation} leads to a vanishing first-order
renormalization. The question we want to address is whether higher
order corrections could give contributions that would change the ground
state properties in a non-trivial way. We have checked through numerical
diagonalization of the three-site problem that in both cases there
are indeed small second-order SU(2)-symmetric interactions with the
side spins of a decimated pair.

\subsection{On the $K^{\left(3\right)}<0$ semi-axis}

The four-spin Hamiltonian can be re-written as

\begin{equation}
\mathcal{H}=V_{1,2}+H_{2,3}^{0}+V_{3,4}
\end{equation}
where $H^{0}$ is the unperturbed Hamiltonian. First-order perturbation
theory gives the following correction to the Hamiltonian

\begin{equation}
\Delta H^{\left(1\right)}=P_{\tilde{S}}V_{1,2}P_{\tilde{S}}+P_{\tilde{S}}V_{3,4}P_{\tilde{S}},
\end{equation}
where $P_{\tilde{S}}$ is the projector onto the ground multiplet,
which in this case is a $\tilde{S}=1$ total angular momentum state.
Applying the Wigner-Eckart theorem, one can easily show that this
correction is zero, since the sum of angular momenta $S=1$ and $S=3$
cannot give $\tilde{S}=1$. The first order effect would be, therefore,
to break the chain into 2 decoupled smaller chains. 

A natural question is what is the lowest-order correction that gives
a non-zero contribution. The second-order correction is given by

\begin{eqnarray}
\Delta H^{\left(2\right)} & = & P_{\tilde{S}}\left(V_{1,2}+V_{3,4}\right)\bar{P}\frac{1}{E_{0}-H_{2,3}^{0}}\bar{P}\left(V_{1,2}+V_{3,4}\right)P_{\tilde{S}},\label{eq:second-order-correc}\\
 & = & \Delta H_{1,2}^{\left(2\right)}+\Delta H_{3,4}^{\left(2\right)}+\Delta H_{\left(1,2\right),\left(3,4\right)}^{\left(2\right)}
\end{eqnarray}
where $\bar{P}=1-P_{\tilde{S}}$ and we have defined

\begin{eqnarray}
\Delta H_{i,i+1}^{\left(2\right)} & = & P_{\tilde{S}}V_{i,i+1}\bar{P}\frac{1}{E_{0}-H_{2,3}^{0}}\bar{P}V_{i,i+1}P_{\tilde{S}},\\
\Delta H_{\left(1,2\right),\left(3,4\right)}^{\left(2\right)} & = & P_{\tilde{S}}V_{1,2}\bar{P}\frac{1}{E_{0}-H_{2,3}^{0}}\bar{P}V_{3,4}P_{\tilde{S}}+\mbox{H.c}.
\end{eqnarray}

We first consider $\Delta H_{\left(1,2\right),\left(3,4\right)}^{\left(2\right)}$.
It gives rise to different types of terms which we call $\Delta\mathcal{V}_{J,\tilde{M}}$,
where $J$ and $\tilde{M}$ correspond to the rank and component of
the IST of $\tilde{\mathbf{S}}$ it contains. There is a next-nearest-neighbor
ferromagnetic interaction independent of $\tilde{\mathbf{S}}$ ($J=\tilde{M}=0$)

\begin{equation}
\Delta\mathcal{V}_{0,0}=-\frac{23\sqrt{\pi}}{35}\hat{O}_{3}\left(\mathbf{S}_{1},\mathbf{S}_{4}\right).
\end{equation}
The other terms are genuine three-body interactions given by

\begin{eqnarray}
\Delta\mathcal{V}_{1,\tilde{M}} & = & \frac{\sqrt{\pi}}{35}\sum_{M=-3}^{3-\tilde{M}}\alpha_{M}^{\left(\tilde{M}\right)}Y_{3,M}\left(\mathbf{S}_{1}\right)Y_{1,\tilde{M}}\left(\tilde{\mathbf{S}}\right)Y_{3,-M-\tilde{M}}\left(\mathbf{S}_{4}\right)\nonumber \\
 & + & \mbox{H.c.},\label{eq:v-cont-1}\\
\Delta\mathcal{V}_{2,\tilde{M}} & = & \frac{\sqrt{\pi}}{35}\sum_{M=-3}^{3-\tilde{M}}\beta_{M}^{\left(\tilde{M}\right)}Y_{3,M}\left(\mathbf{S}_{1}\right)Y_{2,\tilde{M}}\left(\tilde{\mathbf{S}}\right)Y_{3,3-\tilde{M}}\left(\mathbf{S}_{4}\right)\nonumber \\
 & + & \mbox{H.c.}.\label{eq:v-cont-2}
\end{eqnarray}
In Eq.~\eqref{eq:v-cont-1}, $\tilde{M}$ runs from $-1$ to $1$,
whereas in Eq.~\eqref{eq:v-cont-2} the sum is from $-2$ to $2$.
The coefficients $\alpha_{M}^{\left(\tilde{M}\right)}$ and $\beta_{M}^{\left(\tilde{M}\right)}$
are given in Table~\ref{tab:V-coeff}. Through numerical diagonalization
of 4-site chains, we find that all the terms $\Delta\mathcal{V}_{J,\tilde{M}}$
are \emph{non-frustrating}, i. e., the ground state is the same whether
we keep them in the RG procedure or not. For this reason, we will
neglect them in what follows. 

\begin{center}
\begin{table}
\begin{centering}
\begin{tabular}{|c|c|}
\hline 
\selectlanguage{brazil}%
$\tilde{M}$\selectlanguage{american}%
 & \selectlanguage{brazil}%
$\alpha_{M}^{\left(\tilde{M}\right)}$\selectlanguage{american}%
\tabularnewline
\hline 
\hline 
\selectlanguage{brazil}%
0\selectlanguage{american}%
 & \selectlanguage{brazil}%
$\frac{1}{\sqrt{3}}\left(3,-2,1,0,-1,2,-3\right)$\selectlanguage{american}%
\tabularnewline
\hline 
\selectlanguage{brazil}%
1\selectlanguage{american}%
 & \selectlanguage{brazil}%
$\frac{1}{\sqrt{3}}\left(\sqrt{3},-\sqrt{5},\sqrt{6},-\sqrt{6},\sqrt{5}\right)$\selectlanguage{american}%
\tabularnewline
\hline 
\end{tabular}
\par\end{centering}

\begin{centering}
\begin{tabular}{|c|c|}
\hline 
\selectlanguage{brazil}%
$\tilde{M}$\selectlanguage{american}%
 & \selectlanguage{brazil}%
$\beta_{M}^{\left(\tilde{M}\right)}$\selectlanguage{american}%
\tabularnewline
\hline 
\hline 
\selectlanguage{brazil}%
0\selectlanguage{american}%
 & \selectlanguage{brazil}%
$\frac{8}{\sqrt{5}}\left(-5,0,3,-4,3,0,-5\right)$\selectlanguage{american}%
\tabularnewline
\hline 
\selectlanguage{brazil}%
1\selectlanguage{american}%
 & \selectlanguage{brazil}%
$\frac{8}{\sqrt{5}}\left(-5,\sqrt{15},-\sqrt{2},-\sqrt{2},\sqrt{15},-5\right)$\selectlanguage{american}%
\tabularnewline
\hline 
\selectlanguage{brazil}%
2\selectlanguage{american}%
 & \selectlanguage{brazil}%
$\frac{16}{\sqrt{10}}\left(-\sqrt{5},\sqrt{10},-\sqrt{12},\sqrt{10},-\sqrt{5}\right)$\selectlanguage{american}%
\tabularnewline
\hline 
\end{tabular}
\par\end{centering}

\caption{Constants that appear in the three-body interaction terms of Eqs.~\eqref{eq:v-cont-1}
and \eqref{eq:v-cont-2}. The vector components correspond to the
$M$ index, starting at $M=-3$. \label{tab:V-coeff}}
\end{table}

\par\end{center}

We keep, however, the $\Delta H_{i,i+1}^{\left(2\right)}$ terms.
For $i=1$, for example,

\begin{eqnarray}
\frac{\Delta H_{1,2}^{\left(2\right)}}{K_{1}^{\left(J\right)}} & = & \sum_{\left\{ M_{1},M_{2}\right\} =-J}^{J}\left(-1\right)^{M_{1}-M_{2}}\left[Y_{JM_{1}}\left(\mathbf{S}_{1}\right)Y_{J-M_{2}}\left(\mathbf{S}_{1}\right)\right]\times\nonumber \\
 &  & \left[PY_{J-M_{1}}\left(\mathbf{S}_{2}\right)\tilde{P}\frac{1}{E_{0}-H_{2,3}^{0}}\tilde{P}Y_{JM_{2}}\left(\mathbf{S}_{2}\right)P\right].\label{eq:decomp_secondorder-1}
\end{eqnarray}
The term $Y_{JM_{1}}\left(\mathbf{S}_{1}\right)Y_{J-M_{2}}\left(\mathbf{S}_{1}\right)$
can be decomposed as a linear combination that conserves the azimuthal
component of the angular momentum, the same as in Eq.~\textbf{\eqref{eq:2-IST-decomp}}
of Appendix~\eqref{sec:appendix-SU(N)-invariant-Hamil}. For example,
for $J=3$ and $M_{1}=M_{2}=-3$,

\begin{align}
Y_{3,-3}\left(\mathbf{S}_{1}\right)Y_{3,3}\left(\mathbf{S}_{1}\right)= & -\frac{315}{32\sqrt{\pi}}Y_{0,0}\left(\mathbf{S}_{1}\right)\nonumber \\
+\frac{63}{16}\sqrt{\frac{3}{\pi}}Y_{1,0}\left(\mathbf{S}_{1}\right)-\frac{21}{16}\sqrt{\frac{5}{\pi}} & Y_{2,0}\left(\mathbf{S}_{1}\right)+\frac{3}{8}\sqrt{\frac{7}{\pi}}Y_{3,0}\left(\mathbf{S}_{1}\right).
\end{align}

The projection onto the ground state multiplet also conserves the
$M$ values and, therefore, can be decomposed in terms of ISTs of
the effective new degrees of freedom,

\begin{align}
PY_{J-M_{1}}\left(\mathbf{S}_{2}\right)\tilde{P}\frac{1}{E_{0}-H_{2,3}^{0}}\tilde{P}Y_{JM_{2}}\left(\mathbf{S}_{2}\right)P & =\nonumber \\
\frac{1}{K_{2}^{\left(3\right)}}\sum_{J'}\beta_{J',\left(M_{2}-M_{1}\right)}Y_{J',\left(M_{2}-M_{1}\right)}\left(\mathbf{\tilde{S}}=1\right).
\end{align}

\begin{figure}[t]
\begin{centering}
\includegraphics[clip,width=0.6\columnwidth]{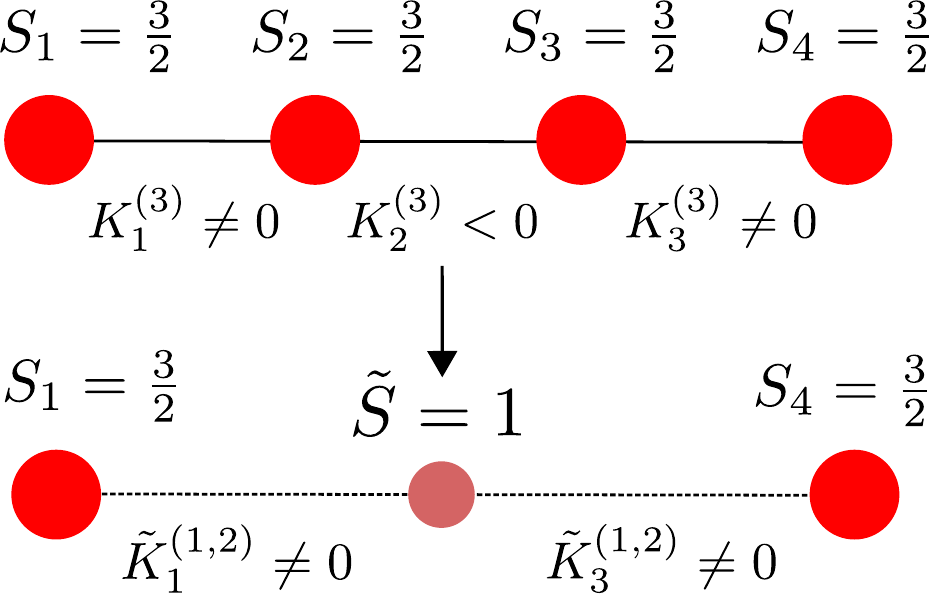}
\par\end{centering}

\caption{(Color online) RG step when the spins are equal to $\frac{3}{2}$
and $K^{\left(3\right)}<0$. The 2-spin ground-state multiplet is
$\tilde{S}=1$ and the first-order perturbation theory yields a vanishing
renormalization. Unlike in the cases discussed in Section \ref{sub:First-Order-Perturbation},
the second-order step generates tensors of ranks that were not present
in the original chain. \label{fig:RG-step-correc}}
\end{figure}

\begin{figure}[b]
\begin{centering}
\includegraphics[clip,width=0.9\columnwidth]{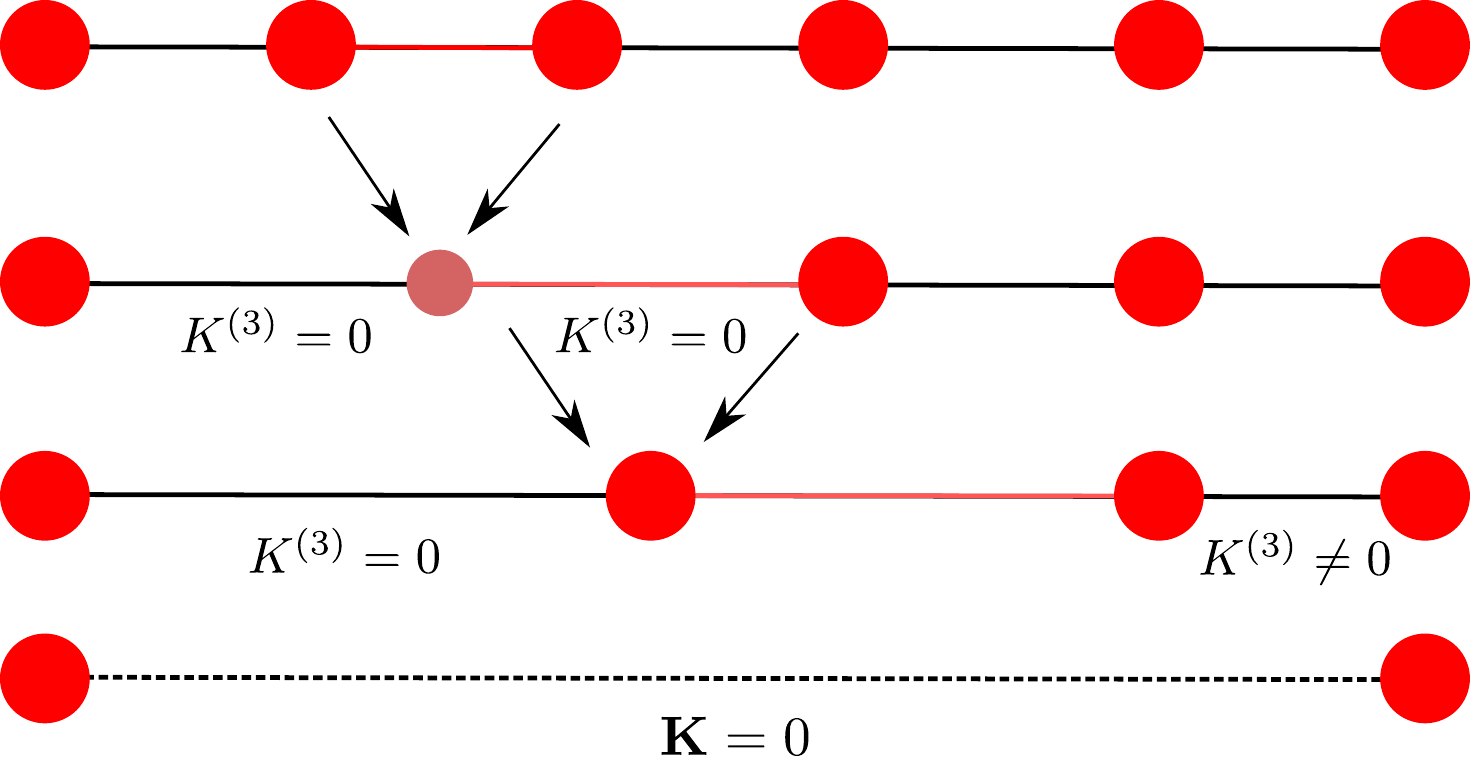}
\par\end{centering}

\caption{(Color online) Schematic representation of the RG decimations after
the decimation procedure shown in this Appendix is implemented, and
the generation of zero couplings after some RG time. The strongest
coupled pair is represented by the red line. The first decimation
generates an effective spin $\tilde{S}=1$ (light red), that is coupled
to its neighbors via $K^{\left(1,2\right)}$, but not $K^{\left(3\right)}$.
The second decimation leads back to an effective spin $\frac{3}{2}$,
with its left neighbor having $K^{\left(3\right)}=0$. The third decimation
is a second-order singlet-formation decimation and the final effective
coupling of the edge spins is zero. This is so because $\tilde{K}^{\left(J\right)}$
involves the product of neighboring couplings of same rank and, whereas
the right neighbor has only $K^{\left(3\right)}\protect\ne0$, the
left neighbor has $K^{\left(3\right)}=0$. \label{fig:RG-step-correc-1}}
\end{figure}

Plugging this decomposition into Eq.~\eqref{eq:decomp_secondorder-1},
we find 

\begin{eqnarray}
PV_{1,2}\tilde{P}\frac{1}{E_{0}-H_{2,3}^{0}}\tilde{P}V_{1,2}P & =\nonumber \\
\frac{\left(K_{1}^{\left(3\right)}\right)^{2}}{\left|K_{2}^{\left(3\right)}\right|}\sum_{J=0}^{2}\gamma_{J}\hat{O}_{J}\left(\mathbf{S}_{1}=\frac{3}{2},\mathbf{\tilde{S}}=1\right).
\end{eqnarray}
For this particular case,

\begin{eqnarray}
\gamma_{0} & = & \frac{2079}{128},~~~\gamma_{1}=\frac{63}{20},~~~\gamma_{2}=\frac{189}{100}.
\end{eqnarray}
Neglecting the constant factor, we find the residual 2-body interaction
between the effective $\tilde{S}=1$ spin and the $S_{1}=\frac{3}{2}$
spin

\begin{equation}
\Delta H_{1,2}^{\left(2\right)}=\frac{\left(K_{1}^{\left(3\right)}\right)^{2}}{\left|K_{2}^{\left(3\right)}\right|}\left(\frac{63}{20}\hat{O}_{1}+\frac{189}{100}\hat{O}_{2}\right),\label{eq:RG-rule-fixed}
\end{equation}
where $\hat{O}_{i}=\hat{O}_{i}\left(\mathbf{S}_{1}=\frac{3}{2},\mathbf{\tilde{S}}=1\right)$.
By symmetry, we obtain the coupling connecting to site 4 by replacing
$1\leftrightarrow4$. Note that the decimation generates couplings
between ISTs that were not coupled in the initial Hamiltonian ($\hat{O}_{1}$
and $\hat{O}_{2}$). A schematic representation of the RG rule Eq.~\eqref{eq:RG-rule-fixed}
is represented in Fig.~\ref{fig:RG-step-correc}. As explained in
Fig.~\ref{fig:RG-step-correc-1}, although the first steps yield
a non-zero renormalization of couplings, in later steps the RG procedure
breaks the chain into two parts.

\subsection{On the $K^{\left(2\right)}>0$ semi-axis}

In this case, the reason why the first-order calculation vanishes
is that the couplings constants are proportional to the following
\emph{6-j} symbol

\begin{equation}
\left\{ \begin{array}{ccc}
2 & 2 & 2\\
\frac{3}{2} & \frac{3}{2} & \frac{3}{2}
\end{array}\right\} =0.
\end{equation}

The steps of this calculation are analogous to the previous case.
The decomposition of $Y_{2,M_{1}}\left(\mathbf{S}_{1}\right)Y_{2,-M_{2}}\left(\mathbf{S}_{1}\right)$
gives terms with $J$ ranging from 0 to 3, except $J=2$. The effective
Hamiltonian that connects a spin $S_{1}=\frac{3}{2}$ with a spin
$\tilde{S}=2$ is, up to an additive constant,

\begin{equation}
\Delta H_{1,2}^{\left(2\right)}=\frac{\left(K_{1}^{\left(2\right)}\right)^{2}}{K_{2}^{\left(2\right)}}\left(\frac{9}{16}\hat{O}_{1}-\frac{3}{56}\hat{O}_{3}\right).\label{eq:low-energy-k2-axis}
\end{equation}
Here, $\hat{O}_{i}=\hat{O}_{i}\left(\mathbf{S}_{1}=\frac{3}{2},\mathbf{\tilde{S}}=2\right)$.
As in the previous subsection, this procedure generates an RG flow
with non-zero couplings in the first steps, but vanishing couplings
are generated later via the same mechanism as the one described in
Fig.~\ref{fig:RG-step-correc-1}. Also analogous to the previous
section are the calculations of all the three-body and long-ranged
effective couplings. 

\bibliographystyle{apsrev4-1}

\begin{thebibliography}{37}%
\makeatletter
\providecommand \@ifxundefined [1]{%
 \@ifx{#1\undefined}
}%
\providecommand \@ifnum [1]{%
 \ifnum #1\expandafter \@firstoftwo
 \else \expandafter \@secondoftwo
 \fi
}%
\providecommand \@ifx [1]{%
 \ifx #1\expandafter \@firstoftwo
 \else \expandafter \@secondoftwo
 \fi
}%
\providecommand \natexlab [1]{#1}%
\providecommand \enquote  [1]{``#1''}%
\providecommand \bibnamefont  [1]{#1}%
\providecommand \bibfnamefont [1]{#1}%
\providecommand \citenamefont [1]{#1}%
\providecommand \href@noop [0]{\@secondoftwo}%
\providecommand \href [0]{\begingroup \@sanitize@url \@href}%
\providecommand \@href[1]{\@@startlink{#1}\@@href}%
\providecommand \@@href[1]{\endgroup#1\@@endlink}%
\providecommand \@sanitize@url [0]{\catcode `\\12\catcode `\$12\catcode
  `\&12\catcode `\#12\catcode `\^12\catcode `\_12\catcode `\%12\relax}%
\providecommand \@@startlink[1]{}%
\providecommand \@@endlink[0]{}%
\providecommand \url  [0]{\begingroup\@sanitize@url \@url }%
\providecommand \@url [1]{\endgroup\@href {#1}{\urlprefix }}%
\providecommand \urlprefix  [0]{URL }%
\providecommand \Eprint [0]{\href }%
\providecommand \doibase [0]{http://dx.doi.org/}%
\providecommand \selectlanguage [0]{\@gobble}%
\providecommand \bibinfo  [0]{\@secondoftwo}%
\providecommand \bibfield  [0]{\@secondoftwo}%
\providecommand \translation [1]{[#1]}%
\providecommand \BibitemOpen [0]{}%
\providecommand \bibitemStop [0]{}%
\providecommand \bibitemNoStop [0]{.\EOS\space}%
\providecommand \EOS [0]{\spacefactor3000\relax}%
\providecommand \BibitemShut  [1]{\csname bibitem#1\endcsname}%
\let\auto@bib@innerbib\@empty
%</preamble>
\bibitem [{\citenamefont {Lee}\ and\ \citenamefont
  {Ramakrishnan}(1985)}]{LeeRMP1985}%
  \BibitemOpen
  \bibfield  {author} {\bibinfo {author} {\bibfnamefont {P.~A.}\ \bibnamefont
  {Lee}}\ and\ \bibinfo {author} {\bibfnamefont {T.~V.}\ \bibnamefont
  {Ramakrishnan}},\ }\href {\doibase 10.1103/RevModPhys.57.287} {\bibfield
  {journal} {\bibinfo  {journal} {Rev. Mod. Phys.}\ }\textbf {\bibinfo {volume}
  {57}},\ \bibinfo {pages} {287} (\bibinfo {year} {1985})}\BibitemShut
  {NoStop}%
\bibitem [{\citenamefont {Igl\'oi}\ and\ \citenamefont
  {Monthus}(2005)}]{igloi-review}%
  \BibitemOpen
  \bibfield  {author} {\bibinfo {author} {\bibfnamefont {F.}~\bibnamefont
  {Igl\'oi}}\ and\ \bibinfo {author} {\bibfnamefont {C.}~\bibnamefont
  {Monthus}},\ }\href {\doibase 10.1016/j.physrep.2005.02.006} {\bibfield
  {journal} {\bibinfo  {journal} {Phys. Rep.}\ }\textbf {\bibinfo {volume}
  {412}},\ \bibinfo {pages} {277} (\bibinfo {year} {2005})}\BibitemShut
  {NoStop}%
\bibitem [{\citenamefont {Vojta}(2006)}]{vojta-review06}%
  \BibitemOpen
  \bibfield  {author} {\bibinfo {author} {\bibfnamefont {T.}~\bibnamefont
  {Vojta}},\ }\href {\doibase 10.1088/0305-4470/39/22/R01} {\bibfield
  {journal} {\bibinfo  {journal} {J. Phys. A: Math. Gen.}\ }\textbf {\bibinfo
  {volume} {39}},\ \bibinfo {pages} {R143} (\bibinfo {year}
  {2006})}\BibitemShut {NoStop}%
\bibitem [{\citenamefont {Fisher}(1994)}]{fisher94-xxz}%
  \BibitemOpen
  \bibfield  {author} {\bibinfo {author} {\bibfnamefont {D.~S.}\ \bibnamefont
  {Fisher}},\ }\href {\doibase 10.1103/PhysRevB.50.3799} {\bibfield  {journal}
  {\bibinfo  {journal} {Phys. Rev. B}\ }\textbf {\bibinfo {volume} {50}},\
  \bibinfo {pages} {3799} (\bibinfo {year} {1994})}\BibitemShut {NoStop}%
\bibitem [{\citenamefont {Ma}\ \emph {et~al.}(1979)\citenamefont {Ma},
  \citenamefont {Dasgupta},\ and\ \citenamefont {Hu}}]{madasguptahu}%
  \BibitemOpen
  \bibfield  {author} {\bibinfo {author} {\bibfnamefont {S.~k.}\ \bibnamefont
  {Ma}}, \bibinfo {author} {\bibfnamefont {C.}~\bibnamefont {Dasgupta}}, \ and\
  \bibinfo {author} {\bibfnamefont {C.~k.}\ \bibnamefont {Hu}},\ }\href
  {\doibase 10.1103/PhysRevLett.43.1434} {\bibfield  {journal} {\bibinfo
  {journal} {Phys. Rev. Lett.}\ }\textbf {\bibinfo {volume} {43}},\ \bibinfo
  {pages} {1434} (\bibinfo {year} {1979})}\BibitemShut {NoStop}%
\bibitem [{\citenamefont {{C. Dasgupta}}\ and\ \citenamefont {{S.-k.
  Ma}}(1980)}]{madasgupta}%
  \BibitemOpen
  \bibfield  {author} {\bibinfo {author} {\bibnamefont {{C. Dasgupta}}}\ and\
  \bibinfo {author} {\bibnamefont {{S.-k. Ma}}},\ }\href {\doibase
  10.1103/PhysRevB.22.1305} {\bibfield  {journal} {\bibinfo  {journal} {Phys.
  Rev. B}\ }\textbf {\bibinfo {volume} {22}},\ \bibinfo {pages} {1305}
  (\bibinfo {year} {1980})}\BibitemShut {NoStop}%
\bibitem [{\citenamefont {Bhatt}\ and\ \citenamefont {Lee}(1982)}]{bhatt-lee}%
  \BibitemOpen
  \bibfield  {author} {\bibinfo {author} {\bibfnamefont {R.~N.}\ \bibnamefont
  {Bhatt}}\ and\ \bibinfo {author} {\bibfnamefont {P.~A.}\ \bibnamefont
  {Lee}},\ }\href {\doibase 10.1103/PhysRevLett.48.344} {\bibfield  {journal}
  {\bibinfo  {journal} {Phys. Rev. Lett.}\ }\textbf {\bibinfo {volume} {48}},\
  \bibinfo {pages} {344} (\bibinfo {year} {1982})}\BibitemShut {NoStop}%
\bibitem [{\citenamefont {Damle}\ and\ \citenamefont
  {Huse}(2002)}]{PhysRevLett.89.277203}%
  \BibitemOpen
  \bibfield  {author} {\bibinfo {author} {\bibfnamefont {K.}~\bibnamefont
  {Damle}}\ and\ \bibinfo {author} {\bibfnamefont {D.~A.}\ \bibnamefont
  {Huse}},\ }\href {\doibase 10.1103/PhysRevLett.89.277203} {\bibfield
  {journal} {\bibinfo  {journal} {Phys. Rev. Lett.}\ }\textbf {\bibinfo
  {volume} {89}},\ \bibinfo {pages} {277203} (\bibinfo {year}
  {2002})}\BibitemShut {NoStop}%
\bibitem [{\citenamefont {Hoyos}\ and\ \citenamefont
  {Miranda}(2004{\natexlab{a}})}]{HoyosMiranda}%
  \BibitemOpen
  \bibfield  {author} {\bibinfo {author} {\bibfnamefont {J.~A.}\ \bibnamefont
  {Hoyos}}\ and\ \bibinfo {author} {\bibfnamefont {E.}~\bibnamefont
  {Miranda}},\ }\href {\doibase 10.1103/PhysRevB.70.180401} {\bibfield
  {journal} {\bibinfo  {journal} {Phys. Rev. B}\ }\textbf {\bibinfo {volume}
  {70}},\ \bibinfo {pages} {180401} (\bibinfo {year}
  {2004}{\natexlab{a}})}\BibitemShut {NoStop}%
\bibitem [{\citenamefont {Fidkowski}\ \emph {et~al.}(2009)\citenamefont
  {Fidkowski}, \citenamefont {Lin}, \citenamefont {Titum},\ and\ \citenamefont
  {Refael}}]{fidkowski-etal-prb09}%
  \BibitemOpen
  \bibfield  {author} {\bibinfo {author} {\bibfnamefont {L.}~\bibnamefont
  {Fidkowski}}, \bibinfo {author} {\bibfnamefont {H.-H.}\ \bibnamefont {Lin}},
  \bibinfo {author} {\bibfnamefont {P.}~\bibnamefont {Titum}}, \ and\ \bibinfo
  {author} {\bibfnamefont {G.}~\bibnamefont {Refael}},\ }\href {\doibase
  10.1103/PhysRevB.79.155120} {\bibfield  {journal} {\bibinfo  {journal} {Phys.
  Rev. B}\ }\textbf {\bibinfo {volume} {79}},\ \bibinfo {pages} {155120}
  (\bibinfo {year} {2009})}\BibitemShut {NoStop}%
\bibitem [{\citenamefont {Quito}\ \emph {et~al.}(2015)\citenamefont {Quito},
  \citenamefont {Hoyos},\ and\ \citenamefont {Miranda}}]{QuitoHoyosMiranda}%
  \BibitemOpen
  \bibfield  {author} {\bibinfo {author} {\bibfnamefont {V.~L.}\ \bibnamefont
  {Quito}}, \bibinfo {author} {\bibfnamefont {J.~A.}\ \bibnamefont {Hoyos}}, \
  and\ \bibinfo {author} {\bibfnamefont {E.}~\bibnamefont {Miranda}},\ }\href
  {\doibase 10.1103/PhysRevLett.115.167201} {\bibfield  {journal} {\bibinfo
  {journal} {Phys. Rev. Lett.}\ }\textbf {\bibinfo {volume} {115}},\ \bibinfo
  {pages} {167201} (\bibinfo {year} {2015})}\BibitemShut {NoStop}%
\bibitem [{\citenamefont {Imambekov}\ \emph {et~al.}(2003)\citenamefont
  {Imambekov}, \citenamefont {Lukin},\ and\ \citenamefont
  {Demler}}]{PhysRevA.68.063602}%
  \BibitemOpen
  \bibfield  {author} {\bibinfo {author} {\bibfnamefont {A.}~\bibnamefont
  {Imambekov}}, \bibinfo {author} {\bibfnamefont {M.}~\bibnamefont {Lukin}}, \
  and\ \bibinfo {author} {\bibfnamefont {E.}~\bibnamefont {Demler}},\ }\href
  {\doibase 10.1103/PhysRevA.68.063602} {\bibfield  {journal} {\bibinfo
  {journal} {Phys. Rev. A}\ }\textbf {\bibinfo {volume} {68}},\ \bibinfo
  {pages} {063602} (\bibinfo {year} {2003})}\BibitemShut {NoStop}%
\bibitem [{\citenamefont {Wu}\ \emph {et~al.}(2003)\citenamefont {Wu},
  \citenamefont {Hu},\ and\ \citenamefont {Zhang}}]{wuetal2003}%
  \BibitemOpen
  \bibfield  {author} {\bibinfo {author} {\bibfnamefont {C.}~\bibnamefont
  {Wu}}, \bibinfo {author} {\bibfnamefont {J.-p.}\ \bibnamefont {Hu}}, \ and\
  \bibinfo {author} {\bibfnamefont {S.-c.}\ \bibnamefont {Zhang}},\ }\href
  {\doibase 10.1103/PhysRevLett.91.186402} {\bibfield  {journal} {\bibinfo
  {journal} {Phys. Rev. Lett.}\ }\textbf {\bibinfo {volume} {91}},\ \bibinfo
  {pages} {186402} (\bibinfo {year} {2003})}\BibitemShut {NoStop}%
\bibitem [{\citenamefont {Garc\'{\i}a-Ripoll}\ \emph
  {et~al.}(2004)\citenamefont {Garc\'{\i}a-Ripoll}, \citenamefont
  {Martin-Delgado},\ and\ \citenamefont {Cirac}}]{PhysRevLett.93.250405}%
  \BibitemOpen
  \bibfield  {author} {\bibinfo {author} {\bibfnamefont {J.~J.}\ \bibnamefont
  {Garc\'{\i}a-Ripoll}}, \bibinfo {author} {\bibfnamefont {M.~A.}\ \bibnamefont
  {Martin-Delgado}}, \ and\ \bibinfo {author} {\bibfnamefont {J.~I.}\
  \bibnamefont {Cirac}},\ }\href {\doibase 10.1103/PhysRevLett.93.250405}
  {\bibfield  {journal} {\bibinfo  {journal} {Phys. Rev. Lett.}\ }\textbf
  {\bibinfo {volume} {93}},\ \bibinfo {pages} {250405} (\bibinfo {year}
  {2004})}\BibitemShut {NoStop}%
\bibitem [{\citenamefont {Wu}(2006)}]{Wumodphys06}%
  \BibitemOpen
  \bibfield  {author} {\bibinfo {author} {\bibfnamefont {C.}~\bibnamefont
  {Wu}},\ }\href {\doibase 10.1142/S0217984906012213} {\bibfield  {journal}
  {\bibinfo  {journal} {Mod. Phys. Lett. B}\ }\textbf {\bibinfo {volume}
  {20}},\ \bibinfo {pages} {1707} (\bibinfo {year} {2006})}\BibitemShut
  {NoStop}%
\bibitem [{\citenamefont {Cazalilla}\ \emph {et~al.}(2009)\citenamefont
  {Cazalilla}, \citenamefont {Ho},\ and\ \citenamefont
  {Ueda}}]{cazalillaetal2009}%
  \BibitemOpen
  \bibfield  {author} {\bibinfo {author} {\bibfnamefont {M.~A.}\ \bibnamefont
  {Cazalilla}}, \bibinfo {author} {\bibfnamefont {A.~F.}\ \bibnamefont {Ho}}, \
  and\ \bibinfo {author} {\bibfnamefont {M.}~\bibnamefont {Ueda}},\ }\href
  {\doibase 10.1088/1367-2630/11/10/103033} {\bibfield  {journal} {\bibinfo
  {journal} {New J. Phys.}\ }\textbf {\bibinfo {volume} {11}},\ \bibinfo
  {pages} {103033} (\bibinfo {year} {2009})}\BibitemShut {NoStop}%
\bibitem [{\citenamefont {Gorshkov}\ \emph {et~al.}(2010)\citenamefont
  {Gorshkov}, \citenamefont {Hermele}, \citenamefont {Gurarie}, \citenamefont
  {Xu}, \citenamefont {Julienne}, \citenamefont {Ye}, \citenamefont {Zoller},
  \citenamefont {Demler}, \citenamefont {Lukin},\ and\ \citenamefont
  {Rey}}]{gorschkovetal2010}%
  \BibitemOpen
  \bibfield  {author} {\bibinfo {author} {\bibfnamefont {A.~V.}\ \bibnamefont
  {Gorshkov}}, \bibinfo {author} {\bibfnamefont {M.}~\bibnamefont {Hermele}},
  \bibinfo {author} {\bibfnamefont {V.}~\bibnamefont {Gurarie}}, \bibinfo
  {author} {\bibfnamefont {C.}~\bibnamefont {Xu}}, \bibinfo {author}
  {\bibfnamefont {P.~S.}\ \bibnamefont {Julienne}}, \bibinfo {author}
  {\bibfnamefont {J.}~\bibnamefont {Ye}}, \bibinfo {author} {\bibfnamefont
  {P.}~\bibnamefont {Zoller}}, \bibinfo {author} {\bibfnamefont
  {E.}~\bibnamefont {Demler}}, \bibinfo {author} {\bibfnamefont {M.~D.}\
  \bibnamefont {Lukin}}, \ and\ \bibinfo {author} {\bibfnamefont {A.~M.}\
  \bibnamefont {Rey}},\ }\href {\doibase 10.1038/nphys1535} {\bibfield
  {journal} {\bibinfo  {journal} {Nature Phys.}\ }\textbf {\bibinfo {volume}
  {6}},\ \bibinfo {pages} {289} (\bibinfo {year} {2010})}\BibitemShut {NoStop}%
\bibitem [{\citenamefont {Zhang}\ \emph {et~al.}(2014)\citenamefont {Zhang},
  \citenamefont {Bishof}, \citenamefont {Bromley}, \citenamefont {Kraus},
  \citenamefont {Safronova}, \citenamefont {Zoller}, \citenamefont {Rey},\ and\
  \citenamefont {Ye}}]{zhangetal2014}%
  \BibitemOpen
  \bibfield  {author} {\bibinfo {author} {\bibfnamefont {X.}~\bibnamefont
  {Zhang}}, \bibinfo {author} {\bibfnamefont {M.}~\bibnamefont {Bishof}},
  \bibinfo {author} {\bibfnamefont {S.~L.}\ \bibnamefont {Bromley}}, \bibinfo
  {author} {\bibfnamefont {C.~V.}\ \bibnamefont {Kraus}}, \bibinfo {author}
  {\bibfnamefont {M.~S.}\ \bibnamefont {Safronova}}, \bibinfo {author}
  {\bibfnamefont {P.}~\bibnamefont {Zoller}}, \bibinfo {author} {\bibfnamefont
  {A.~M.}\ \bibnamefont {Rey}}, \ and\ \bibinfo {author} {\bibfnamefont
  {J.}~\bibnamefont {Ye}},\ }\href {\doibase 10.1126/science.1254978}
  {\bibfield  {journal} {\bibinfo  {journal} {Science}\ }\textbf {\bibinfo
  {volume} {345}},\ \bibinfo {pages} {1467} (\bibinfo {year}
  {2014})}\BibitemShut {NoStop}%
\bibitem [{\citenamefont {Scazza}\ \emph {et~al.}(2014)\citenamefont {Scazza},
  \citenamefont {Hofrichter}, \citenamefont {H\"{o}fer}, \citenamefont
  {De~Groot}, \citenamefont {Bloch},\ and\ \citenamefont
  {S.~F\"{o}lling}}]{scazzaetal2014}%
  \BibitemOpen
  \bibfield  {author} {\bibinfo {author} {\bibfnamefont {F.}~\bibnamefont
  {Scazza}}, \bibinfo {author} {\bibfnamefont {C.}~\bibnamefont {Hofrichter}},
  \bibinfo {author} {\bibfnamefont {M.}~\bibnamefont {H\"{o}fer}}, \bibinfo
  {author} {\bibfnamefont {P.~C.}\ \bibnamefont {De~Groot}}, \bibinfo {author}
  {\bibfnamefont {I.}~\bibnamefont {Bloch}}, \ and\ \bibinfo {author}
  {\bibfnamefont {S.}~\bibnamefont {S.~F\"{o}lling}},\ }\href {\doibase
  10.1038/nphys3061} {\bibfield  {journal} {\bibinfo  {journal} {Nature Phys.}\
  }\textbf {\bibinfo {volume} {10}},\ \bibinfo {pages} {779} (\bibinfo {year}
  {2014})}\BibitemShut {NoStop}%
\bibitem [{\citenamefont {Taie}\ \emph {et~al.}(2012)\citenamefont {Taie},
  \citenamefont {Yamazaki}, \citenamefont {Sugawa},\ and\ \citenamefont
  {Takahashi}}]{taieetal2012}%
  \BibitemOpen
  \bibfield  {author} {\bibinfo {author} {\bibfnamefont {S.}~\bibnamefont
  {Taie}}, \bibinfo {author} {\bibfnamefont {R.}~\bibnamefont {Yamazaki}},
  \bibinfo {author} {\bibfnamefont {S.}~\bibnamefont {Sugawa}}, \ and\ \bibinfo
  {author} {\bibfnamefont {Y.}~\bibnamefont {Takahashi}},\ }\href {\doibase
  10.1038/nphys2430} {\bibfield  {journal} {\bibinfo  {journal} {Nature Phys.}\
  }\textbf {\bibinfo {volume} {8}},\ \bibinfo {pages} {825} (\bibinfo {year}
  {2012})}\BibitemShut {NoStop}%
\bibitem [{\citenamefont {{E. Westerberg}}\ \emph {et~al.}(1997)\citenamefont
  {{E. Westerberg}}, \citenamefont {{A. Furusaki}}, \citenamefont {{M.
  Sigrist}},\ and\ \citenamefont {{P. A. Lee}}}]{westerbergetal}%
  \BibitemOpen
  \bibfield  {author} {\bibinfo {author} {\bibnamefont {{E. Westerberg}}},
  \bibinfo {author} {\bibnamefont {{A. Furusaki}}}, \bibinfo {author}
  {\bibnamefont {{M. Sigrist}}}, \ and\ \bibinfo {author} {\bibnamefont {{P. A.
  Lee}}},\ }\href {\doibase 10.1103/PhysRevB.55.12578} {\bibfield  {journal}
  {\bibinfo  {journal} {Phys. Rev. B}\ }\textbf {\bibinfo {volume} {55}},\
  \bibinfo {pages} {12578} (\bibinfo {year} {1997})}\BibitemShut {NoStop}%
\bibitem [{\citenamefont {Yang}\ and\ \citenamefont {Bhatt}(1998)}]{YangBhatt}%
  \BibitemOpen
  \bibfield  {author} {\bibinfo {author} {\bibfnamefont {K.}~\bibnamefont
  {Yang}}\ and\ \bibinfo {author} {\bibfnamefont {R.~N.}\ \bibnamefont
  {Bhatt}},\ }\href {\doibase 10.1103/PhysRevLett.80.4562} {\bibfield
  {journal} {\bibinfo  {journal} {Phys. Rev. Lett.}\ }\textbf {\bibinfo
  {volume} {80}},\ \bibinfo {pages} {4562} (\bibinfo {year}
  {1998})}\BibitemShut {NoStop}%
\bibitem [{\citenamefont {Edmonds}(1996)}]{Edmondsbook}%
  \BibitemOpen
  \bibfield  {author} {\bibinfo {author} {\bibfnamefont {A.~R.}\ \bibnamefont
  {Edmonds}},\ }\href@noop {} {\emph {\bibinfo {title} {Angular Momentum in
  Quantum Mechanics}}}\ (\bibinfo  {publisher} {Princeton University Press},\
  \bibinfo {year} {1996})\BibitemShut {NoStop}%
\bibitem [{\citenamefont {Ciobanu}\ \emph {et~al.}(2000)\citenamefont
  {Ciobanu}, \citenamefont {Yip},\ and\ \citenamefont {Ho}}]{CiobanuPRA2000}%
  \BibitemOpen
  \bibfield  {author} {\bibinfo {author} {\bibfnamefont {C.~V.}\ \bibnamefont
  {Ciobanu}}, \bibinfo {author} {\bibfnamefont {S.-K.}\ \bibnamefont {Yip}}, \
  and\ \bibinfo {author} {\bibfnamefont {T.-L.}\ \bibnamefont {Ho}},\ }\href
  {\doibase 10.1103/PhysRevA.61.033607} {\bibfield  {journal} {\bibinfo
  {journal} {Phys. Rev. A}\ }\textbf {\bibinfo {volume} {61}},\ \bibinfo
  {pages} {033607} (\bibinfo {year} {2000})}\BibitemShut {NoStop}%
\bibitem [{\citenamefont {Alexander}\ and\ \citenamefont
  {Bernasconi}(1979)}]{alexander-bernasconi-jpc79}%
  \BibitemOpen
  \bibfield  {author} {\bibinfo {author} {\bibfnamefont {S.}~\bibnamefont
  {Alexander}}\ and\ \bibinfo {author} {\bibfnamefont {J.}~\bibnamefont
  {Bernasconi}},\ }\href@noop {} {\bibfield  {journal} {\bibinfo  {journal}
  {Journal of Physics C: Solid State Physics}\ }\textbf {\bibinfo {volume}
  {12}},\ \bibinfo {pages} {L1} (\bibinfo {year} {1979})}\BibitemShut {NoStop}%
\bibitem [{\citenamefont {Ziman}(1982)}]{ziman-prl82}%
  \BibitemOpen
  \bibfield  {author} {\bibinfo {author} {\bibfnamefont {T.~A.~L.}\
  \bibnamefont {Ziman}},\ }\href {\doibase 10.1103/PhysRevLett.49.337}
  {\bibfield  {journal} {\bibinfo  {journal} {Phys. Rev. Lett.}\ }\textbf
  {\bibinfo {volume} {49}},\ \bibinfo {pages} {337} (\bibinfo {year}
  {1982})}\BibitemShut {NoStop}%
\bibitem [{\citenamefont {Theodorou}(1982)}]{theodorou-jpc82}%
  \BibitemOpen
  \bibfield  {author} {\bibinfo {author} {\bibfnamefont {G.}~\bibnamefont
  {Theodorou}},\ }\href@noop {} {\bibfield  {journal} {\bibinfo  {journal}
  {Journal of Physics C: Solid State Physics}\ }\textbf {\bibinfo {volume}
  {15}},\ \bibinfo {pages} {L1315} (\bibinfo {year} {1982})}\BibitemShut
  {NoStop}%
\bibitem [{\citenamefont {Cieplak}\ and\ \citenamefont
  {Ismail}(1987)}]{cieplak-ismail-jpc87}%
  \BibitemOpen
  \bibfield  {author} {\bibinfo {author} {\bibfnamefont {M.}~\bibnamefont
  {Cieplak}}\ and\ \bibinfo {author} {\bibfnamefont {G.}~\bibnamefont
  {Ismail}},\ }\href@noop {} {\bibfield  {journal} {\bibinfo  {journal}
  {Journal of Physics C: Solid State Physics}\ }\textbf {\bibinfo {volume}
  {20}},\ \bibinfo {pages} {1309} (\bibinfo {year} {1987})}\BibitemShut
  {NoStop}%
\bibitem [{\citenamefont {Evangelou}\ and\ \citenamefont
  {Katsanos}(1992)}]{evangelou-katsanos-pla92}%
  \BibitemOpen
  \bibfield  {author} {\bibinfo {author} {\bibfnamefont {S.}~\bibnamefont
  {Evangelou}}\ and\ \bibinfo {author} {\bibfnamefont {D.}~\bibnamefont
  {Katsanos}},\ }\href {\doibase
  http://dx.doi.org/10.1016/0375-9601(92)90114-2} {\bibfield  {journal}
  {\bibinfo  {journal} {Physics Letters A}\ }\textbf {\bibinfo {volume}
  {164}},\ \bibinfo {pages} {456 } (\bibinfo {year} {1992})}\BibitemShut
  {NoStop}%
\bibitem [{\citenamefont {Hikihara}\ \emph {et~al.}(1999)\citenamefont
  {Hikihara}, \citenamefont {Furusaki},\ and\ \citenamefont
  {Sigrist}}]{hikihara}%
  \BibitemOpen
  \bibfield  {author} {\bibinfo {author} {\bibfnamefont {T.}~\bibnamefont
  {Hikihara}}, \bibinfo {author} {\bibfnamefont {A.}~\bibnamefont {Furusaki}},
  \ and\ \bibinfo {author} {\bibfnamefont {M.}~\bibnamefont {Sigrist}},\ }\href
  {\doibase 10.1103/PhysRevB.60.12116} {\bibfield  {journal} {\bibinfo
  {journal} {Phys. Rev. B}\ }\textbf {\bibinfo {volume} {60}},\ \bibinfo
  {pages} {12116} (\bibinfo {year} {1999})}\BibitemShut {NoStop}%
\bibitem [{\citenamefont {Hoyos}(2008)}]{hoyos08}%
  \BibitemOpen
  \bibfield  {author} {\bibinfo {author} {\bibfnamefont {J.~A.}\ \bibnamefont
  {Hoyos}},\ }\href {\doibase 10.1103/PhysRevE.78.032101} {\bibfield  {journal}
  {\bibinfo  {journal} {Phys. Rev. E}\ }\textbf {\bibinfo {volume} {78}},\
  \bibinfo {pages} {032101} (\bibinfo {year} {2008})}\BibitemShut {NoStop}%
\bibitem [{\citenamefont {Refael}\ \emph {et~al.}(2002)\citenamefont {Refael},
  \citenamefont {Kehrein},\ and\ \citenamefont {Fisher}}]{RefaelFisher}%
  \BibitemOpen
  \bibfield  {author} {\bibinfo {author} {\bibfnamefont {G.}~\bibnamefont
  {Refael}}, \bibinfo {author} {\bibfnamefont {S.}~\bibnamefont {Kehrein}}, \
  and\ \bibinfo {author} {\bibfnamefont {D.~S.}\ \bibnamefont {Fisher}},\
  }\href {\doibase 10.1103/PhysRevB.66.060402} {\bibfield  {journal} {\bibinfo
  {journal} {Phys. Rev. B}\ }\textbf {\bibinfo {volume} {66}},\ \bibinfo
  {pages} {060402} (\bibinfo {year} {2002})}\BibitemShut {NoStop}%
\bibitem [{\citenamefont {Saguia}\ \emph {et~al.}(2003)\citenamefont {Saguia},
  \citenamefont {Boechat},\ and\ \citenamefont {Continentino}}]{Saguia2003}%
  \BibitemOpen
  \bibfield  {author} {\bibinfo {author} {\bibfnamefont {A.}~\bibnamefont
  {Saguia}}, \bibinfo {author} {\bibfnamefont {B.}~\bibnamefont {Boechat}}, \
  and\ \bibinfo {author} {\bibfnamefont {M.~A.}\ \bibnamefont {Continentino}},\
  }\href {\doibase 10.1103/PhysRevB.68.020403} {\bibfield  {journal} {\bibinfo
  {journal} {Phys. Rev. B}\ }\textbf {\bibinfo {volume} {68}},\ \bibinfo
  {pages} {020403} (\bibinfo {year} {2003})}\BibitemShut {NoStop}%
\bibitem [{\citenamefont {Carlon}\ \emph {et~al.}(2004)\citenamefont {Carlon},
  \citenamefont {Lajk\'o}, \citenamefont {Rieger},\ and\ \citenamefont
  {Igl\'oi}}]{Carlon2004}%
  \BibitemOpen
  \bibfield  {author} {\bibinfo {author} {\bibfnamefont {E.}~\bibnamefont
  {Carlon}}, \bibinfo {author} {\bibfnamefont {P.}~\bibnamefont {Lajk\'o}},
  \bibinfo {author} {\bibfnamefont {H.}~\bibnamefont {Rieger}}, \ and\ \bibinfo
  {author} {\bibfnamefont {F.}~\bibnamefont {Igl\'oi}},\ }\href {\doibase
  10.1103/PhysRevB.69.144416} {\bibfield  {journal} {\bibinfo  {journal} {Phys.
  Rev. B}\ }\textbf {\bibinfo {volume} {69}},\ \bibinfo {pages} {144416}
  (\bibinfo {year} {2004})}\BibitemShut {NoStop}%
\bibitem [{\citenamefont {Jones}(1998)}]{Jonesbook}%
  \BibitemOpen
  \bibfield  {author} {\bibinfo {author} {\bibfnamefont {H.}~\bibnamefont
  {Jones}},\ }\href@noop {} {\emph {\bibinfo {title} {Groups, Representations
  and Physics}}}\ (\bibinfo  {publisher} {CRC Press},\ \bibinfo {year}
  {1998})\BibitemShut {NoStop}%
\bibitem [{\citenamefont {Getelina}\ \emph {et~al.}(2016)\citenamefont
  {Getelina}, \citenamefont {Alcaraz},\ and\ \citenamefont
  {Hoyos}}]{getelina2015}%
  \BibitemOpen
  \bibfield  {author} {\bibinfo {author} {\bibfnamefont {J.~C.}\ \bibnamefont
  {Getelina}}, \bibinfo {author} {\bibfnamefont {F.~C.}\ \bibnamefont
  {Alcaraz}}, \ and\ \bibinfo {author} {\bibfnamefont {J.~A.}\ \bibnamefont
  {Hoyos}},\ }\href {\doibase 10.1103/PhysRevB.93.045136} {\bibfield  {journal}
  {\bibinfo  {journal} {Phys. Rev. B}\ }\textbf {\bibinfo {volume} {93}},\
  \bibinfo {pages} {045136} (\bibinfo {year} {2016})}\BibitemShut {NoStop}%
\bibitem [{\citenamefont {Hoyos}\ and\ \citenamefont
  {Miranda}(2004{\natexlab{b}})}]{Hoyosladders}%
  \BibitemOpen
  \bibfield  {author} {\bibinfo {author} {\bibfnamefont {J.~A.}\ \bibnamefont
  {Hoyos}}\ and\ \bibinfo {author} {\bibfnamefont {E.}~\bibnamefont
  {Miranda}},\ }\href {\doibase 10.1103/PhysRevB.69.214411} {\bibfield
  {journal} {\bibinfo  {journal} {Phys. Rev. B}\ }\textbf {\bibinfo {volume}
  {69}},\ \bibinfo {pages} {214411} (\bibinfo {year}
  {2004}{\natexlab{b}})}\BibitemShut {NoStop}%
\end{thebibliography}
\end{document}